\documentclass[prd,amsmath,amsfonts,floatfix,nofootinbib,preprintnumbers]{revtex4}

\usepackage{graphicx}
\usepackage{slashed}
\usepackage{color}
\usepackage{amsmath, amssymb}
\usepackage{amssymb}

\def\beq{\begin{equation}}
\def\eeq{\end{equation}}
\def\bea{\begin{eqnarray}}
\def\eea{\end{eqnarray}}
\def\barr{\begin{array}}
\def\earr{\end{array}}

\def\exponential{{\mathrm{e}}}

\begin{document}

\begin{titlepage}


\vskip 0.5cm

\begin{center}
{\Large \bf Lattice study of infrared behaviour in SU(3) gauge
  theory\\with twelve massless flavours} 
\vskip1cm {\large\bf 
C.-J.~David~Lin$^{a,b}$,
Kenji~Ogawa$^{a}$,
Hiroshi~Ohki$^{c}$,
Eigo~Shintani$^{d}$
}\\ \vspace{.5cm}
{\normalsize {\sl 
$^{a}$ Institute of Physics, National Chiao-Tung University, Hsinchu 300, Taiwan\\
$^{b}$ Division of Physics, National Centre for Theoretical Sciences, Hsinchu 300, Taiwan\\
$^{c}$ Kobayashi-Maskawa Institute for the Origin of Particles and the Universe (KMI), Nagoya University, Nagoya, Aichi 464-8602, Japan\\
$^{d}$ RIKEN-BNL Research Centre, Brookhaven National Laboratory, Upton, NY 11973, USA
}}

\vskip1.0cm {\large\bf Abstract:\\[10pt]} \parbox[t]{\textwidth}{{
We present details of a lattice study of infrared behaviour in SU(3) gauge theory
with twelve massless fermions in the fundamental representation.
Using the step-scaling method, we compute the coupling constant in
this theory over a large range of scale.  The renormalisation scheme
in this work is defined by the ratio of Polyakov loops in the
directions with different boundary conditions.  We closely examine 
systematic effects, and find that they are dominated by
errors arising from the continuum extrapolation.  Our
investigation suggests that SU(3) gauge theory with twelve flavours
contains an infrared fixed point.}}
\end{center}
\end{titlepage}

\section{Introduction}
The origin of electroweak (EW) symmetry breaking is
one of the most important research topics in physics.  With the
progress of experiments at the Large Hadron Collider (LHC), it is
urgent for a theoretical understanding for the mechanism of the mass
generation and its relation to EW symmetry breaking.  One appealing scenario for this mechanism is the technicolour
models~\cite{Weinberg:1975gm,Susskind:1978ms}.  These models involve new
asymptotically-free gauge theories in which the coupling constants
become strong at the TeV scale.  The strong coupling can induce
condensates to generate mass gaps, and asymptotic freedom leads to
the absence of the naturalness problem.   In order to dynamically suppress the
flavour-changing neutral currents (FCNC), and to evade the constraints
from precision EW data, it is important that the candidate theories
exhibit the ``walking'' (quasi-conformal)  behaviour and contain large anomalous
dimension for the technifermion mass term~\cite{Holdom:1984sk,Yamawaki:1985zg,Appelquist:1986an}.

In recent years, there has been a significant amount of work in search
of gauge theories viable for walking-technicolour model building.   The most
important task in this endeavour is the determination of the critical
number of massless fermions, given the gauge group and the fermion
representation, above which a theory is conformal in the infrared
(IR).  For theories involving fermions in the fundamental
representation, this is denoted as the critical number of flavours,
$N_{f}^{\rm cr}$.  For $N_{f}^{\rm cr} \le N_{f} < N_{f}^{\rm AF}$
($N_{f}^{\rm AF}$ is the number of flavours above which asymptotic
freedom is lost), the theory contains an infrared fixed point (IRFP).
A candidate walking-technicolour theory with fundamental fermions is
believed to have the number of flavours just below $N_{f}^{\rm cr}$.
This makes the determination of $N_{f}^{\rm cr}$ a task with
phenomenological significance, in addition to its importance in
field-theoretic studies.
Since the couplings must be strong at low energies in these theories, nonperturbative
methods, such as the Schwinger-Dyson equation and gauge-gravity
duality, have to be employed.  Amongst these, lattice gauge theory is
the only first-principle tool, and has been applied by many groups in
this research
avenue~\cite{Aoyama:2011ry,Appelquist:2007hu,Appelquist:2009ty,Appelquist:2009ka,Appelquist:2011dp,Appelquist:2012sm,Appelquist:2012nz,Bursa:2009we,Bursa:2010xn,Bursa:2011ru,Catterall:2007yx,Catterall:2008qk,Catterall:2011zf,Cheng:2011ic,deForcrand:2012se,DeGrand:2008kx,DeGrand:2009mt,DeGrand:2009hu,DeGrand:2010na,DeGrand:2011qd,DeGrand:2011cu,DelDebbio:2009fd,DelDebbio:2010hx,DelDebbio:2010hu,DelDebbio:2010ze,DelDebbio:2010jy,Deuzeman:2008sc,Deuzeman:2009mh,Fodor:2009wk,Fodor:2009ar,Fodor:2011tw,Fodor:2011tu,Giedt:2011kz,Giedt:2012rj,Hasenfratz:2010fi,Hasenfratz:2011xn,Hayakawa:2010yn,Hietanen:2008mr,Hietanen:2009az,Karavirta:2011zg,Miura:2011mc,Shamir:2008pb}\footnote{There
have also been many works on walking-technicolour model building using
the gauge/gravity duality, as reviewed in Ref.~\cite{Piai:2010ma}.}.  

Of all the theories which have been investigated using the lattice
technique, the value of $N_{f}^{\rm cr}$ for SU(3) gauge theories with
fundamental-representation fermions remains a controversy.  Although
several
groups~\cite{Appelquist:2007hu,Appelquist:2009ty,Appelquist:2009ka,Appelquist:2011dp,Appelquist:2012sm,DeGrand:2011cu,Deuzeman:2009mh,Miura:2011mc}
found evidence that SU(3) gauge theory with $N_{f}=12$ is conformal in
the IR, authors of
Refs.~\cite{Fodor:2009wk,Fodor:2011tu} argued that chiral symmetry
is broken in this theory.  In this paper, we report our study of
this theory, using the step-scaling method to compute the running
coupling constant.  We adopt the 
Twisted Polyakov Loop (TPL)
scheme~\cite{Luscher:1985wf,deDivitiis:1993hj,deDivitiis:1994yp}.
This article complements the letter~\cite{Aoyama:2011ry} which was 
released in 2011 with other colleagues on this collaboration, and contains more details
of our simulations and improved analysis using more data.  
In Ref.~\cite{Aoyama:2011ry}, we concentrated on the
analysis with the step size, $s$, set to 1.5, while here we emphasise the
case in which $s$ equals two.   Furthermore, in the improved analysis
with new data, as
presented in this paper, we significantly reduce the correlation
between data for the step-scaling functions on different lattice volumes.
This makes the continuum extrapolation simpler and better controlled,
compared to the analysis published in Ref.~\cite{Aoyama:2011ry}.
We will discuss this in detail in Sec.~\ref{sec:L7_interpolation}.
Related to this work and Ref.~\cite{Aoyama:2011ry},
we have also published conference
proceedings~\cite{Bilgici:2009nm,Itou:2010we,Ogawa:2011ki}, 
as well as for a similar
project on SU(2) gauge theory with eight flavours~\cite{Ohki:2010sr}.

In addition to computing the running coupling constant, we also obtain
the ratio between the step-scaling function and the coupling
constant, which becomes one when the $\beta{-}$function is zero.
To claim the discovery of the IRFP in an asymptotically-free gauge
theory, we have to demonstrate that this ratio is indeed one in the
ultraviolet (UV)
and the IR,
while being obviously different from this value between these two regimes.  Our
study suggests that SU(3) gauge theory with $N_{f}=12$ contains an 
IRFP around the TPL-scheme coupling constant,
%
\beq
\label{eq:g_star_intro}
g^{2}_{\ast} \sim 2.0 .
\eeq
%
%
%
Amongst systematic effects that we estimate, 
errors arising from the continuum extrapolation dominate.  We also
notice that some of our procedures in performing this extrapolation
lead to weaker evidence for the existence of the IRFP.
Details of the estimation of
systematic errors will be presented in Secs.~\ref{sec:analysis} and 
\ref{sec:final_results}.

Our finding for the evidence of the existence of the IRFP agrees with the result of
Refs.~\cite{Appelquist:2007hu,Appelquist:2009ty}, where the
Schr\"{o}dinger-functional (SF) scheme~\cite{Luscher:1992zx,Luscher:1993gh} was used in defining the
coupling constant, and the
calculation was performed using the same gauge and fermion actions.  
The values of $g_{\ast}$ are different because of scheme dependence.   
Here we also stress that the lattice artefacts can be very different in
these two schemes.  In particular, the SF scheme contains $O(a)$ ($a$ is the
lattice spacing) lattice artefacts through the introduction of the
boundary terms\footnote{In Refs.~\cite{Appelquist:2007hu,Appelquist:2009ty},
  it was found that such $O(a)$ lattice artefacts can be numerically
very small.}, while in the TPL scheme the lattice artefacts remain
of $O(a^{2})$, making the continuum extrapolation more
reliable.

This paper is organised in the following way.
In Sec.~\ref{sec:twbc}, we review twisted boundary conditions and the Twisted Polyakov Loop scheme.
Section~\ref{sec:simulation_setting} contains the details of our
simulation strategy and parameters.  We describe our analysis
procedure in Sec.~\ref{sec:analysis}, give our results and discussion
in Sec.~\ref{sec:final_results}, and conclude in
Sec.~\ref{sec:conclusion}.  Appendix~\ref{sec:eigenvalues} contains the study of the
eigenvalue spectrum of the Dirac operator used in this work.  Values of plaquette and
the raw data for the TPL-scheme coupling constants are presented in App.~\ref{sec:plaquette_values_raw_data}.
\section{Twisted Polyakov Loop Scheme}
\label{sec:twbc}
In this section, we give the details of
our definition of the renormalised coupling constant in the
twisted-Polyakov-loop (TPL) scheme~\cite{deDivitiis:1993hj,deDivitiis:1994yp}.   This scheme makes use of
twisted boundary condition (TBC)~\cite{'tHooft:1979uj}, which is implemented on the link
variables, $U_{\mu}(\hat{n})$ ($\mu = x,y,z,t$ is the Lorentz index and
$\hat{n}$ is the position of a lattice site),  through
\beq
\label{eq:TBC_gauge}
U_\mu( \hat{n} + \hat \nu L_\nu/a )
= \Omega_\nu U_\mu( \hat{n} ) \Omega_\nu^\dagger , 
\eeq
where $L_{\nu}$ is the (dimensionful) box size in the $\nu$ direction
(with $\hat{\nu}$ denoting the unit vector),
and $a$ is the lattice spacing.  The ``twisting matrices'',
$\Omega_{\nu}$, act in the colour space.  In this work, we apply
TBC for $\nu = x,y$, while maintaining periodic boundary
condition (PBC) for the other two directions.  This means
\beq
 \Omega_{z} = \Omega_{t} = {\mathbf{1}} .
\eeq
For SU(3), the twisting matrices, $\Omega_{x,y}$, satisfy
\bea
&& \Omega_{x} \Omega_{y}
= \exponential^{i2\pi/3} \Omega_{y} \Omega_{x} ,
\nonumber \\
\label{eq:omega_12_relations}
&& 
\Omega_\mu \Omega_\mu^\dagger = 1 , \mbox{ }
\left ( \Omega_\mu \right )^3 = 1 , \mbox{ }
\mathrm{Tr} 
\left [ \Omega_\mu \right ] = 0 , \mbox{ } {\mathrm{for}} \mbox{ } 
\mu = x,y .
\eea
In this work, we explicitly implement~\cite{Trottier:2001vj}
\beq
\label{eq:omega_12_explicit}
\Omega_{x} = \left ( \begin{array}{ccc} 0 & 1 & 0\\
                                                                   0 & 0 & 1\\
                                                                   1 & 0 & 0\\
                                     \end{array} \right ) ,\mbox{ }
\Omega_{y} = \left ( \begin{array}{ccc} \exponential^{-2 \pi i /3} & 0 & 0\\
                                        0 & \exponential^{2 \pi i /3} & 0\\
                                                                  0 & 0 & 1\\
                                     \end{array} \right ) .
\eeq
%
%
%

The inclusion of fermions is not straightforward when TBC,
Eq.~(\ref{eq:TBC_gauge}), is imposed on gauge fields.  In order to
maintain gauge invariance and single-valuedness of the fermion field,
$\psi(\hat{n}+\hat{x}L_{x}/a +\hat{y} L_{y}/a)$, under the application
of two boundary twistings (in $\hat{x}$ and $\hat{y}$ directions) with
different orderings, it is necessary  to introduce the ``smell''
degrees of freedom~\cite{Parisi:1984cy}.  This quantum number is carried by fermions.
The number of smells, $N_{s}$, is equal to the
number of colours, $N_{c}$.
Twisted boundary condition on fermion fields is given by,
\beq
\psi^a_\alpha( \hat{n} + \hat{\nu} L_{\nu}/a) = e^{i\pi/3} \Omega_\nu^{ab}
\psi^b_\beta (\hat{n}) \left ( \Omega_\nu \right )^\dagger_{\beta
  \alpha} ,
\eeq
where $a$ and $b$ are colour indices, and the twisting matrices, $\Omega_{\mu}$, have been generalised to
act on the smell degrees of freedom (indices $\alpha$ and $\beta$).
The factor $\exponential^{i\pi/3}$ is introduced only for $\nu= x,y$, to remove the zero-momentum modes in these directions.
For $\nu = z, t$ directions, we implement ordinary PBC, $\psi(
\hat{n}+ \hat{\nu} L_{\nu}/a ) = \psi( \hat{n} )$. 
Since the smell quantum number is not carried by the gauge fields, it can be considered as additional flavours.
Therefore the number of flavours in simulations involving dynamical fermions with TBC has to be a multiple of $N_{s}$ (=$N_{c}$).

The Polyakov loops in the twisted directions, $\nu = x, y$, are
\bea
P_{x} ( \hat{n}_{x} , \hat{n}_{y} , \hat{n}_{t} )
&=&
\mathrm{Tr}
\left ( 
\left [
\prod_j
U_{x}
\left (
\hat{n}_{x} = j , \hat{n}_{y} , \hat{n}_{z} , \hat{n}_{t}
\right )
\right ]
\Omega_{x} \exponential^{i 2 \pi \hat{n}_{y} a / ( 3 L_{y} ) }
\right ) ,
\nonumber\\
P_{y} ( \hat{n}_{x} , \hat{n}_{z} , \hat{n}_{t} )
&=&
\mathrm{Tr}
\left ( 
\left [
\prod_j
U_{y}
\left (
\hat{n}_{x}, \hat{n}_{y} = j, \hat{n}_{z} , \hat{n}_{t}
\right )
\right ]
\Omega_{y} \exponential^{i 2 \pi \hat{n}_{x} a / ( 3 L_{x} ) }
\right ) .
\label{eq:PL_twisted}
\eea
The extra factors outside the square brackets are introduced to maintain gauge and translation invariance.
The renormalised coupling constant can be defined via the ratio between correlators of Polyakov loops in the twisted and periodic directions,
\beq
\label{eq:TPL_scheme_def}
\frac{ 
\langle 
P_{x}( \hat{n}_{t} = 0 )^\dagger P_{x}( \hat{n}_{t} = L_{t}/(2 a) ) 
\rangle
}{ 
\langle 
P_{z}( \hat{n}_{t} = 0 )^\dagger P_{z}( \hat{n}_{t} = L_{t}/(2 a))
\rangle
} = k \bar{g}^{2}_{\rm latt},
\eeq
where $P_{z,t}$ are ordinary Polyakov loops in the directions with PBC.
In this study, we always use hypercubic lattice $L_x/a=L_y/a=L_z/a=L_t/a=L/a$.
The proportionality factor $k$ can be extracted by computing the above ratio in perturbation theory to $O(g^{2}_{0})$,
where $g_{0}$ is the bare coupling constant. Using lattice perturbation theory, one obtains the lattice 
version of this factor~\cite{Aoyama:2011ry},
\beq
\label{eq:k_latt}
 k^{{\mathrm{latt}}} = 0.03184 + 0.00453 \left ( \frac{a}{L} \right )^{2} + O \left [ \left ( \frac{a}{L}\right )^{4} \right ].
\eeq
The coupling, $\bar{g}_{\rm latt}$, defined in Eq.~(\ref{eq:TPL_scheme_def}) contains lattice artefacts,
therefore depends on the lattice spacing as well as the volume.  Its continuum-limit counterpart at fixed physical volume is defined as, 
\beq
 \bar{g}_{\rm c} = \lim_{a\rightarrow 0} \bar{g}_{\rm latt} , {\rm
   ~at~fixed~} L .
\label{eq:g_TPL_cont}
\eeq

The TPL scheme, as defined in Eq.~(\ref{eq:TPL_scheme_def}), contains the feature that 
the renormalised coupling constant has the fixed value $\sqrt{1/k} \sim 5.6$ in the IR limit ($L \rightarrow \infty$).
Therefore, in order to firmly establish the existence of the IR fixed
point, we have to show that $\bar{g}_{\rm c}$ is significantly 
different from this value at the fixed point.  

Contrary to the SF scheme, the $O(a)$ lattice
artefacts are absent in the TPL scheme.  As explained in the following
sections, it is important to control the continuum extrapolation
in the step-scaling study of the running coupling constant.  This makes the use of the TPL scheme very desirable.  
In Sec.~\ref{sec:analysis}, we will show the lattice-spacing
dependence of the TPL-scheme coupling constant.

\section{Simulation setting}
\label{sec:simulation_setting}
We give the details of our lattice simulation in this section.  As discussed in Sec.~\ref{sec:twbc}, the number of flavours in our
calculation must be a multiple of $N_{s} = N_{c} = 3$.  Since we are using staggered fermions, it also has to be proportional to the number
of tastes, $N_{t} = 4$.  In this work, we investigate the SU(3) gauge theory coupled to twelve flavours, which is allowed by these constraints.  

\subsection{Step scaling}
\label{sec:step_scaling}
Our goal is to measure the evolution of the running coupling constant
over a wide range of scale.  Given that the lattice imposes 
infrared (the volume) and ultraviolet (the lattice spacing) scales, the most convenient way to achieve this goal is the step-scaling technique.  
In this approach, we first measure the renormalised coupling constant, $\bar{g}_{\rm latt}$, on the lattice in the scheme 
defined in Eq.~(\ref{eq:TPL_scheme_def}).  Since we perform
computation at vanishing fermion mass, $\bar{g}_{\rm latt}$ only depends on the 
lattice spacing and the lattice volume, $L/a$.  Choosing a few values
of $L/a$, we then simulate at a wide range of $\beta \equiv
6/g^{2}_{0}$, where $g_{0}$ is the lattice bare coupling constant.
This enables us to tune $\beta$ (lattice spacing) to obtain the renormalised coupling in the continuum limit,
\beq
\label{eq:continuum_g_input}
 \bar{g}_{\rm c} \left ( L \right ) = \bar{g}_{\rm latt} \left ( \beta_{1}, L/a_{1} \right ) = \bar{g}_{\rm latt} \left ( \beta_{2}, L/a_{2} \right ) = 
  \ldots = \bar{g}_{\rm latt} \left ( \beta_{n_{0}}, L/a_{n_{0}} \right ) ,
\eeq
where $n_{0}$ is the number of choices of $L/a$.  Since $\bar{g}_{\rm c}$ is independent of the lattice spacing, it is renormalised at the length scale
$L$.  In this work, we perform lattice simulations at 
\beq
\label{eq:choices_of_L_over_a}
 L/a = 6, 8, 10 .
\eeq

Using the combinations of $(\beta, L/a)$ which lead to the same
$\bar{g}_{\rm c}(L)$ (or $u=\bar{g}^{2}_{\rm c}$), we compute the lattice step-scaling function,
\beq
\label{eq:lattice_step_scaling_function}
 \Sigma \left ( \beta_{i}, L/a_{i}, u, s \right ) \equiv \bar{g}^{2}_{\rm latt}\left . 
   \left ( \beta_{i}, sL/a_{i}\right ) \right |_{u = \bar{g}^{2}_{\rm latt} \left ( \beta_{i}, L/a_{i}\right )} ,
\eeq
where $i = 1,2,\ldots, n_{0}$ as in Eq.~(\ref{eq:continuum_g_input}), and $s$ is the step size.  Since we can obtain $n_{0}$ results for $\Sigma$ at the same physical
volume, $L$, with different lattice spacings, this allows us to determine the continuum-limit step-scaling function,
\beq
\label{eq:step_scaling_function}
 \sigma \left ( u, s \right ) \equiv \bar{g}^{2}_{\rm c} \left
   . \left ( sL \right ) \right |_{u=\bar{g}^{2}_{\rm c}\left ( L \right )} 
  = \lim_{a \rightarrow 0} \Sigma \left ( \beta_{i}, L/a_{i}, u, s \right ) .
\eeq
In this work, we choose the step size $s=2$, leading to the need for
simulations performed on the lattice volumes,
\beq
\label{eq:sL_over_a}
 sL/a = 12, 16, 20, \mbox{ }{\rm with~} s=2.
\eeq
For convenience, we define
\beq
\label{eq:sigma_at_s2}
 \sigma \left ( u \right ) \equiv \sigma \left ( u, s = 2 \right ) .
\eeq

The step-scaling function is a scheme-dependent quantity, since it is simply the renormalised coupling constant computed at a
certain scale.  To facilitate a better method in demonstrating the existence of the IRFP, we compute the ratio
\beq
\label{eq:r_sigma_def}
 r_{\sigma} \left ( u \right ) \equiv \frac{\sigma \left ( u \right )}{u} .
\eeq
This ratio becomes one at the zeros of the $\beta$-function.  The existence of such zeros is independent of the renormalisation scheme
used in the calculation.  In order to show that the gauge theory under investigation does contain an IR fixed point, we have to 
verify that $r_{\sigma}(u)$ is one at both UV and IR regimes, while deviating from this value in between.

A major source of systematic errors in the step-scaling method is the
continuum extrapolation.   It is a challenging task to properly
address this issue.  In order to have more information regarding this
extrapolation and its possible systematic effects, we also perform
simulation with $L/a=14$, and resort to an interpolation procedure to
obtain data for the TPL-scheme renormalised coupling on the lattice
size $L/a=7$.  This enables us to carry out the investigation with
\beq
\label{eq:step_scaling_including_L7}
 \left ( L/a = 6, 7, 8, 10  \right ) \longrightarrow \left ( 2 L/a =
   12, 14, 16, 20 \right ) .
\eeq
Here we stress that staggered fermions are used in this work,
therefore it is not possible to have data directly on the $L/a=7$
lattice.  The interpolation procedure for obtaining such data is 
explained in detail in Sec.~\ref{sec:L7_interpolation}.  This
interpolation in volume can introduce systematic effects, although it
may result in more information regarding the continuum limit.  
Therefore, we only use the 4-point step-scaling analysis in 
Eq.~(\ref{eq:step_scaling_including_L7})
as a means to estimate errors in the continuum extrapolation.

\subsection{Details of simulation parameters}
\label{sec:simulation_details}
Our calculation is performed using the Wilson plaquette action for
the gauge fields, and unimproved staggered fermions.   We implement the
standard Hybrid Monte Carlo (HMC) algorithm using the Omelyan
integrator with multi-time steps~\cite{Sexton:1992nu,Hasenbusch:2001ne}.
To compute the inversion of the lattice
fermion operator, biCGstab solver with convergence condition that the residue is
smaller than $10^{-16}$ for molecular dynamics, and the accuracy of 
$10^{-24}$ for the Metropolis tests, are used.  A significant fraction of of our
simulations were carried out on Graphics Processing Units (GPU's),
where a mixed-precision solver with defect correction was implemented. 
The GPU codes were developed with CUDA~\cite{cuda}.

To thermalise configurations in the Markov chains, we have used two
procedures.  In the first procedure, we start a
simulation from a trivial gauge-field configuration
\beq
\label{eq:u_eq_1}
U_{\mu}(\hat{n}_{x},\hat{n}_{y},\hat{n}_{z},\hat{n}_{t}) = 1 ,
\eeq
with fermion mass
\beq
\label{eq:init_fermion_mass}
 a m_{f} \sim 0.5 .
\eeq
Then, we gradually decrease the mass to zero.  In this process, we
monitor the Polyakov loops in the untwisted directions, and make certain
that the imaginary parts are non-vanishing.
This ensures that the Markov chains progress mostly near the
true vacua~\cite{Aoyama:2011ry}.  In the second procedure, we start with a configuration,
\bea
\label{eq:conf_ntv}
U_{z}( \hat{n}_{x},\hat{n}_{y},\hat{n}_{z}=1,\hat{n}_{t}) &=& e^{ - 2 i \pi / 3 } 
~,~U_{t}(\hat{n}_{x},\hat{n}_{y},\hat{n}_{z},\hat{n}_{t}=1) = e^{ + 2 i \pi / 3 }~,\\ \nonumber
U_{\mu}(\hat{n}_{x},\hat{n}_{y},\hat{n}_{z},\hat{n}_{t}) &=& 1 \rm{~~elsewhere } ,
\eea
which always results in non-zero imaginary parts in the Polyakov loops
in the untwisted directions.   
It also produces the largest gap in the vicinity of zero in the fermion matrix.
In this case, we can start the simulation directly with zero fermion mass,
making this procedure significantly more efficient than the one
implemented with the initial conditions of Eqs.~(\ref{eq:u_eq_1}) and (\ref{eq:init_fermion_mass}).
In both cases, we observe that the
simulations always stay near the true vacua, and 
tunnelling amongst these vacua occur occasionally.
We will discuss
this issue in more detail in Sec.~\ref{sec:vacuum}.

In order to implement the step-scaling investigation of the running
coupling constant as discussed in Sec.~\ref{sec:step_scaling}, we
carry out simulations at the lattice volumes,
\beq
\label{eq:lattice_volumes}
 L/a = 6, 8, 10, 12, 14, 16, 20 .
\eeq
For each volume, we simulate at several $\beta$ values between 4 and
99, in the gauge action.  Since the running is expected to be slow in
SU(3) gauge theory with twelve flavours, this large range 
of $\beta$ is necessary to trace the coupling constant from the UV to
the IR regimes.  We aim at determining the Polyakov-loop correlators
with statistical errors around $2.5\%$ or smaller.  For this purpose, a
significant amount of gauge-field ensembles have to be generated.  
The raw
data for the TPL-scheme renormalised coupling, as defined in
Eq.~(\ref{eq:TPL_scheme_def}), are given in App.~\ref{sec:plaquette_values_raw_data}.

\section{Plaquette, Polyakov loop and the vacuum structure}
\label{sec:plaquette_and_ploop}
In this work, we perform several detailed checks on the simulations,
in order to ensure that we are estimating the autocorrelation and
performing the continuum extrapolations reliably.  These checks
include the lowest-lying eigenvalue spectrum of the Dirac operator,
the plaquette values, and the phases of the Polyakov loops.  The
computation of the lowest-lying eigenvalues is presented in 
App.~\ref{sec:eigenvalues}, while in this section we address the other
two topics.

\subsection{Plaquette}
\label{sec:plaquette}
\begin{figure}[t]
   \hspace{0.8cm}\includegraphics*[width=0.55\textwidth,angle=0]{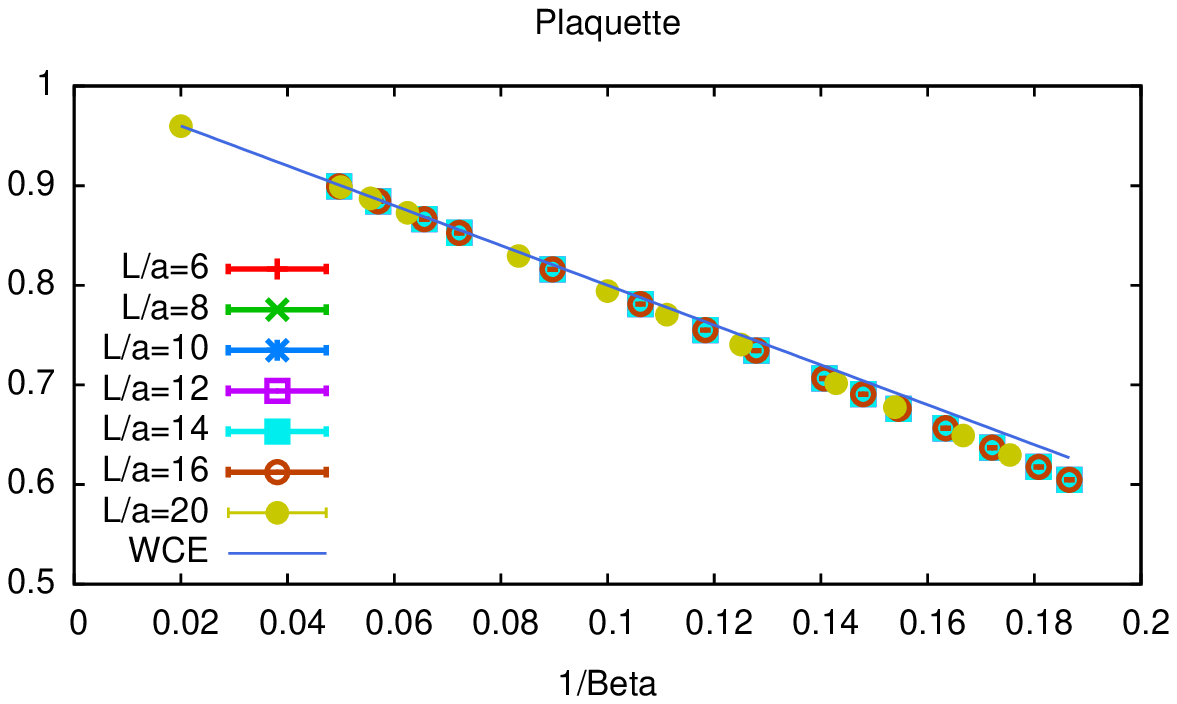}
   \includegraphics*[width=0.58\textwidth,angle=0]{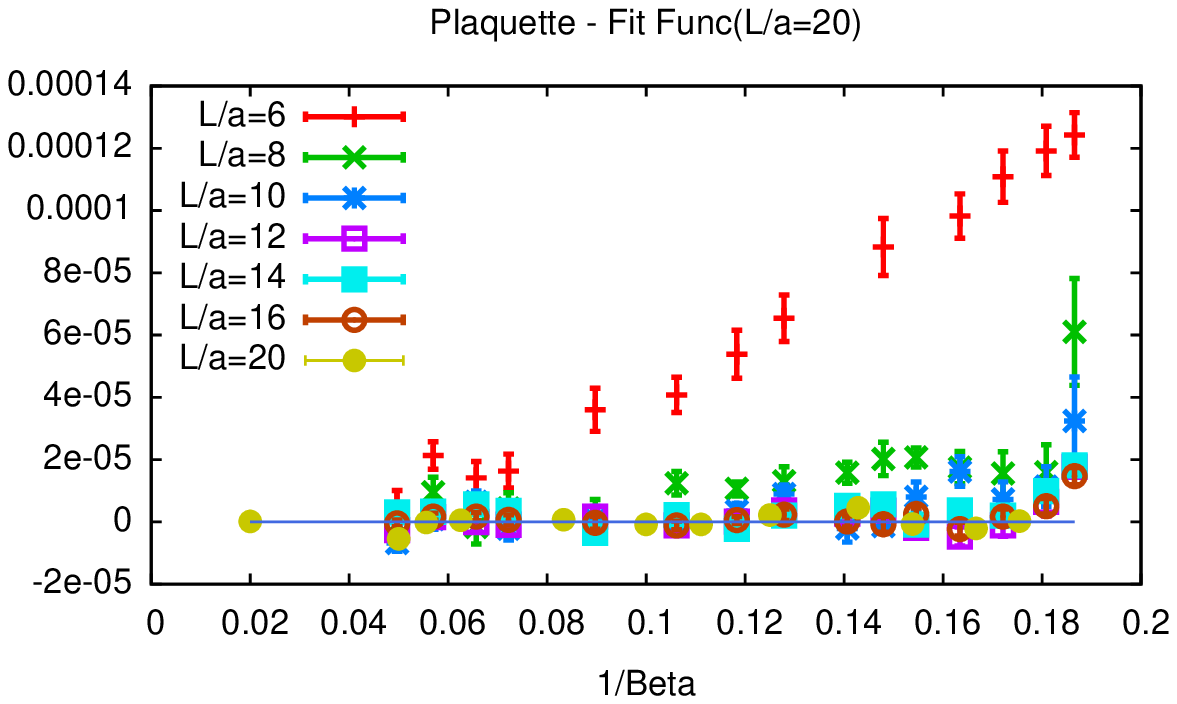}
\caption{Upper panel: The values of plaquette as a function of $1/\beta$.
  ``WCE'' stands for ``weak coupling expansion''.
  Lower panel: The difference between the data points and the
  function, Eq~(\ref{eq:L20_plaquette_wce}), fitted to the $L/a=20$ data
  points only.}
\label{fig:plaquette}
\end{figure}
As shown 
in App.~\ref{sec:plaquette_values_raw_data}, some
of our simulations are performed at small $\beta$ values (coarse
lattice spacings).   It is
necessary to check that these simulations are still in the
weak-coupling phase, in order to make certain that at these $\beta$
values, the theory is still in the same universality class as that with
high${-}\beta$ (fine lattice spacings).  This is essential in order to
ensure that the continuum limit can be reliably taken in our calculations.

For the above purpose, we examine the expectation values of the
plaquette for many of our HMC simulations.  The results are summarised 
in Table~\ref{tab:plaquette} in App.~\ref{sec:plaquette_values_raw_data}.
These expectation values are plotted in the Upper panel of
Fig.~\ref{fig:plaquette}, where we also show the predictions from the
weak coupling expansion for pure Yang-Mills theory,
\beq
 {\rm plaquette}  \approx 1 - \frac{2}{\beta} \mbox{ }({\rm
   weak~coupling~expansion}) .
\label{eq:plaquette_weak_exp}
\eeq
By comparing our data with this function, it is evident that all
our simulations are in the weak-coupling phase, and are safe from
being in the novel phase observed in 
Ref.~\cite{Cheng:2011ic}\footnote{We thank David Schaich for private
  communications regarding this issue.}.
We have also studied the volume dependence of the plaquette, by first
fitting the data obtained on the largest lattice, $L/a=20$, to a
weak-coupling expansion formula ($p_{i}$ are the fit parameters),
\begin{equation}
f(\beta) = p_{0} + \frac{p_{1}}{\beta} +  \frac{p_{2}}{\beta^{2}} +
\frac{p_{3}}{\beta^{3}} +  \frac{p_{4}}{\beta^{4}} +
\frac{p_{5}}{\beta^{5}} ,
\label{eq:L20_plaquette_wce}
\end{equation}
then computing the difference between the data points to this curve.
The result of this investigation is shown in the lower panel of
Fig.~\ref{fig:plaquette}.  This shows that finite-size effects are
minor in the computation of the plaquette in this
work.

\subsection{Polyakov loops and vacuum tunneling}
\label{sec:vacuum}
The study of the plaquettes in the last section confirms that
our simulations have been carried out in the weak coupling phase.  
In this phase, 
as pointed out in Ref.~\cite{Aoyama:2011ry}, the true vacua in 
SU(3) gauge theory with fermions are always those in which the vacuum expectation values
of the Polyakov loops in the untwisted directions are non-vanishing
and complex.    On
the other hand, in the vicinity of the false vacua, the untwisted
Polyakov loops are real.  As for the Polyakov loops in the twisted
directions, we expect that they will scatter around zero,
configuration by configuration.

Markov
chains in our simulations can be trapped in the false vacua.  However,
by using the above property of the untwisted Polyakov loops, we can
monitor the simulations and ensure that they
are mostly progressing near the true vacua.

\begin{figure}
\includegraphics[scale=0.6]{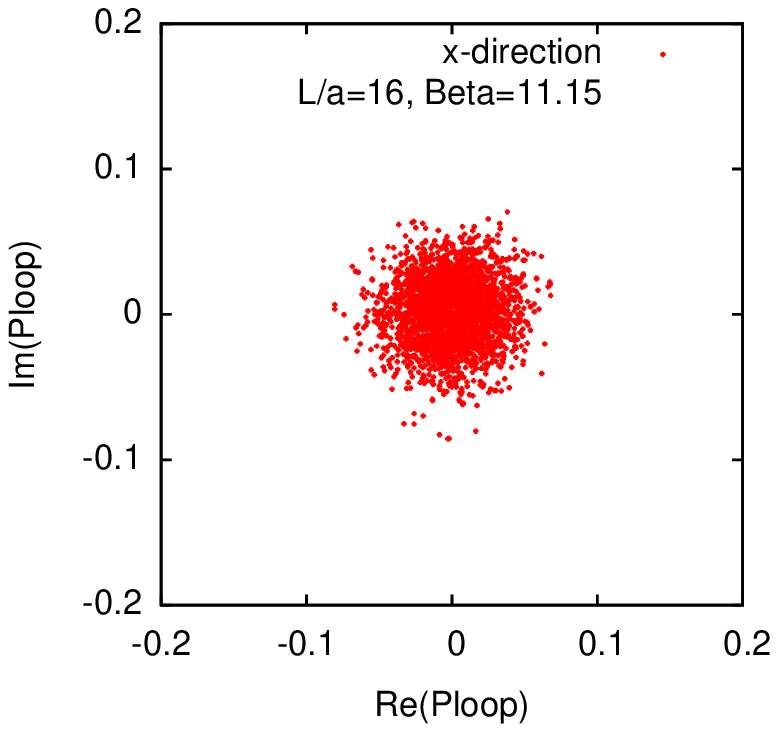}
\includegraphics[scale=0.6]{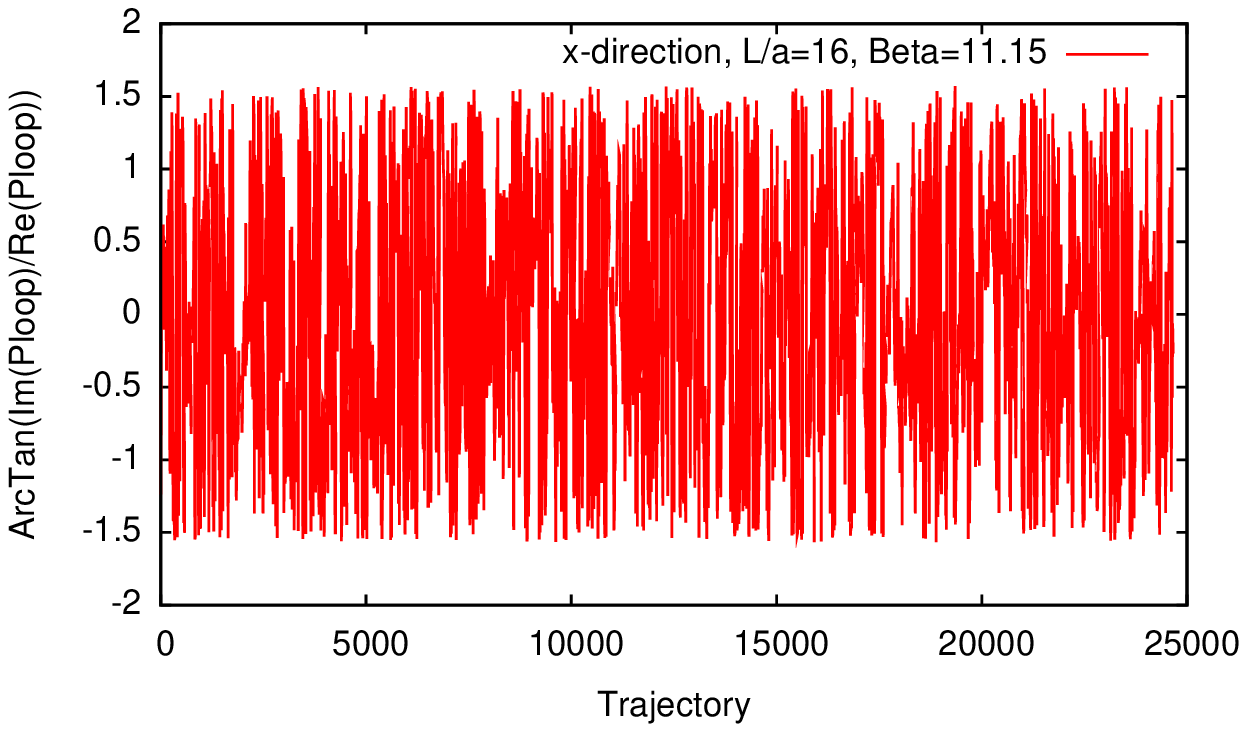}
\includegraphics[scale=0.6]{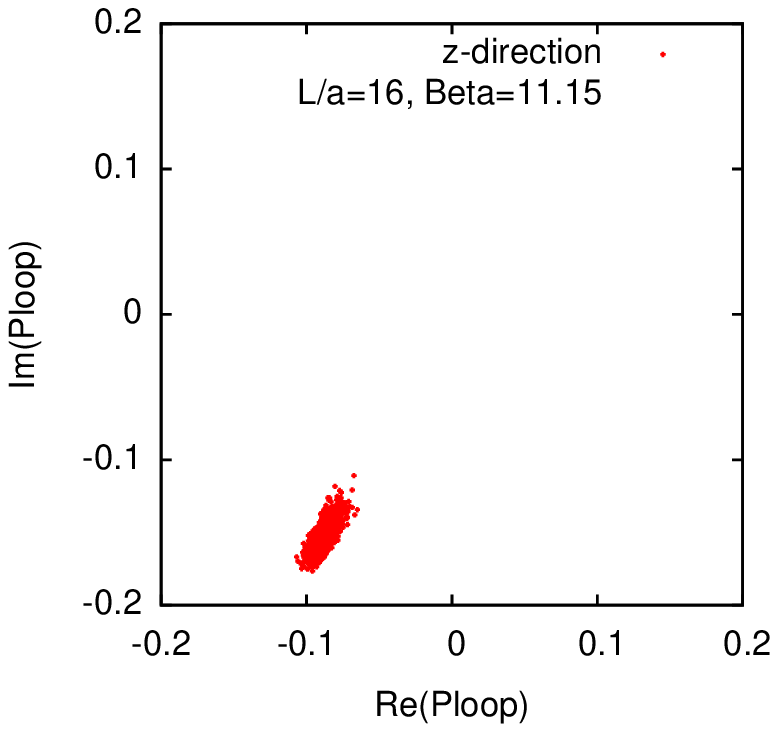}
\includegraphics[scale=0.6]{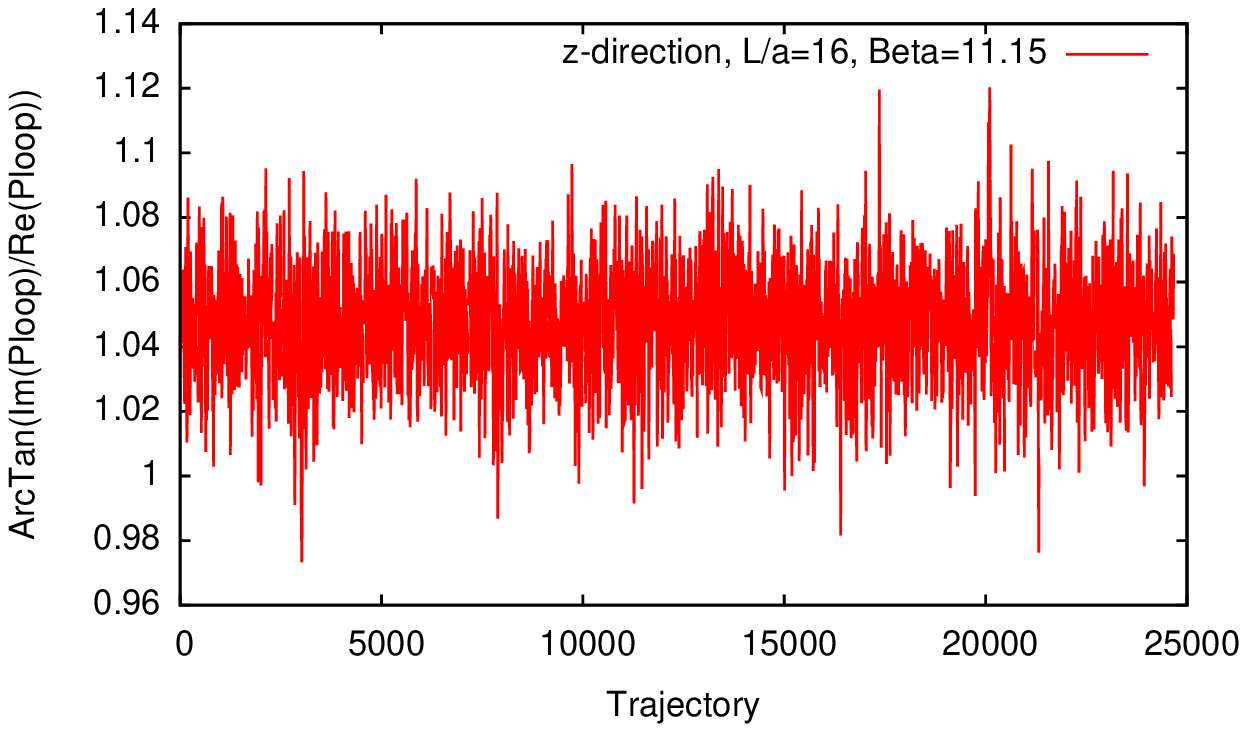}
\caption{Complex values (left panels) and the ratios between the real and
  imaginary parts (right panels) for Polyakov loops in the twisted
  (upper panels) and untwisted (lower panels)
  directions, in the first 25000 trajectories in the simulation
  performed at $\beta = 11.15$ and $L/a = 16$.}
\label{fig:ploop_scatter_phase_no_tunnelling}
\end{figure}

\begin{figure}
\includegraphics[scale=0.6]{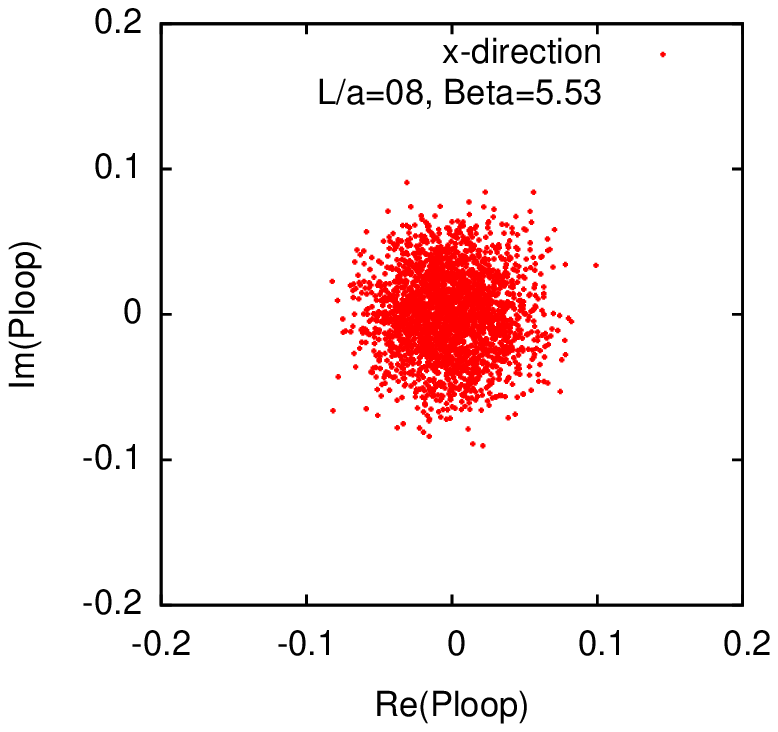}
\includegraphics[scale=0.6]{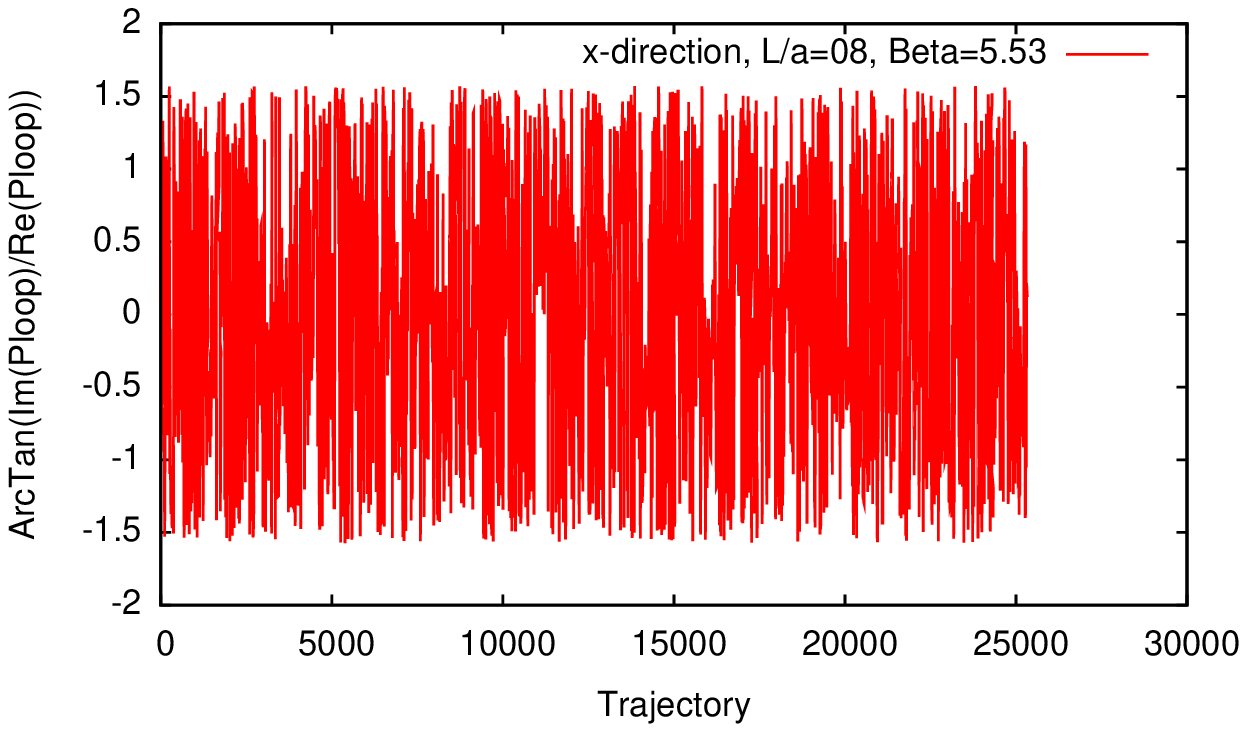}
\includegraphics[scale=0.6]{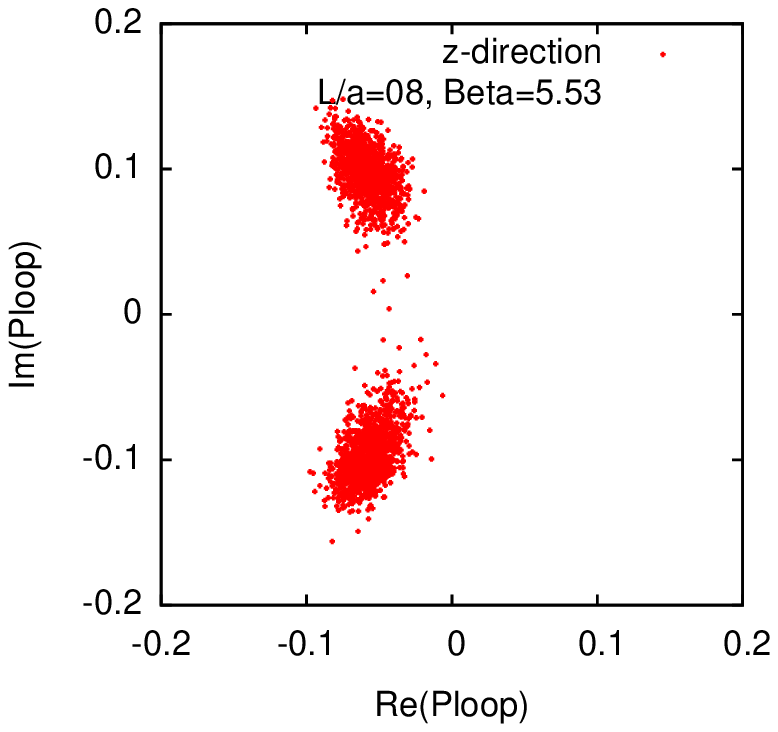}
\includegraphics[scale=0.6]{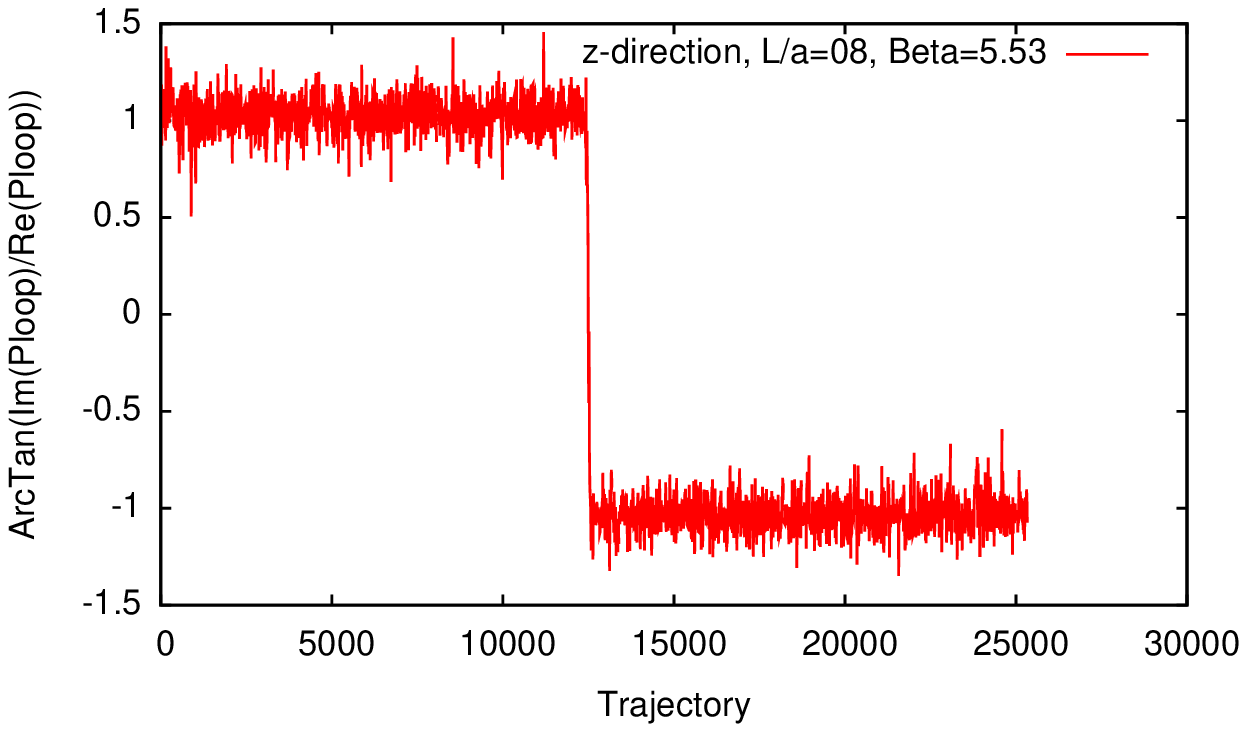}
\caption{Complex values (left panel) and the ratios between the real and
  imaginary parts (right panel) for Polyakov loops in the twisted
  (upper panels) and untwisted (lower panels)
  directions, in the first 25000 trajectories in the simulation
  performed at $\beta = 5.53$ and $L/a = 8$.}
\label{fig:ploop_scatter_phase_with_tunnelling}
\end{figure}

Investigating the Polyakov loops trajectory by trajectory, we first
confirm that in the twisted directions, they are fluctuating around
zero for all simulations in this work.  This is shown for two typical
cases in the upper panels of
Figs.~\ref{fig:ploop_scatter_phase_no_tunnelling} and
\ref{fig:ploop_scatter_phase_with_tunnelling}.   

\begin{figure}
\includegraphics[scale=0.6]{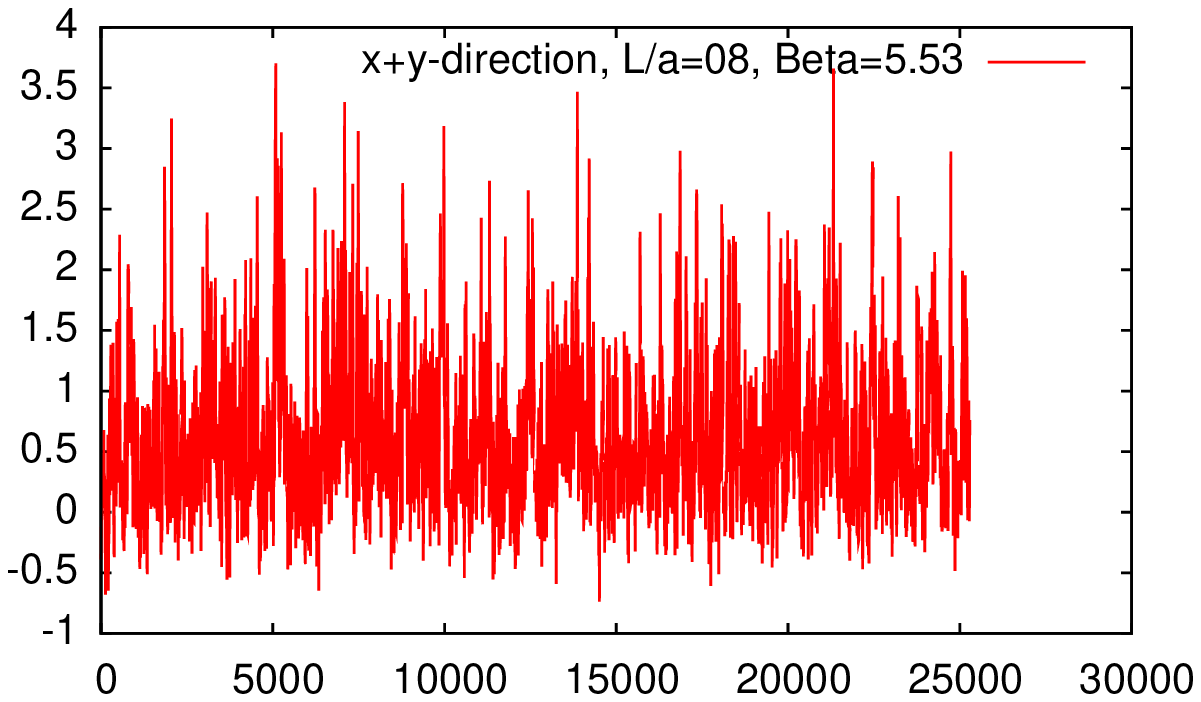}
\includegraphics[scale=0.6]{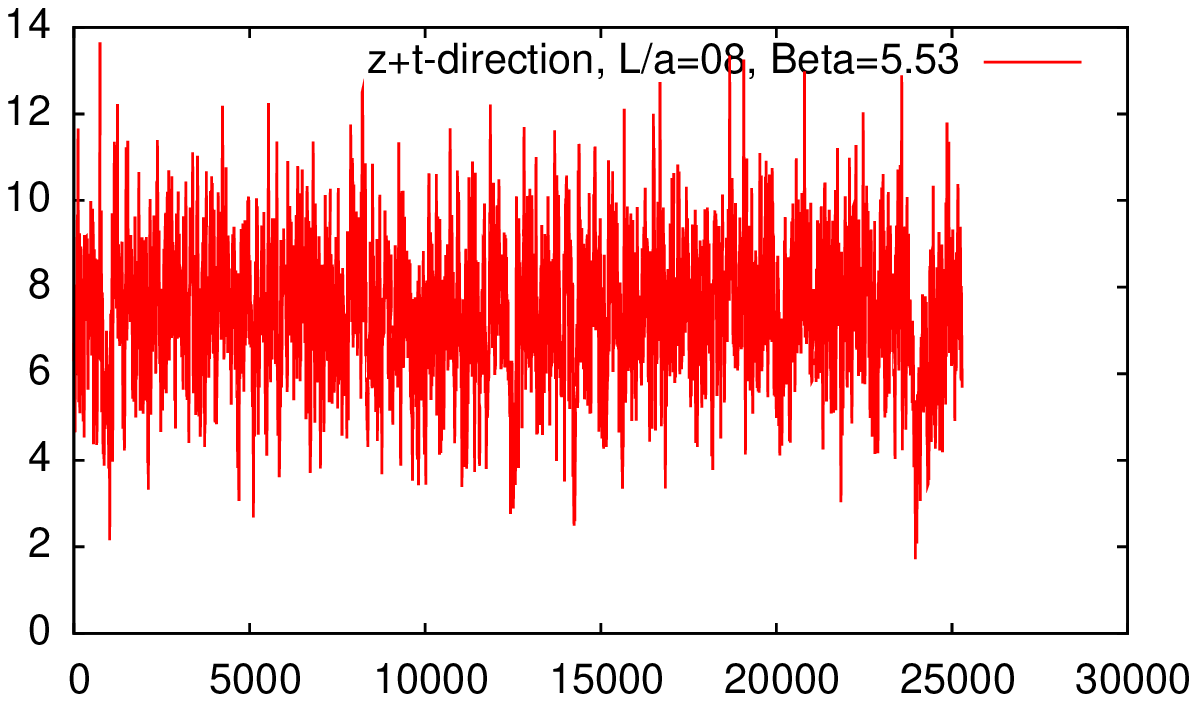}
\caption{Polyakov loop correlators in the twisted (left panel) and
  untwisted (right panel) directions, 
 in the first 25000 trajectories in the simulation
  performed at $\beta = 5.53$ and $L/a = 8$.}
\label{fig:ploop_correlators_with_tunnelling}
\end{figure}

Next, we study the Polyakov loops in the untwisted directions.   In
all our simulations, their values are non-vanishing and complex in all
trajectories.  The complex phase fluctuates around $\pm 2\pi/3$,
indicating that the Markov chains are progressing near the true
vacua.  The lower panels of
Fig.~\ref{fig:ploop_scatter_phase_no_tunnelling} 
demonstrate a case ($L/a=16, \beta=11.15$) in
which the simulation stays near the vacuum with the phase of Polyakov loop
being $-2\pi/3$.  For simulations performed at smaller $L/a$ (fewer
total degrees of freedom) and larger $\beta$ (stronger coupling),
tunnelling between the two true vacua may occur.   One of such cases
is shown in the lower panels of Fig.~\ref{fig:ploop_scatter_phase_with_tunnelling}.
Every time this takes place, we then investigate the Polyakov loop
correlators trajectory by trajectory, ensuring that these correlators
do not exhibit any ``discontinuous'' behaviour when the tunnelling
happens.  In Fig.~\ref{fig:ploop_correlators_with_tunnelling}, we show
the result of this study for the corresponding simulation presented 
in Fig.~\ref{fig:ploop_scatter_phase_with_tunnelling}.
From these plots for the Polyakov loop correlators in the twisted and
untwisted directions, we conclude that tunnelling between the true
vacua does not result in artefacts which complicate the estimation of
autocorrelation time.

\section{Analysis details}
\label{sec:analysis}
In this section, we explain the details of our analysis.  
The statistical analysis in this work is performed using
the bootstrap procedure, in which 1000 bootstrap samples are generated
for each $(L/a,\beta)$.

\subsection{Autocorrelation and data binning}
\label{sec:autocorrelation}
%
%
As presented in App.~\ref{sec:plaquette_values_raw_data}, we perform our calculations
with a large number of HMC trajectories.  The first step in our
analysis is the binning of the raw data.   In order to make certain that
the binning procedure is reasonable, we study the autocorrelation of
the ratio, appearing in the left-hand side of
Eq.~(\ref{eq:TPL_scheme_def}), 
between the Polyakov loop correlators.
To describe our investigation, we start from the autocorrelation function of primary quantities,
~\cite{Schroers:2001fw,Luscher:2005rx,Wolff:2003sm}
%
%
\begin{equation}
\Gamma_{\hat \alpha \hat \beta}( \tau ) = \frac{1}{N -\tau} \sum_{i=1}^{N-\tau} 
\left( {\cal O}_{\hat \alpha}(i) - \bar {\cal O}_{\hat \alpha} \right ) 
\left( {\cal O}_{\hat \beta}(i+\tau) - \bar {\cal O}_{\hat \beta} \right ).
\label{eq:gamma_alpha_beta}
\end{equation}
Here, $\hat \alpha$ and $\hat \beta$ label the types of primary quantities. 
In our case, ${\cal O}_1(i)$ and ${\cal O}_2(i)$ are Polyakov loop
correlators of the $i$-th sample in the twisted and in the periodic
directions, respectively.   The quantity $\bar {\cal O}_{\hat \alpha}$ is the average of  ${\cal O}_{\hat \alpha}(i)$,
$
\bar {\cal O}_{\hat \alpha} = (1/N) \sum_i^{N} {\cal O}_{\hat \alpha}(i).
$

By using $\Gamma_{\hat \alpha \hat \beta}$, the autocorrelation
function of the Polyakov loop ratio, as in the left-hand side of
Eq.~(\ref{eq:TPL_scheme_def}), can be written as,
\begin{equation}
\Gamma( \tau ) = \sum_{\hat \alpha, \hat \beta=1}^2 f_{\hat{\alpha}} f_{\hat{\beta}} \Gamma_{\hat{\alpha}\hat{\beta}}(\tau)
\end{equation}
with,
\beq
f_1= \frac{\partial}{\partial \bar {\cal O}_1}\left ( \frac{\bar {\cal O}_1}{\bar {\cal O}_2} \right )
= \frac{1}{\bar {\cal O}_2}
,~~~ 
f_2= \frac{\partial}{\partial \bar {\cal O}_2}\left ( \frac{\bar {\cal O}_1}{\bar {\cal O}_2} \right )
= - \frac{\bar {\cal O}_1}{{\bar {\cal O}_2}^2}.
\eeq 

We define the normalised autocorrelation function,
\begin{equation}
\rho ( \tau ) = \frac{ \Gamma(  \tau ) }{ \Gamma ( 0 ) }
\end{equation}
which is normally assumed to behave as,
\beq
\rho (\tau) \sim {\rm e}^{-\frac{\tau}{\tau_{\rm A}}} .
\label{eq:autocorr_single_e}
\eeq

The quantity $\tau_{\rm A}$ is the autocorrelation time of single exponential autocorrelation. 

Since the integrated autocorrelation function is less noisy than $\rho(\tau)$,
we use it to estimate the autocorrelation time between
Polyakov-loop-correlator ratios. Upon integrating over $\tau$, we obtain
\beq
\int_{0}^{\tau} \rho (\tau^{\prime}) d\tau^{\prime} \sim \tau_{\rm
   A} 
\left ( 1 - {\rm e}^{-\tau / \tau_{\rm A}}\right ) \sim \tau_{\rm A}
{\rm ~when~} \tau \gg \tau_{\rm A}.
\eeq
The single-exponential form in Eq.~(\ref{eq:autocorr_single_e}) is
often a poor approximation to $\rho(\tau)$, when the system contains
degrees of freedom that are characterised by very different
autocorrelation times. 
In general, the autocorrelation function can be multi-exponential,
\beq
\rho(\tau) \sim \sum_k a_k \, e^{-\tau/\tau_A^{(k)}} {\rm~with~}
\sum_k a_k = 1 .
\label{eq:autocorr_multi_e}
\eeq
The integrated autocorrelation is
\beq
\int_{0}^{\tau} \rho (\tau^{\prime}) d\tau^{\prime}
\sim \sum_k \tau_A^{(k)} \, a_k \, \left ( 1 -  e^{-\tau/\tau_A^{(k)}}
\right ) .
\eeq
This function reaches a plateau $ \sum_k \tau_A^{(k)} \, a_k $ when $ \tau \gg \tau_A^{(k)} $ for all $k$.
We use this criteria for the estimation of autocorrelation without
explicitly determining $\tau_{A}^{(k)}$ and $a_{k}$.  A more detailed
study of autocorrelation times for conformal field theories will be
reported in a separate paper~\cite{AC_in_prep}.

\begin{figure}
\includegraphics[scale=0.6]{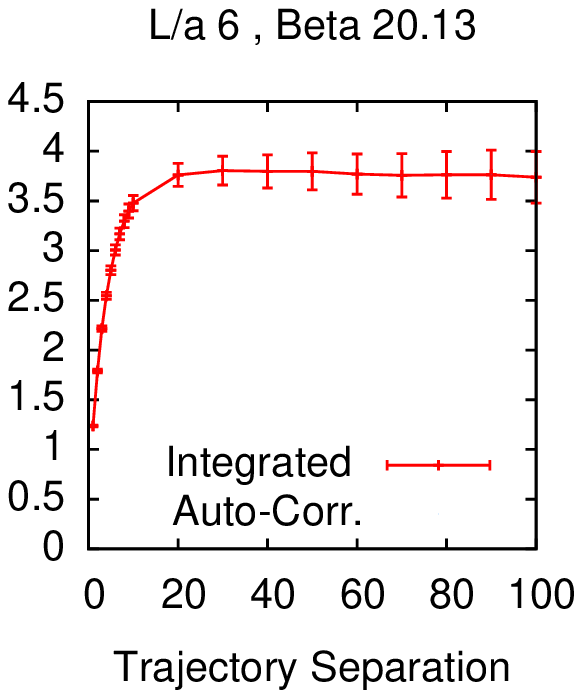}
\hspace{-3cm}
\includegraphics[scale=0.6]{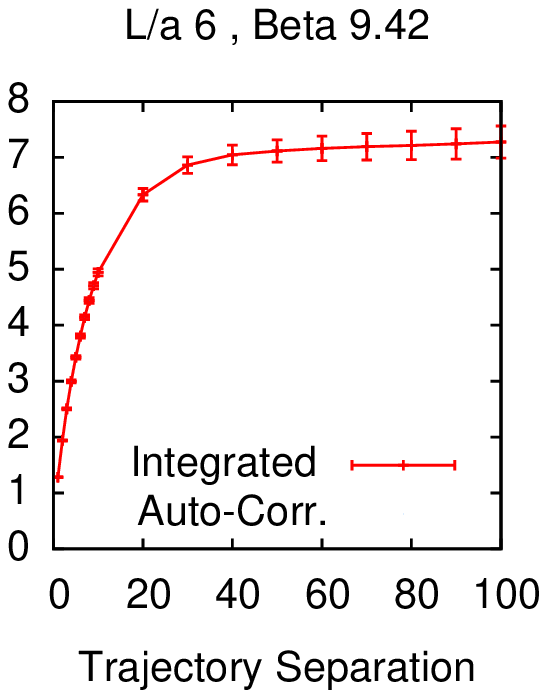}
\hspace{-3cm}
\includegraphics[scale=0.6]{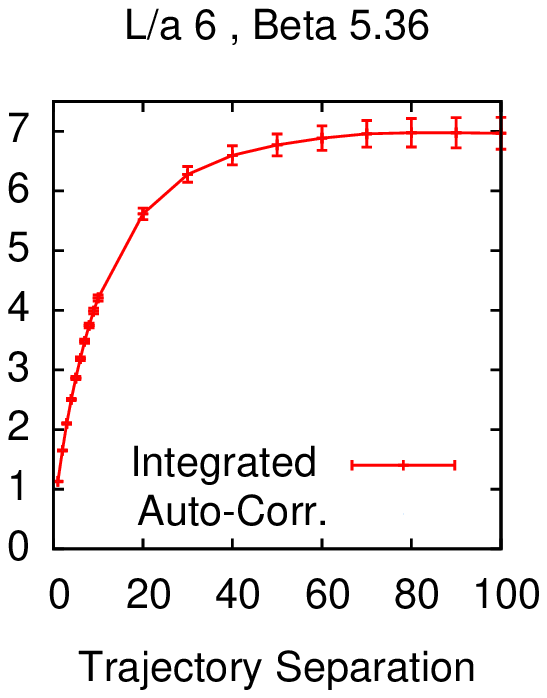}
\hspace{-3cm}
\includegraphics[scale=0.6]{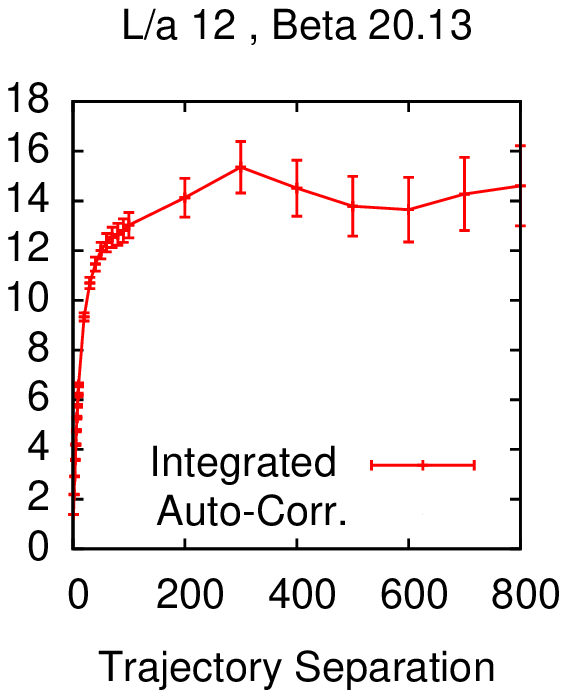}
\hspace{-3cm}
\includegraphics[scale=0.6]{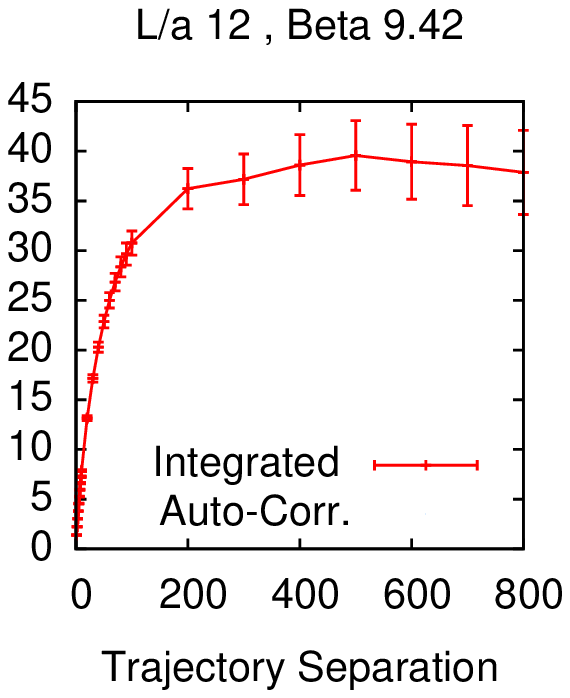}
\hspace{-3cm}
\includegraphics[scale=0.6]{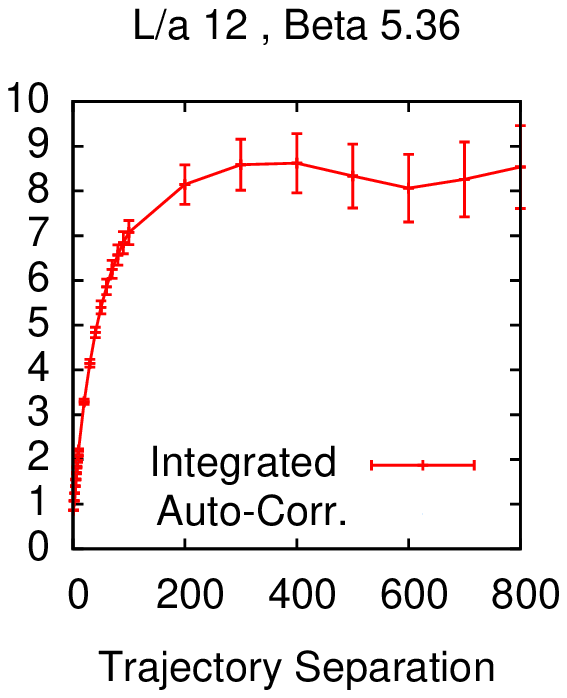}
\hspace{-3cm}
\includegraphics[scale=0.6]{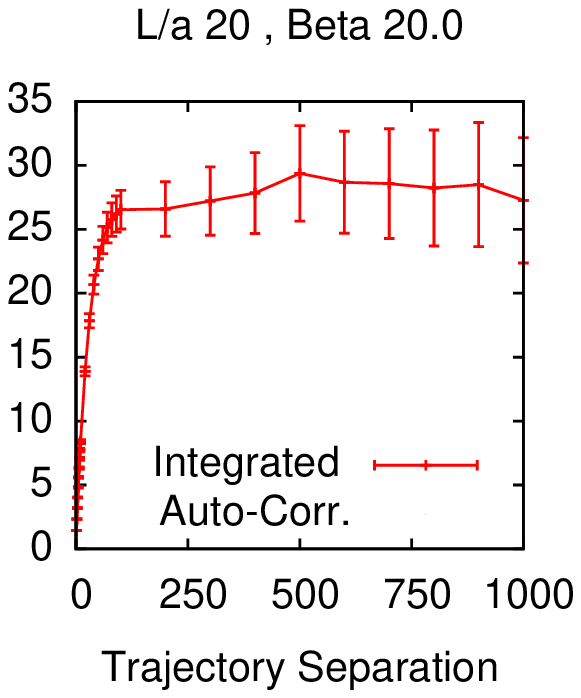}
\hspace{-3cm}
\includegraphics[scale=0.6]{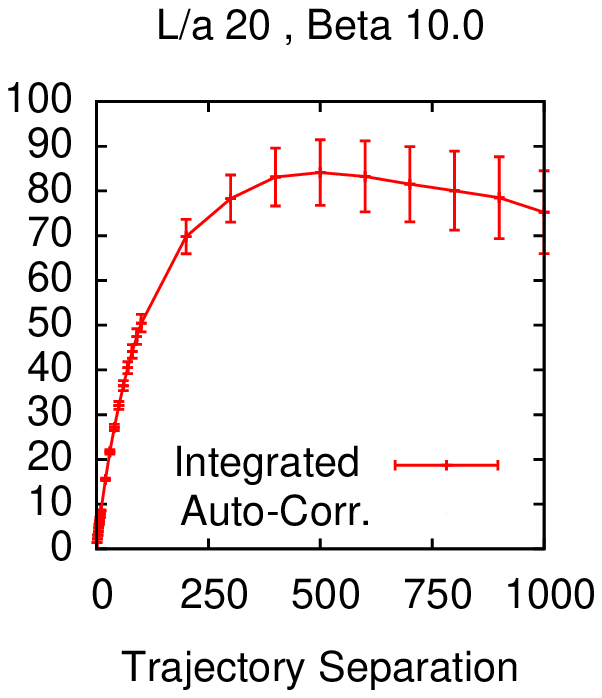}
\hspace{-3cm}
\includegraphics[scale=0.6]{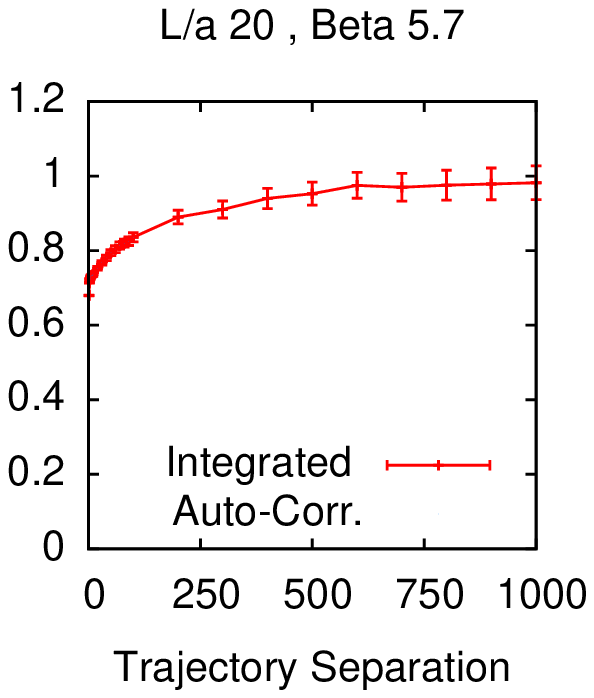}
\caption{Representative plots for the integrated autocorrelation of
  the ratio of Polyakov loop correlators at
  various values of $L/a$ and $\beta$.  The physical volume increases
  from the top-left to the right-bottom corners.}
\label{fig:autocorr}
\end{figure}

In our numerical calculation, the integrated autocorrelation is defined as,
\begin{equation}
\Theta (\tau) = \frac{1}{2} + \sum_{\tau^{\prime} = 1}^{\tau} \rho(\tau^{\prime}) .
\end{equation}
To estimate error in $\Theta(\tau)$, we apply the Madras-Sokal formula~\cite{Madras:1988ei},
\begin{equation}
( \Delta \Theta (\tau) )^2 = \frac{4 \tau + 2 }{N} \Theta( \tau )^2.
\end{equation}
Figure~\ref{fig:autocorr} shows $\Theta (\tau)$ for the representative cases in this work.   
%
%
%
%
%
%
%
The separation between two decorrelated trajectories can be estimated
by investigating the plateau of $ \Theta(\tau)$.  
As demonstrated in Fig.~\ref{fig:autocorr}, this separation depends on
the physical volume, $L$.  It is around 20 on the smallest volumes, and
about a few hundred to 1000 on the largest volumes.

\begin{figure}
\begin{center}
\includegraphics[scale=0.6]{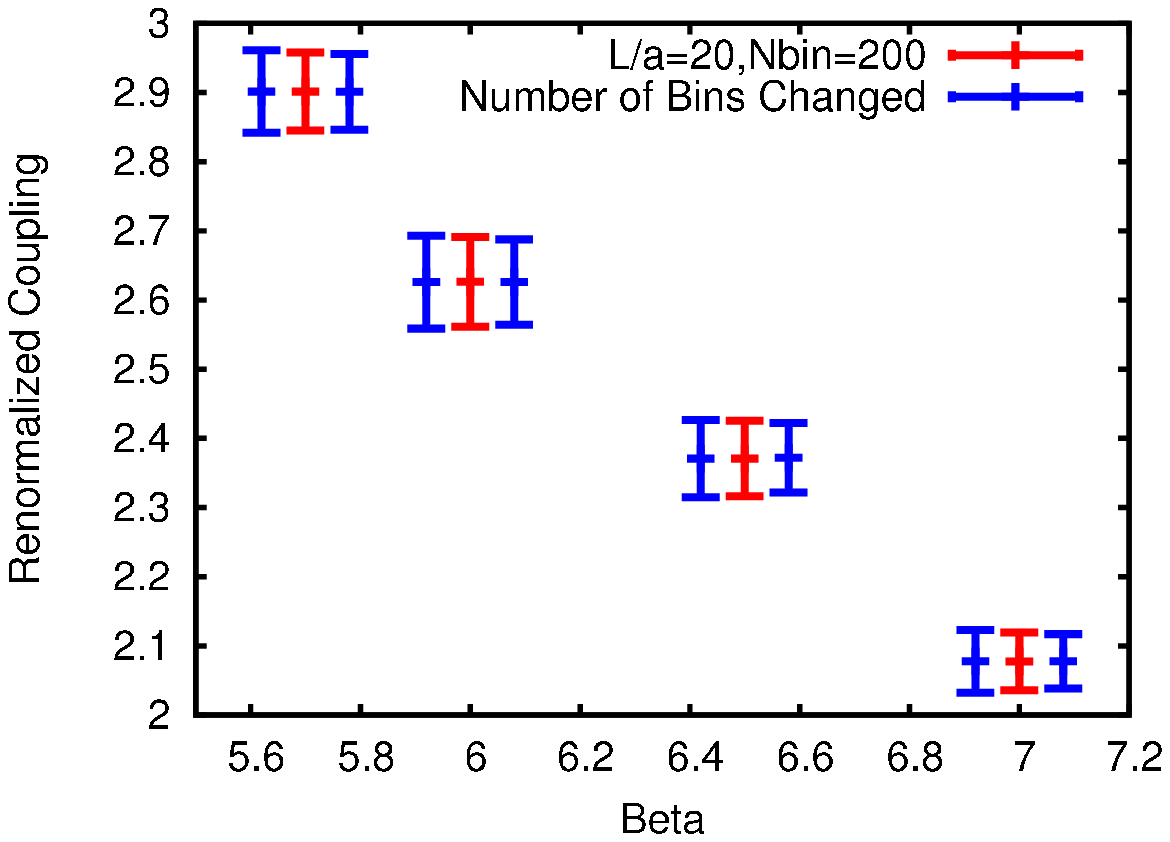}
\includegraphics[scale=0.6]{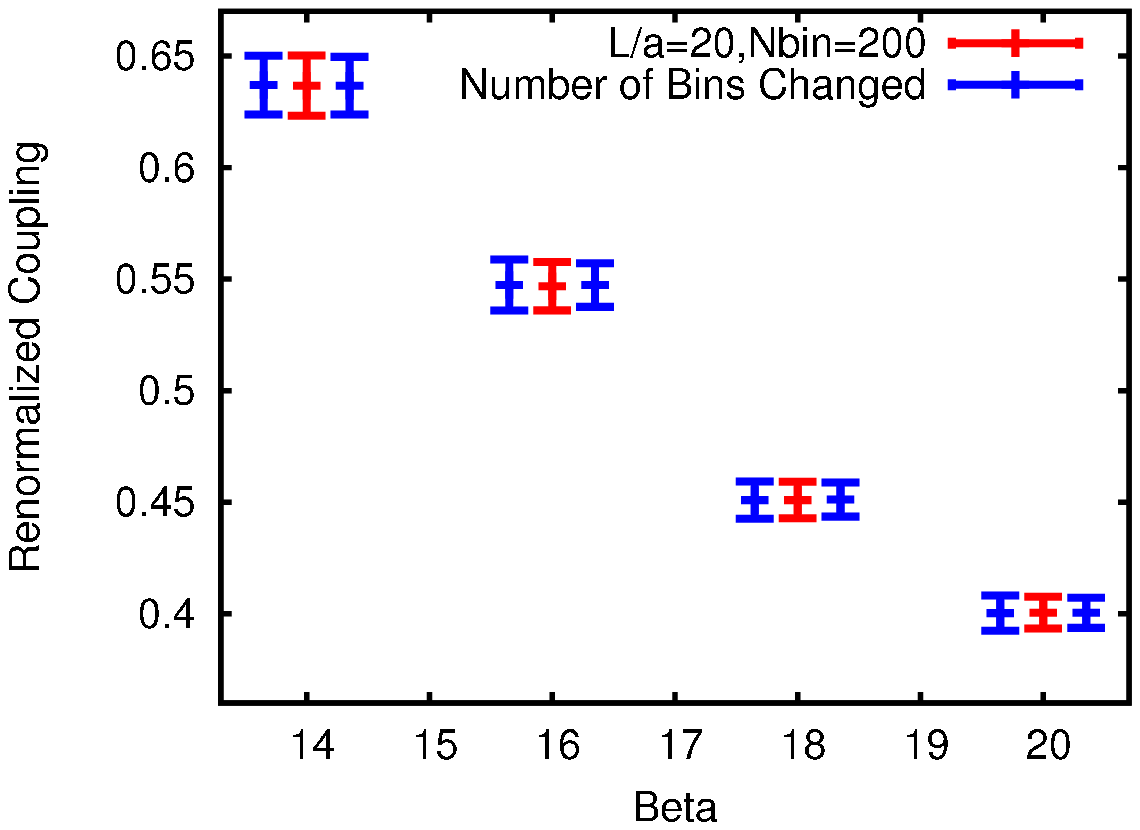}
\caption{
Dependence of the TPL-scheme renormalised coupling on the number of
bins at various $\beta$
values on the $L/a=20$ lattice.  Each group of data points contains the results of the
jackknife analysis with the numbers of bins set to 100, 200 and 500.  
The red point in the centre of each group is the result of using 200
bins, as chosen in our bootstrap procedure.  The $\beta$ values for
the two blue points are slightly shifted for the purpose of
presentation.  They correspond to choosing the numbers of bins to be
100 (left) and 500 (right).
}
\label{fig:jackknife}
\end{center}
\end{figure}

In App.~\ref{sec:plaquette_values_raw_data},
we show the details for the numbers of HMC trajectories in our simulations.
For each choice of $(L/a, \beta)$, we divide the trajectories evenly into $\sim 200$ bins
by averaging over them in each bin.  
These bins are then used to create 1000 bootstrap samples.
From the result presented in this section, 
it is evident that our bin sizes are large enough compared to
the autocorrelation times.
This ensures that the data amongst
these bins are decorrelated.  We have also confirmed this with the
Jackknife analysis using these and larger bin sizes.   The
statistical errors in this approach are almost the same as those in 
our bootstrap analysis, and they are stable against the change of the bin sizes.
Figure~\ref{fig:jackknife} shows some examples for this Jackknife check for
the TPL-scheme renormalised coupling computed at various $\beta$
values on the $L/a=20$ lattice.
In this figure, each group of data points contains the results of the
jackknife analysis with the numbers of bins set to 100, 200 and 500.  
The red point in the centre of each group is the result of using 200
bins, as chosen in our bootstrap procedure.  The $\beta$ values for
the two blue points are slightly shifted for the purpose of
presentation.  They correspond to choosing the numbers of bins to be
100 (left) and 500 (right).  From these plots, it is apparent that
having 200 bins leads to enough trajectories in each bin, in order to
correctly estimate statistical errors.

\subsection{Interpolation in $\beta$ (bare coupling constant)}
\label{sec:beta_interpolation}

In the step-scaling study of the running coupling constant, we first
have to perform the tuning of the $\beta$ values in
Eq.~(\ref{eq:continuum_g_input}), for the lattice volumes $L/a =
6,8,10$.  In principle, this can be achieved by repeatedly adjusting $\beta$ and carrying out new
simulations, until Eq.~(\ref{eq:continuum_g_input}) is satisfied to
high accuracy.   As discussed at the end of
Sec.~\ref{sec:step_scaling}, we also want to obtain 
the TPL-scheme renormalised coupling on the $L/a=7$ lattice through
interpolation in volume, in order to estimate systematic errors in the continuum extrapolation.  
For this purpose, one has to tune a different set
of  $\beta$ values for $L/a = 6, 8, 10$ and interpolate to $L/a=7$ at
each step of this tuning.

The above procedure is very time-consuming, and becomes
impractical for studies in which one has to trace the coupling
constant across a large range of length scale.  This is the case in
the current work.  Therefore we resort to a variation of the above
method.  That is, we simulate at many $\beta$ values for each $L/a$,
and perform interpolations in $\beta$ for the renormalised coupling constant, 
volume by volume.  The choices
of these $\beta$ values are presented 
in App.~\ref{sec:plaquette_values_raw_data}.  The use of this interpolation method
inevitably introduces systematic effects in our calculation.  We will
address this issue in this section.

Since we are simulating at a large range of bare coupling constant, it
is a challenging task to have a well-inspired interpolation function in
$\beta$.   One reasonable way to proceed is to note that in the
large${-}\beta$ (small bare-coupling) regime,  one-loop perturbation
theory has to be valid, and therefore at fixed $L/a$,
\beq
 u_{\rm latt} \equiv \bar{g}^{2}_{\rm latt} (\beta, L/a) \approx
 \frac{6}{\beta} = g^{2}_{0}\mbox{
 }\mbox{ }(\rm for~\beta >> 1) ,
\label{eq:high_beta_interp}
\eeq
where $g_{0}$ is the bare gauge coupling.
This motivates the use of polynomial functions in $1/\beta$ to
perform the interpolation.  Since we have data for many $\beta$
values (see App.~\ref{sec:plaquette_values_raw_data}) 
for each $L/a$, it is in principle
possible to have high degrees of polynomials for these fits.  Such
high-degree polynomials will generally fit all the data points.  On
the other hand, the Runge phenomenon may occur in this procedure,
resulting in artificial oscillatory behaviour of the fit functions.
In order to avoid this artefact in the $\beta{-}$interpolation, we
note that the renormalised coupling should always be non-decreasing with growing
lattice spacing ({\it i.e.}, decreasing $\beta$) at fixed $L/a$, otherwise
the theory will be in the strong-coupling phase and the continuum
extrapolation cannot be reliably performed.  

\begin{figure}[t]
\hspace*{5mm}\\
\hspace*{83mm}\vspace*{-7mm}\\
\hspace*{2mm}
\scalebox{0.65}{\includegraphics{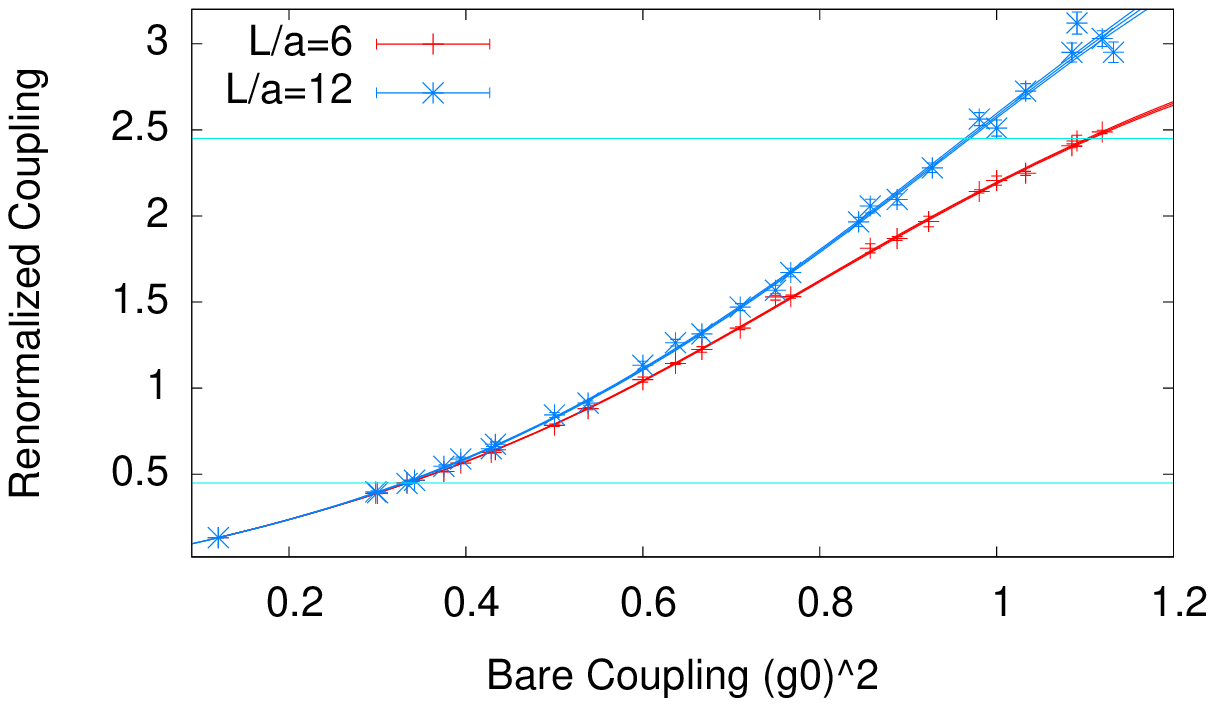}}
\hspace*{3mm}
\scalebox{0.65}{\includegraphics{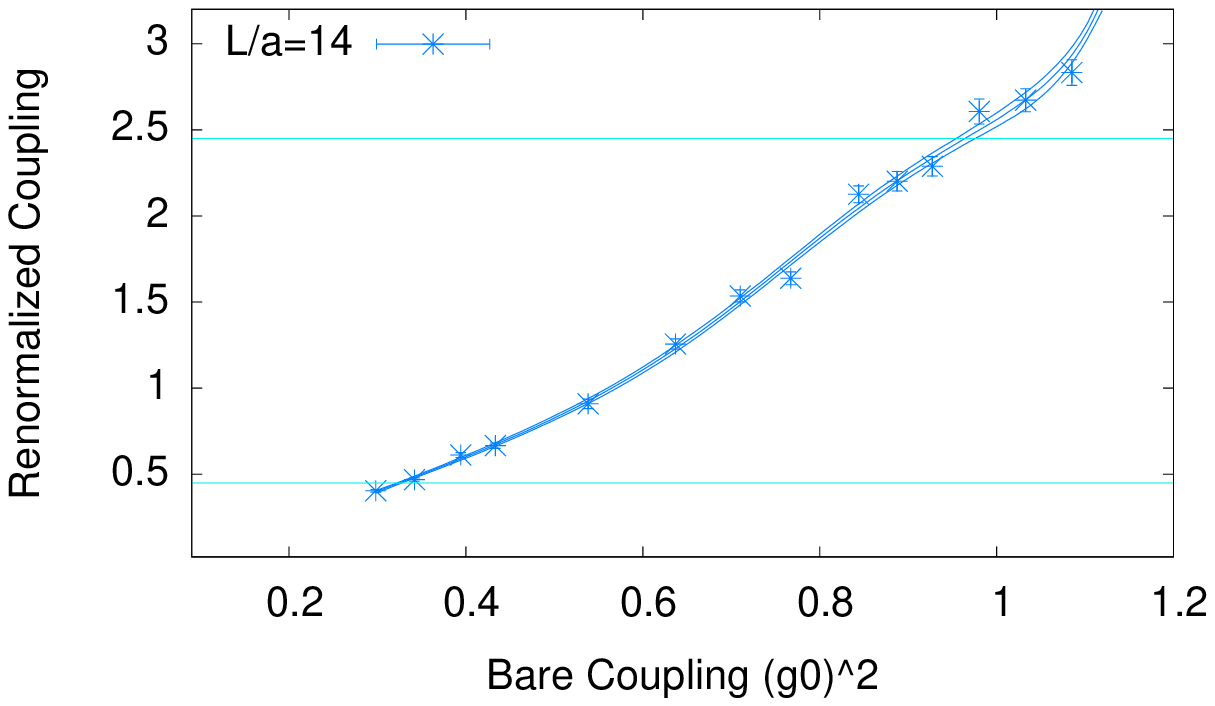}}
\vspace*{1mm}\\
\hspace*{83mm}\vspace*{-7mm}\\
\hspace*{2mm}
\scalebox{0.65}{\includegraphics{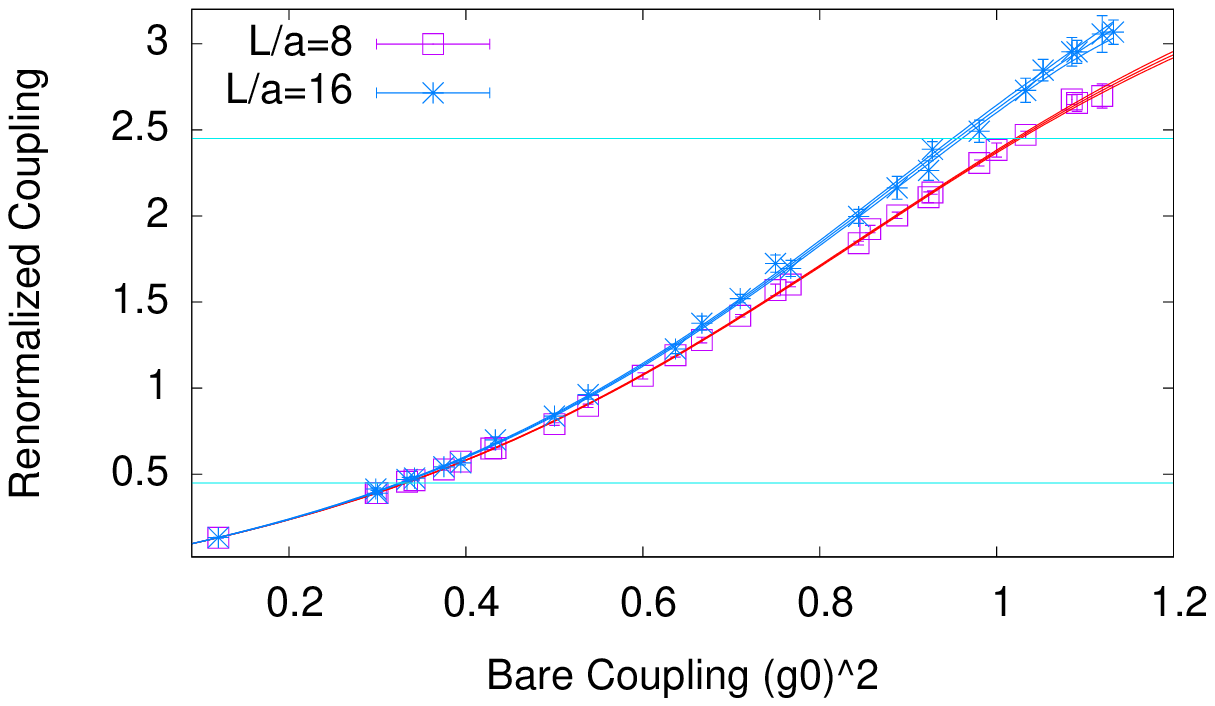}}
\hspace*{3mm}
\scalebox{0.65}{\includegraphics{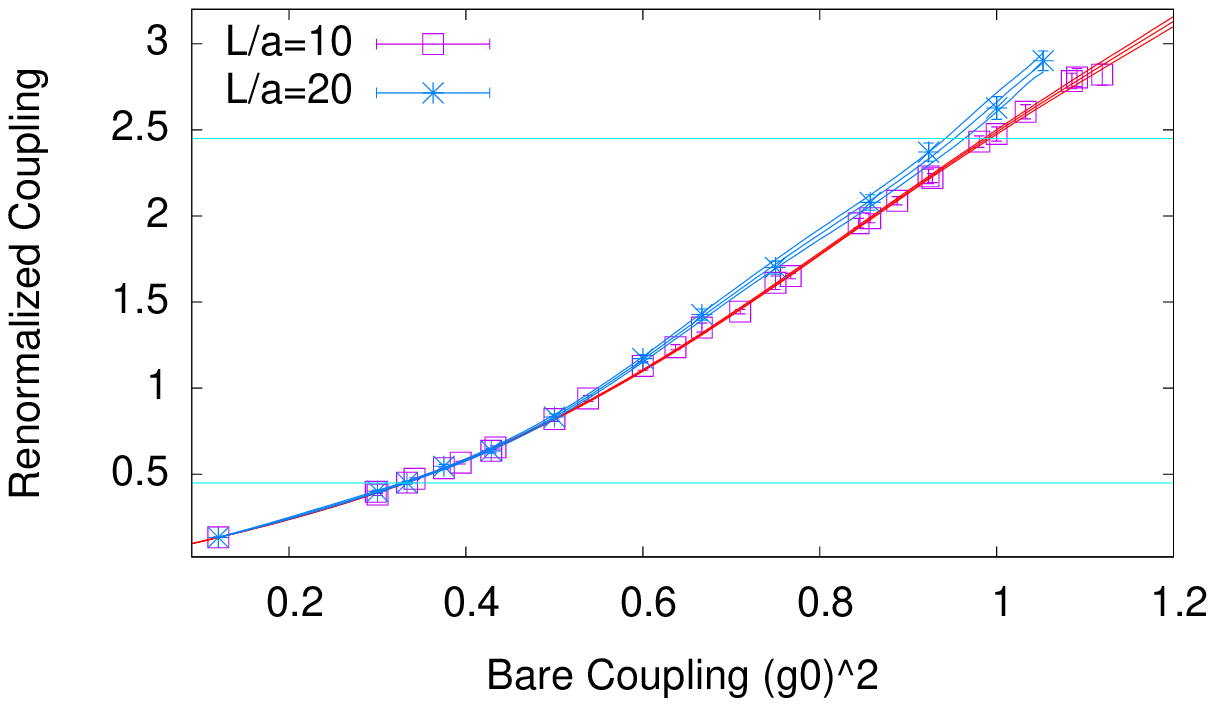}}
\caption{
The renormalized coupling, $u_{\rm latt}$, from the simulations on the $L/a=6,8,10,12,14,16,20$ are shown by points with error bars. 
Fit functions, Eq.~(\ref{eq:non_decr_poly}),  are shown as curves. 
}
\label{fig:coupling}
\end{figure}

From our study of the plaquettes in Sec.~\ref{sec:plaquette}, it is
evident that our simulations are all carried out in the weak-coupling
regime.   This is also reflected on the data points plotted in
Fig.~\ref{fig:coupling}, in which we see that all our renormalised
couplings, $u_{\rm latt}$, are non-decreasing when $\beta$ decreases.
This leads to the use of the non-decreasing polynomial,
\beq
u_{\rm latt} = f(u_{0}) = \int du_{0} \mbox{ }\left (
  \sum_{m=0}^{N_{\rm deg}} c_{m} u_{0}^{m} \right )^{2} = \sum_{n=0}^{N_{h}}
h_{n} u_{0}^{n} \mbox{ }\mbox{ }\left ( {\rm where~}u_{0} \equiv \frac{1}{\beta} =
  \frac{g^{2}_{0}}{6} \right ) ,
\label{eq:non_decr_poly}
\eeq
%
in the $\beta{-}$interpolation procedure at fixed $L/a$.   We
implement the constraint from perturbation theory,
Eq.~(\ref{eq:high_beta_interp}), which results in,
\beq
 h_{0} = 0 , \mbox{ } h_{1}=6 \mbox{ } ({\rm then~}c_{0} = \sqrt{6}) .
\label{eq:non_decr_constraint}
\eeq
%
%
This constraint leads to the number of fit parameters,
\beq
 N_{\rm param} = N_{\rm deg} = \frac{N_{h}-1}{2} ,
\label{eq:nparam_ndeg}
\eeq
where $N_{\rm deg}$ and $N_{h}$ are defined in Eq.~(\ref{eq:non_decr_poly}).
The use of the non-decreasing polynomial ansatz makes the Runge
phenomenon milder compared to the simple polynomial fits.   The
inverse of the fit
function in Eq.~(\ref{eq:non_decr_poly}) is also single-valued.
This is essential in the step-scaling method.  The results of applying
this (uncorrelated)  fitting procedure in the $\beta$ interpolation are shown in Fig.~\ref{fig:coupling}.
The optimal choices of $N_{\rm param}$, leading to the best (smallest)
$\chi^{2}/{\rm d.o.f.}$, are listed in Table~\ref{tab:chisq_beta_interpolation}.

In order to estimate systematic error resulting from the
interpolation in $\beta$, we change the fit function from
Eq.~(\ref{eq:non_decr_poly}) to a simple polynomial function,
\beq
u_{\rm latt} = \tilde{f}(u_{0}) = \sum_{m=0}^{\tilde{N}_{\rm deg}} \tilde{c}_{m} u_{0}^{m} ,
\label{eq:simple_polyn_beta}
\eeq
with the constraint,
\beq
 \tilde{c}_{0} = 0,\mbox{ }\tilde{c}_{1} = 6 .
\label{eq:pt_constraint_simple_polyn}
\eeq
from the validity of perturbation theory at high${-}\beta$.  This
constraint results in the number of fit parameters,
\beq
 \tilde{N}_{\rm param} = \tilde{N}_{\rm deg} -1 .
\label{eq:naparam_ndeg_simple_polyn}
\eeq
The values of $\tilde{N}_{\rm param}$ for the best $\chi^{2}/{\rm d.o.f.}$ are
presented in Table~\ref{tab:chisq_beta_interpolation}.
\begin{table}
\begin{tabular}{|c|c|c|}\hline
\multicolumn{3}{|c|}{Non-decreasing Polynomial}\\ \hline
L/a & $N_{\rm param}$ & $\chi^{2}/{\rm
  d.o.f.}$ \\ \hline
6  & 7 &   1.654259 \\ \hline
8  & 5 &   0.837240 \\ \hline
10 & 5 &   0.828201 \\ \hline
12 & 4 &   1.597743 \\ \hline
14 & 4 &   2.498352 \\ \hline
16 & 4 &   0.834323 \\ \hline
20 & 7 &   0.685983 \\ \hline
\end{tabular}
\hspace{20mm}
\begin{tabular}{|c|c|c|}\hline
\multicolumn{3}{|c|}{Simple Polynomial}\\ \hline
L/a & $\tilde{N}_{\rm param}$ & $\chi^{2}/{\rm
  d.o.f.}$ \\ \hline
6  & 8   &    1.580600\\ \hline
8  & 11  &    0.652351\\ \hline
10 & 5   &    0.819650\\ \hline
12 & 4   &   1.612676\\ \hline
14 & 6   &   2.608492\\ \hline
16 & 4   &     0.837765\\ \hline
20 & 7   &    0.689820\\ \hline
\end{tabular}
\caption{
Left: The $\chi^{2}/{\rm d.o.f.}$ of the $\beta$
  interpolation using Eq.~(\ref{eq:non_decr_poly}). $N_{\rm param} = N_{\rm deg} = \frac{N_{h} -1}{2}$ is the number of fit parameters.  
Right: The $\chi^{2}/{\rm d.o.f.}$ of the $\beta$
  interpolation using Eq.~(\ref{eq:simple_polyn_beta}). $\tilde{N}_{\rm param} = \tilde{N}_{\rm deg} - 1$ is the number of fit parameters.}
\label{tab:chisq_beta_interpolation}
\end{table}
\subsection{Interpolation for $L/a=7$}
\label{sec:L7_interpolation}
As indicated at the end of Sec.~\ref{sec:step_scaling},  it is
desirable to gain more information regarding systematic errors in 
the continuum extrapolation.   In view of the fact that our reference
input renormalised couplings are computed on $L/a = 6, 8, 10$, a
practical way to proceed is to have data for $L/a=7$.  This enables us
to attempt the step-scaling study,
\beq
\label{eq:step_arrow}
 (L/a = 6, 7, 8, 10) \longrightarrow (sL/a = 12, 14, 16, 20) ,
 {\rm~where~} s = 2,
\eeq
without having to perform simulations on large lattices, such as $L/a=24$.

Since staggered fermions are used in this work, we have to use an
interpolating procedure to obtain $u_{\rm latt}$ for $L/a=7$. 
To have a well-motivated method for this interpolation, we
resort to the $\beta{-}$function of the theory. 
It is well-established that the coupling constant in SU(3) gauge
theory with twelve flavours runs slowly compared to, {\it e.g.}, QCD.
This is reflected on the fact that a small change in the renormalised
coupling has to result from a significant variation of the scale.  As
shown in Fig.~\ref{fig:coupling}, this is indeed the
case.  Namely, enlarging the box size by a
factor of two induces very little changes in $u_{\rm latt}$, and one can
{\it locally} approximate the $\beta{-}$function using a linear form
\beq
\label{eq:linear_beta_function}
 L \frac{d u_{\rm latt}}{d L} = \beta(u_{\rm latt}) \approx a_{l} +
 b_{l}\mbox{ } u_{\rm latt} ,
\eeq
where $a_{l}$ and $b_{l}$ are unknown parameters.  We stress that this
approximated form is not based on perturbation theory, and is only valid within a
small range of $u_{\rm latt}$.  That is, in different ranges of
$u_{\rm latt}$, the
parameters, $a_{l}$ and $b_{l}$, have different
values.

\begin{figure}
   \includegraphics*[width=0.40\textwidth,angle=0]{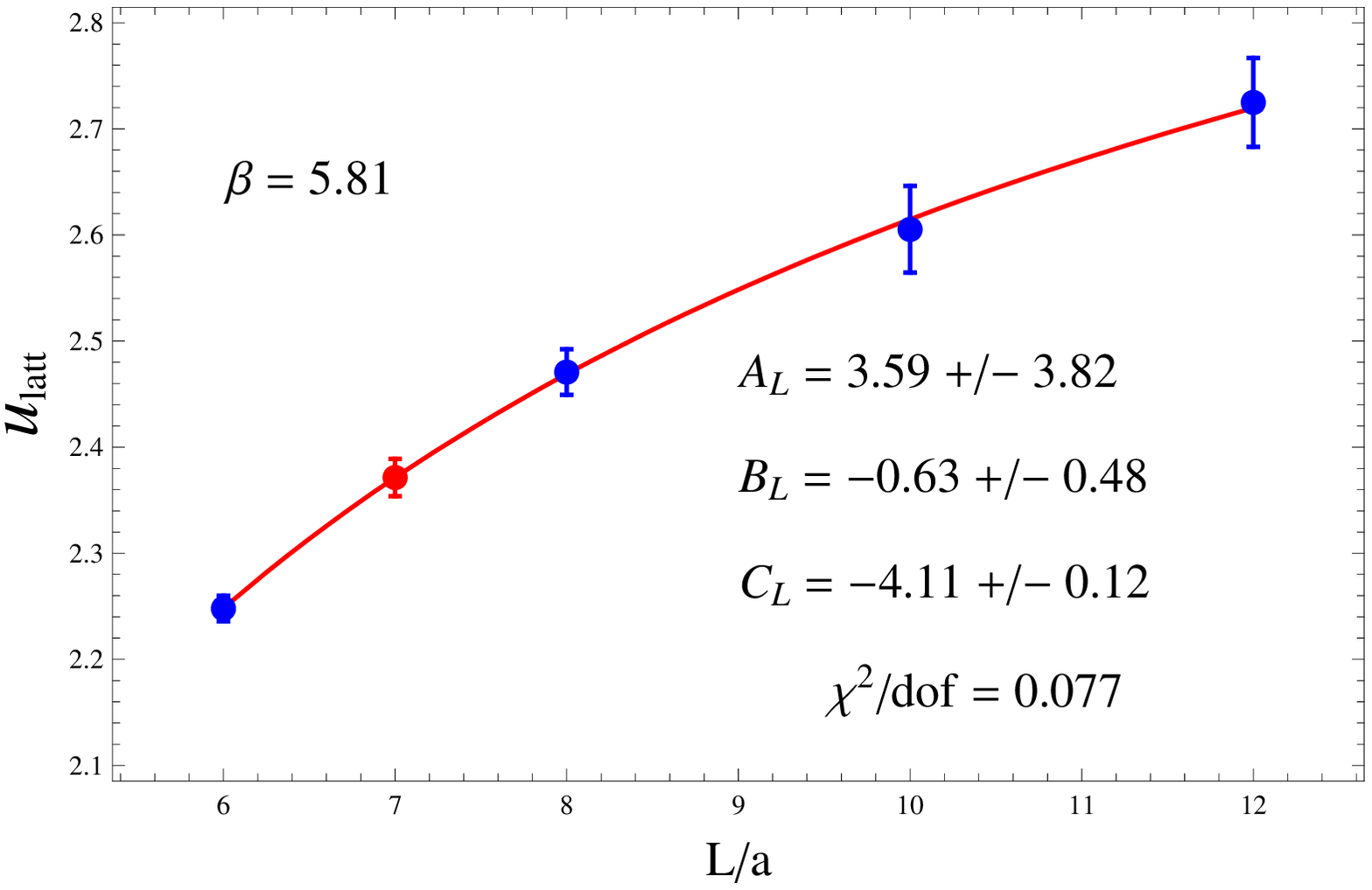}
   \includegraphics*[width=0.40\textwidth,angle=0]{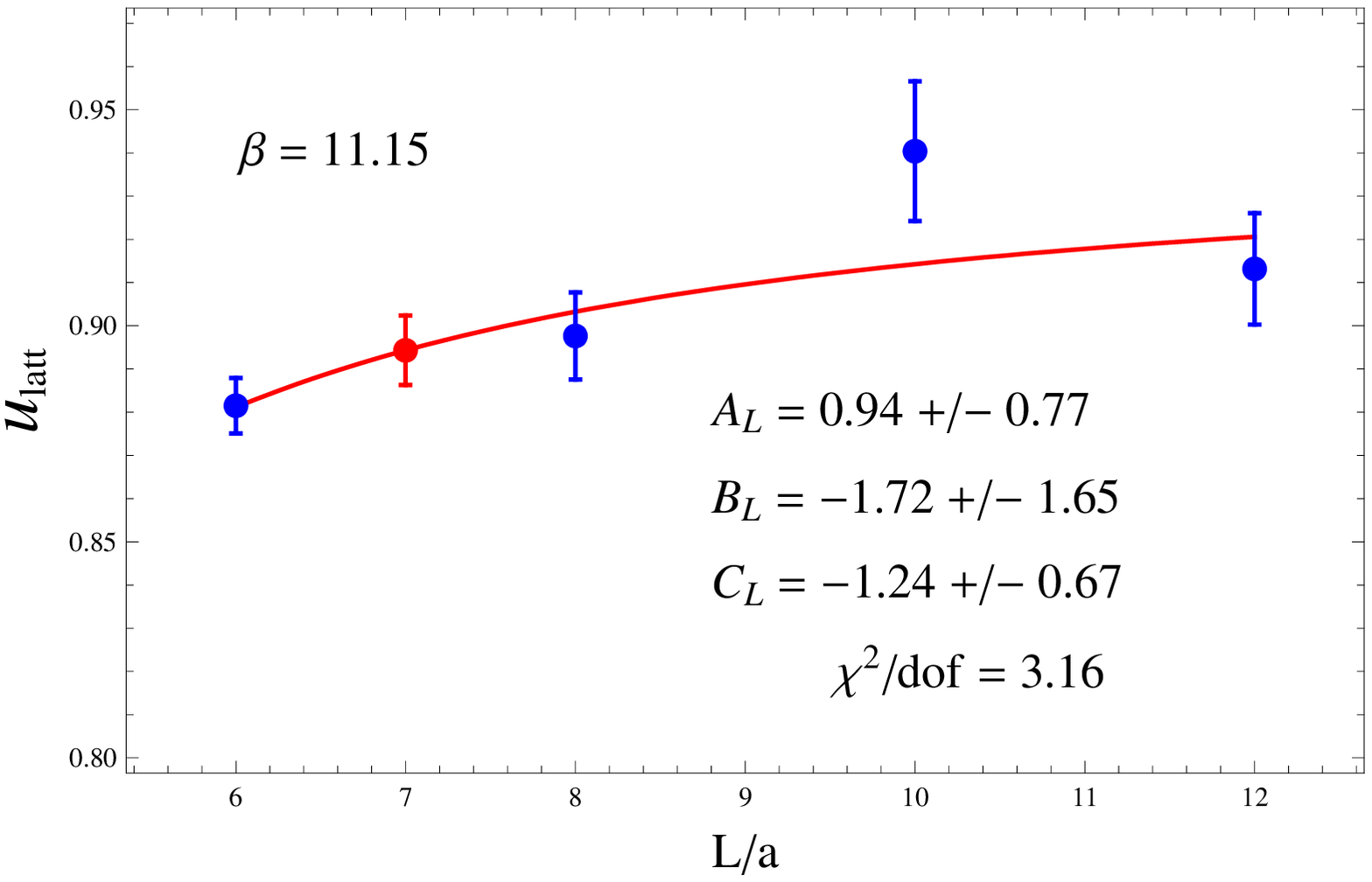}
\caption{The interpolation for obtaining the running coupling constant
at $L/a = 7$ at two fixed lattice spacings using Eq.~(\ref{eq:L7_interpolation_function}).}
\label{fig:L7_interpolation}
\end{figure}

To determine $u_{\rm latt}$ on the $L/a = 7$ lattice, we use our data on the $L/a
= 6, 8, 10, 12$ lattices, and interpolate with the 
function, 
\beq
\label{eq:L7_interpolation_function}
 u_{\rm latt} = A_{L} + C_{L} \left ( \frac{L}{a} \right )^{B_{L}} ,
\eeq
at fixed lattice spacing.  The unknown coefficients, $A_{L}$, $B_{L}$, and $C_{L}$ are related to $a_{l}$ and
$b_{l}$, and the integration constant in solving
Eq.~(\ref{eq:linear_beta_function}).  
Figure~\ref{fig:L7_interpolation} shows two representative plots for
the interpolation using Eq.~(\ref{eq:L7_interpolation_function}).   It
is obvious that the interpolation is smooth, and
the values of the coefficients, $A_{L}$, $B_{L}$, and $C_{L}$, can
vary significantly in different ranges of $u_{\rm latt}$.
The fits presented in Fig.~\ref{fig:L7_interpolation} are performed on
the $L/a = 6, 8, 10, 12$ data without $\beta{-}$interpolation.   
%

Equations~(\ref{eq:linear_beta_function}) and
(\ref{eq:L7_interpolation_function}) are used to motivate an
interpolation function in $L/a$ at fixed $a$ ($\beta$ value).
However, the the effects of the lattice spacing can appear as powers
of $(a/L)^{2}$ in our simulations.  This means the data points used in each of this
volume interpolation may have different lattice artefacts, leading to
systematic effects introduced in this procedure.  In view of this, we
do not include the $L/a=7$ data in our central analysis procedure, and
only use them to perform the step-scaling investigation in
Eq.~(\ref{eq:step_arrow}) as a means to estimate errors in the
continuum extrapolation.
 


Another issue in this volume-interpolation method for obtaining the
$L/a=7$ data is statistical correlation.  The procedure is carried out using uncorrelated
fits in this work.  However, it is
natural to expect that there will be correlation between the $L/a=7$
(interpolated) data and those extracted directly from independent
simulations on $L/a=6,8,10,12$.  This
correlation has to be closely examined,  since all these data are used
in the investigation of the continuum extrapolation, as discussed in
Sec.~\ref{sec:continuum_extrapolation}.
For this purpose, we study the likelihood function,
\beq
 L(u_{i}, u_{j}) = \frac{1}{2\pi \sqrt{{\rm det}\left ( {\rm
         Cov}\right )}} {\rm exp}\left \{ \frac{-1}{2} \left ( u_{i} -
   \bar{u}_{i}  \right ) \left [ {\rm Cov}^{-1}\right ]_{ij} \left ( u_{j}
   - \bar{u}_{j} \right
)\right \} ,
\eeq
where $u_{i}$ denotes $u_{\rm latt}$ computed on the
lattice volume $L/a = i$.  Here $u_{i}$ is kept as a variable, and 
$\bar{u}_{i}$ is its central value for this quantity from
our simulation.  The symbol Cov is the covariance matrix which can be computed
from the bootstrap samples of $u_{i}$ and $u_{j}$ obtained from
numerical calculations.

\begin{figure}[t]
\includegraphics[scale=0.64]{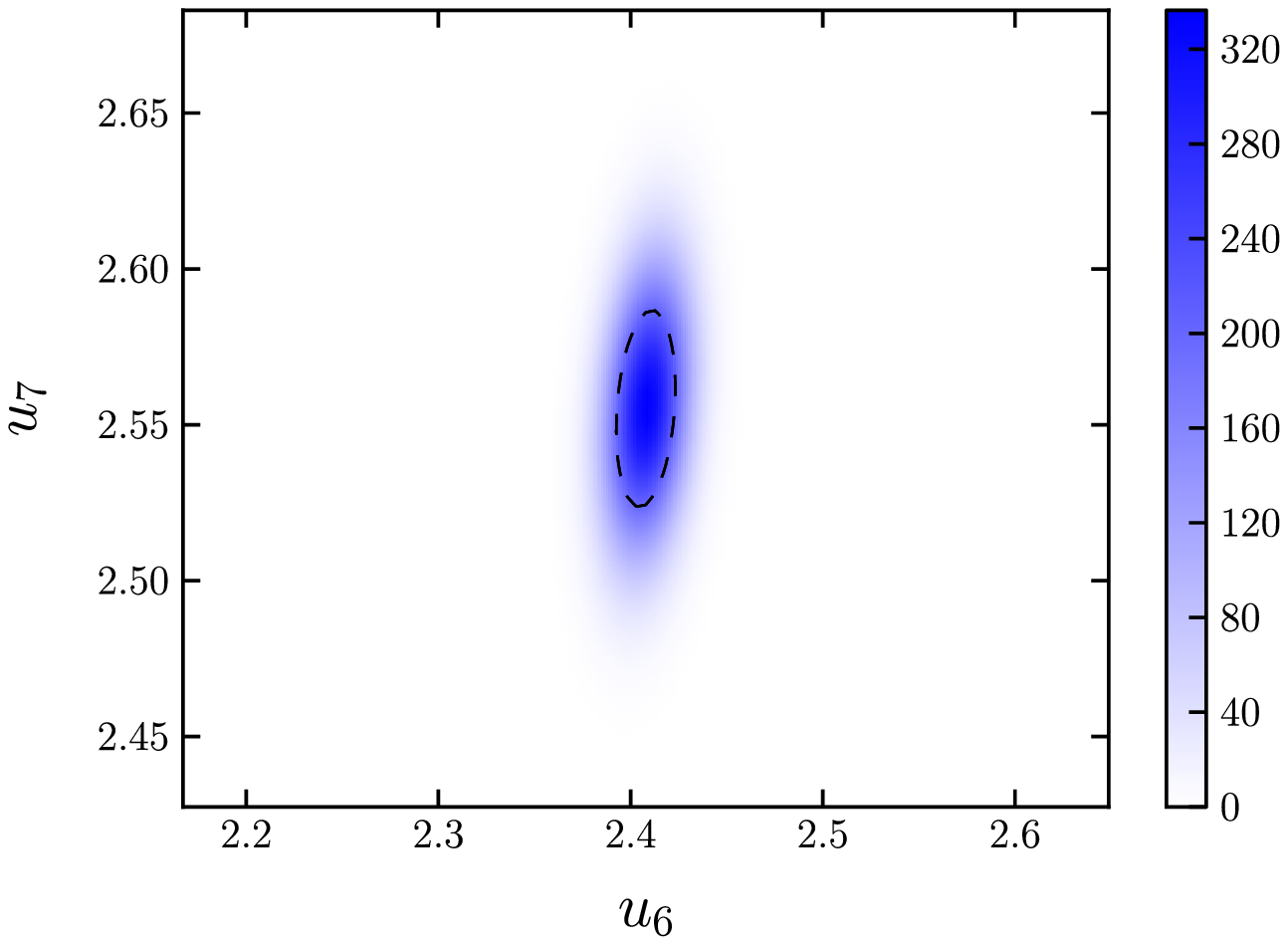}
\includegraphics[scale=0.64]{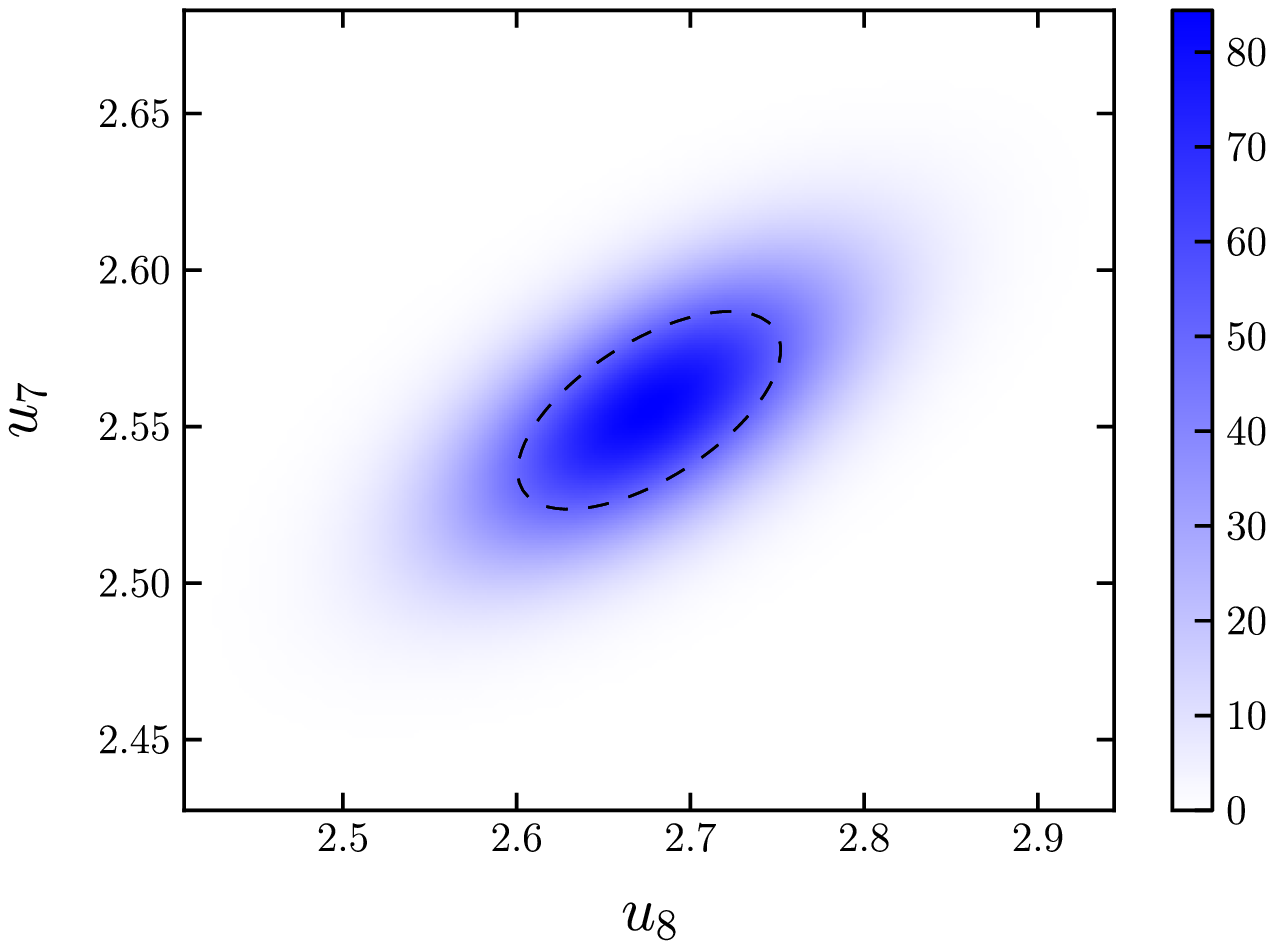} 
\includegraphics[scale=0.64]{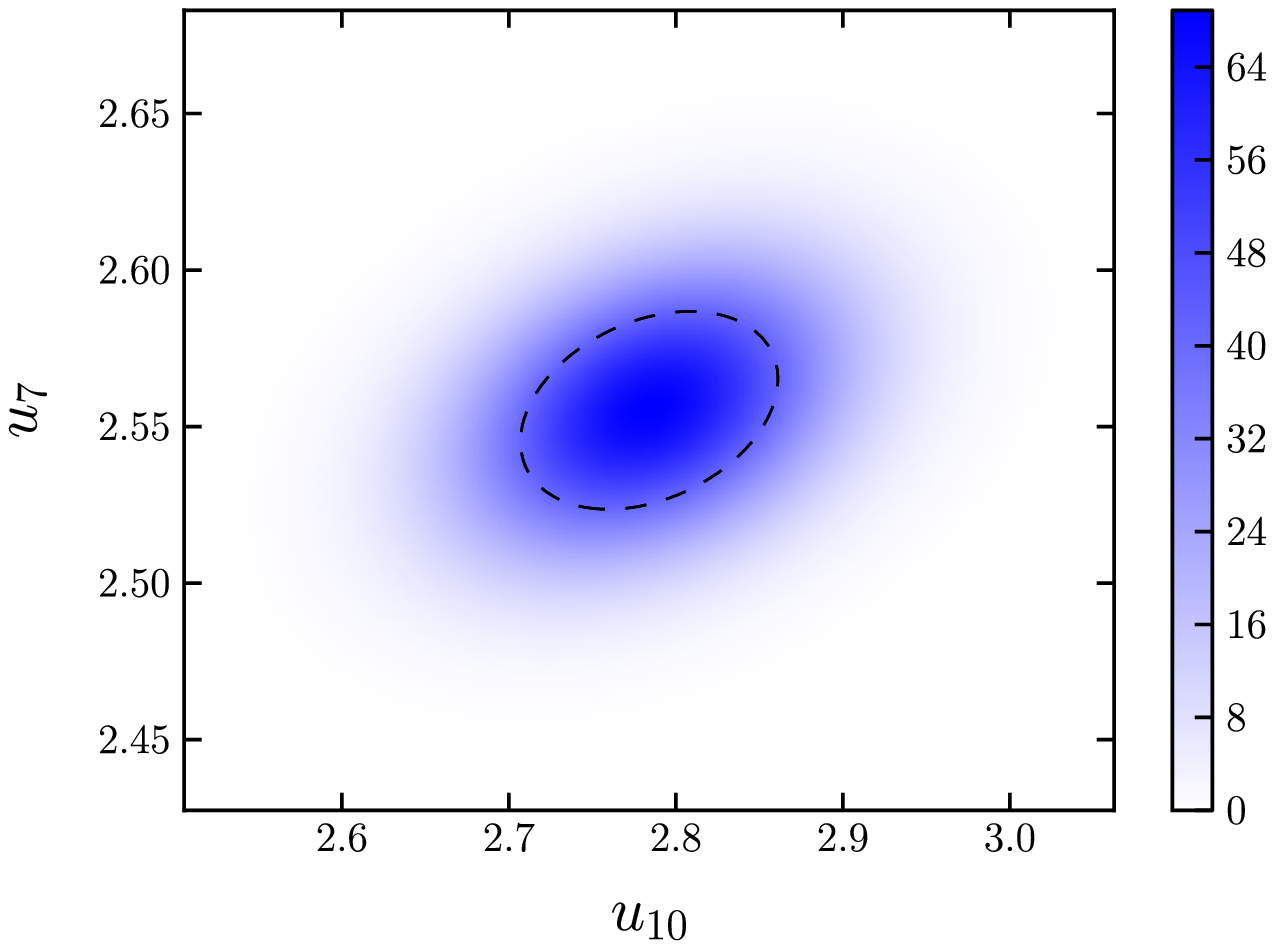}
\includegraphics[scale=0.64]{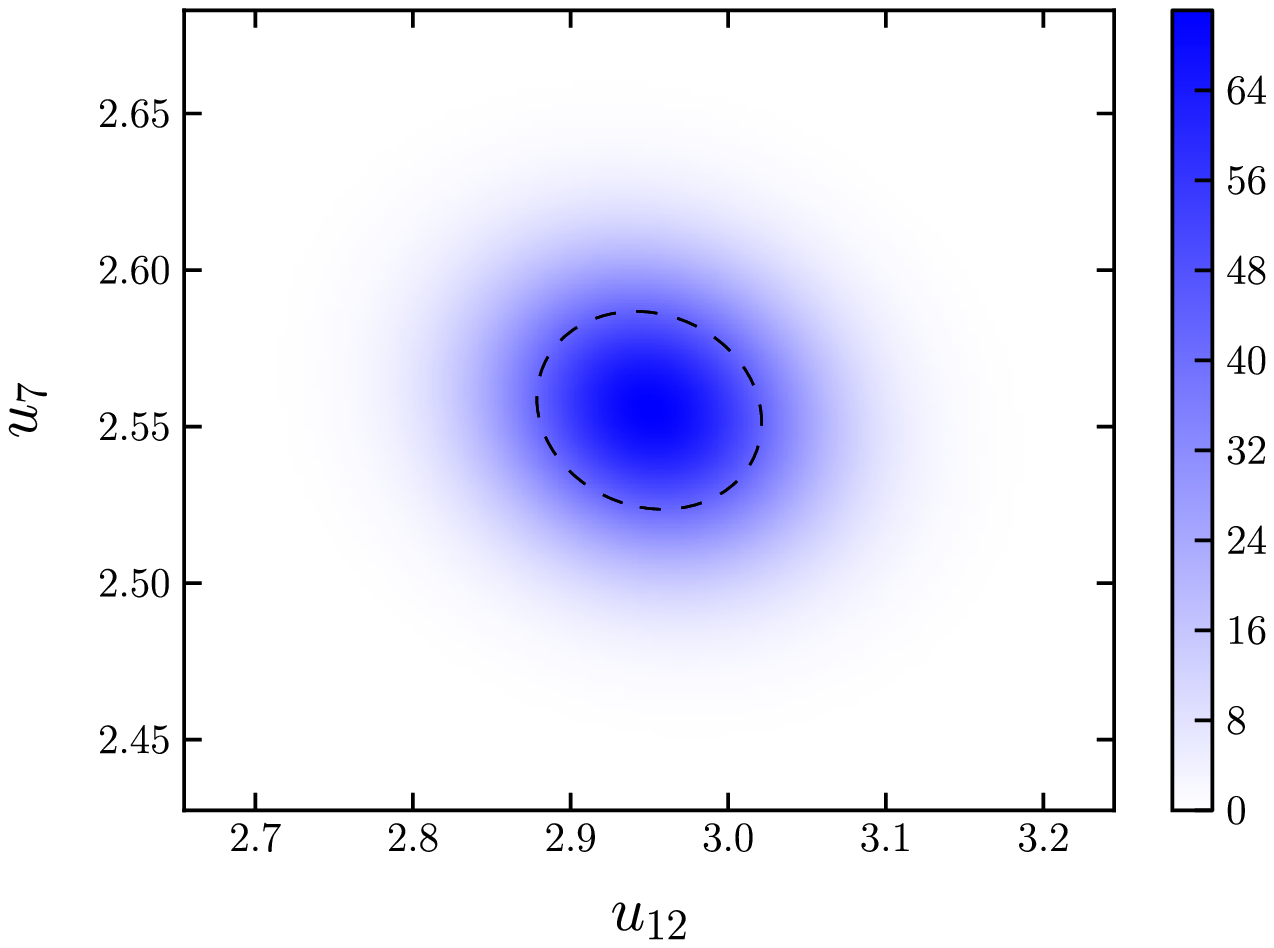}
\caption{The likelihood function plotted against ($u_{6}$, $u_{7}$),
  ($u_{8}$, $u_{7}$), ($u_{10}$, $u_{7}$), and ($u_{12}$, $u_{7}$) at
  $\beta = 5.53$.  The dashed curves indicate the standard error ellipses.}
\label{fig:L7_L12_L10_L8_L6_corr}
\end{figure}

Our investigation shows that, although the coupling constant on the
$L/a = 7$ lattice is interpolated using those on the $L/a=6,8,10,12$
lattices, it only shows significant correlation with that on the $L/a=8$
lattice.  In Fig.~\ref{fig:L7_L12_L10_L8_L6_corr}, we display an example of
this likelihood-function study performed for $\beta = 5.53$.  
It is obvious from this figure that coupling constants on $L/a = 7$  
and $L/a=6, 10, 12$ exhibit very small correlation, while it is the opposite
between $L/a=7$ and $L/a=8$.  The corresponding covariance matrices
for the example in Fig.~\ref{fig:L7_L12_L10_L8_L6_corr}  are

\bea
 {\rm Cov}^{(\beta=5.53)} &=& \left ( \begin{array}{cc}
0.000235 & 0.000104 \\
0.000104 & 0.000999 
                       \end{array}\right )  {\rm ~for~} u_{6}{-}u_{7}
                     , \nonumber \\
 {\rm Cov}^{(\beta=5.53)} &=& \left ( \begin{array}{cc}
0.005731 & 0.001473 \\
0.001473 & 0.000999  
                       \end{array}\right )  {\rm ~for~} u_{8}{-}u_{7}
                     , \nonumber \\
 {\rm Cov}^{(\beta=5.53)} &=& \left ( \begin{array}{cc}
0.005945 & 0.000776 \\
0.000776 & 0.000999 
                       \end{array}\right )  {\rm ~for~} u_{10}{-}u_{7}
                     , \nonumber \\
\label{eq:example_cov_matrices}
 {\rm Cov}^{(\beta=5.53)} &=& \left ( \begin{array}{cc}
0.005090 & −0.000276  \\
−0.000276 &  0.000999
                       \end{array}\right )  {\rm ~for~} u_{12}{-}u_{7} .
\eea
The volume, $L/a=7$, is one of the ``small''
lattices, on which we compute the reference coupling instead of
the step-scaling function.  The importance of the above study is the
demonstration that there is negligible correlation between data on
this lattice and that on the ``large'' lattice, $L/a=12$, from which we
compute the step-scaling function.  Furthermore, the statistical
errors of the data obtained on all our small lattices are small.  In view of this,
it is reasonable to expect that this
correlation between the $L/a=7$ and $L/a=8$ TPL coupling constants does not necessitate
correlated fits in the continuum extrapolation.

The above study of the data correlation also leads to the conclusion that
one has to be very cautious about interpolating $u_{\rm latt}$ in
$L/a$.  In certain analysis procedures, such as the one we adopted in 
Ref.~\cite{Aoyama:2011ry} by setting the step size to 1.5,  large
correlation amongst data used in the continuum extrapolation can occur.

\subsection{Continuum extrapolation for the step-scaling function}
\label{sec:continuum_extrapolation}
The last step in our analysis is the continuum extrapolation for the
step-scaling function, $\sigma(u)$, defined in Eqs.~(\ref{eq:step_scaling_function})
and (\ref{eq:sigma_at_s2}).   Since unimproved staggered fermions and
the Wilson plaquette action
are used in this work, we will investigate $(a/L)^{2}$ dependence in
the lattice step-scaling function,
$\Sigma (\beta, L/a, u, s=2)$.  
\begin{figure}[t]
\includegraphics[scale=0.95]{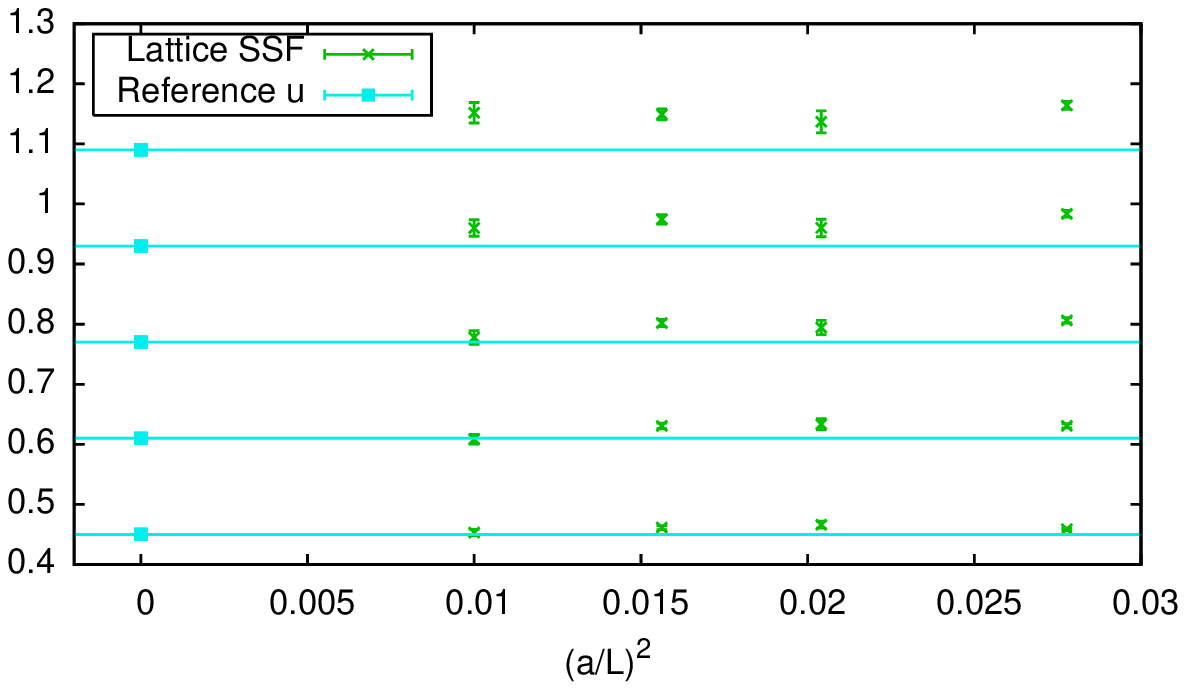}
\includegraphics[scale=0.95]{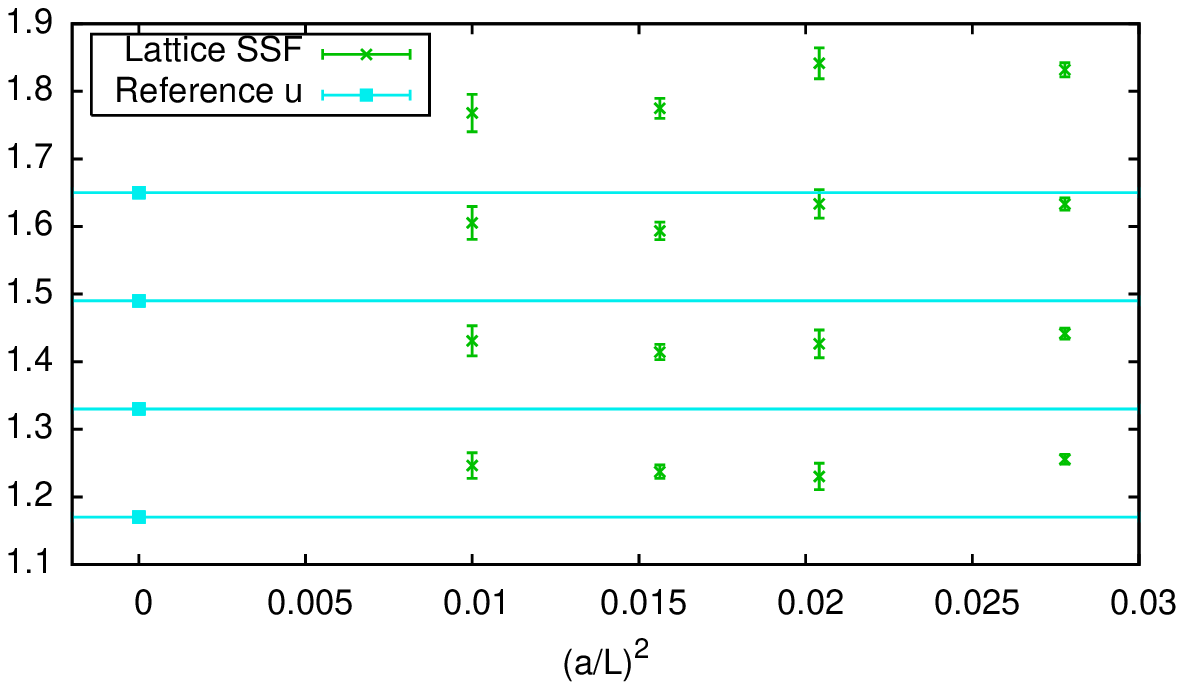}
\includegraphics[scale=0.95]{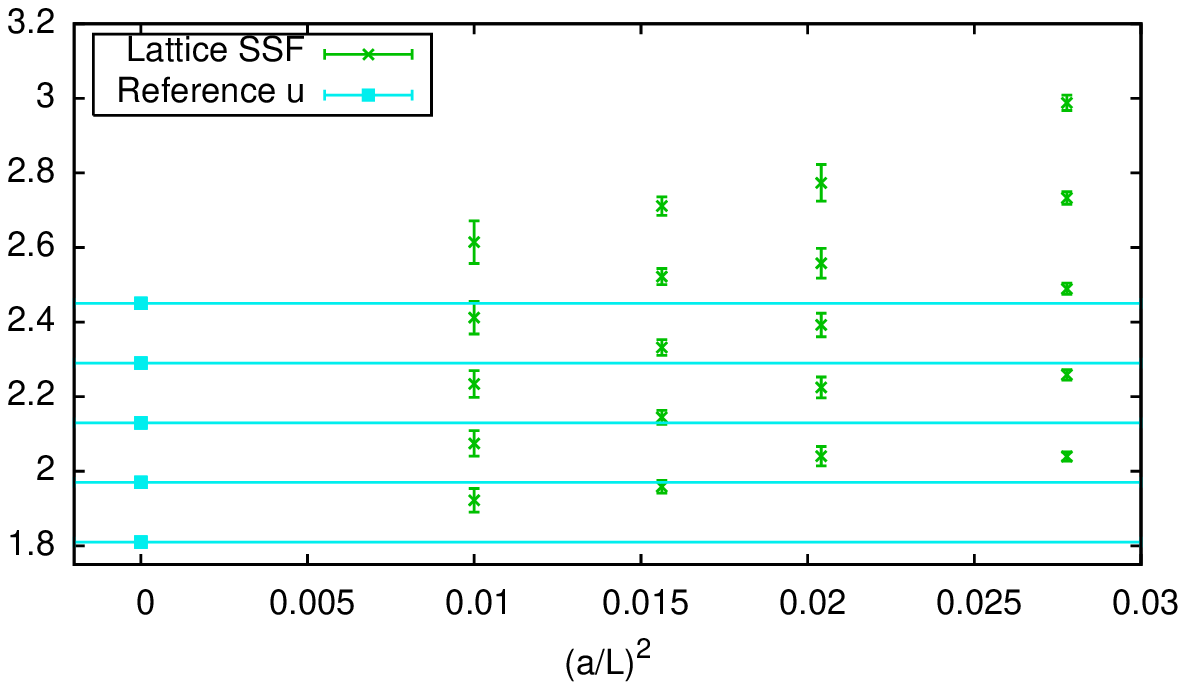}
\caption{Lattice-spacing dependence of the step-scaling function
  (SSF) in weak, intermediate and strong coupling regimes (from the
  top).  The horizontal lines indicate the central values of the input
reference $u$.}
\label{fig:ssf_a_over_L_sq}
\end{figure}
In Fig.~\ref{fig:ssf_a_over_L_sq}, this dependence is displayed at representative
values of $u$ in the regimes of weak, intermediate and strong
coupling.  From this figure, it is obvious that effects of the lattice
artefacts grow with increasing $u$, as expected.  In the region $u <
0.8$, we see that the step-scaling functions show insignificant
dependence on the lattice spacing, and are almost consistent with the
input reference coupling.  On the other hand, in the strong-coupling
regime, the $a{-}$dependence in $\Sigma$ becomes noticeable,
necessitating good control of the continuum extrapolation in the
investigation of the existence of the IRFP.
It is worth noting that the lattice artefacts tend to make
the step-scaling function larger than its continuum-limit
counterpart, especially in the strong-coupling regime.  This feature
is different from what was discovered in the Sch\"{o}dinger-functional 
scheme~\cite{Appelquist:2007hu,Appelquist:2009ty}.  

In performing the continuum extrapolation for our central analysis procedure, we use our
simulation results for $\Sigma (\beta, L/a, u, s=2)$, obtained at 
$sL/a=12,16, 20$, and carry out  the linear fit ($\sigma_{l}(u)$ and $A_{l}$ are
the fit parameters),
\beq
\label{eq:linear_continuum_extrap}
 \Sigma (\beta, L/a, u, s=2) = \sigma_{l} (u) + A_{l} (u) \left ( \frac{a}{L}
 \right )^{2} ,
\eeq
with the $\beta{-}$values
for various $L/a$ determined by tuning the coupling, $u$,
to be the same on the corresponding small lattices ($L/a = 6, 8,
10$).   This procedure does not include the $L/a=7$ data which are
extracted with an additional volume-interpolation, as detailed in
Sec.~\ref{sec:L7_interpolation}.  

To estimate systematic errors in the continuum extrapolation, we
include the volume-interpolated, $L/a=7$ data, as well as the
step-scaling functions computed on the lattice, $s L/a=14$.  We first 
perform the quadratic fit ($\sigma_{q}(u)$, $A_{q}$ and $B_{q}$ are the
fit parameters),
\beq
\label{eq:quadratic_continuum_extrap}
 \Sigma (\beta, L/a, u, s=2) = \sigma_{q} (u) + A_{q} (u) \left ( \frac{a}{L}
 \right )^{2} + B_{q} (u) \left ( \frac{a}{L} \right )^{4},
\eeq
to implement the 4-point step-scaling method in Eq.~(\ref{eq:step_scaling_including_L7}).

In order to further account for systematic effects arising from the continuum
extrapolation, we perform two additional linear fits:
\begin{enumerate}
\item Using the data for the step-scaling functions from
   $sL/a=14, 16, 20$ ($L/a = 7, 8, 10$).
 \item Using the data for the step-scaling functions from
   $sL/a=12, 14, 16, 20$ ($L/a = 6, 7, 8, 10$).
\end{enumerate}
\begin{figure}[t]
\includegraphics[scale=0.65]{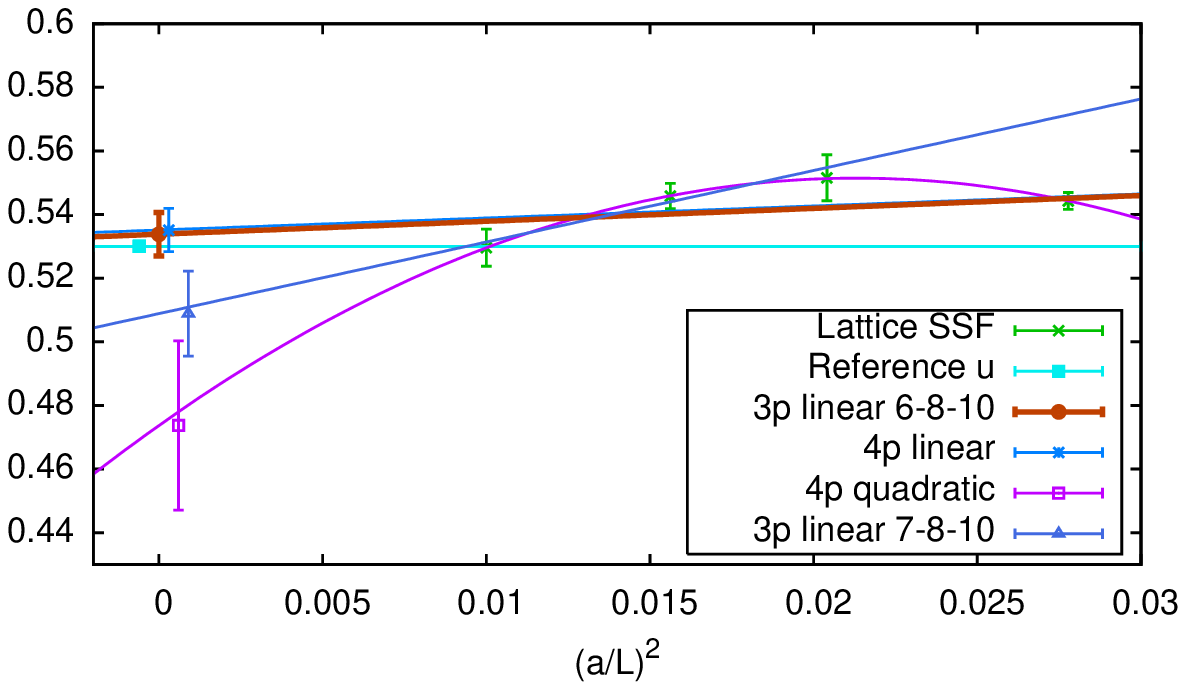}
\includegraphics[scale=0.65]{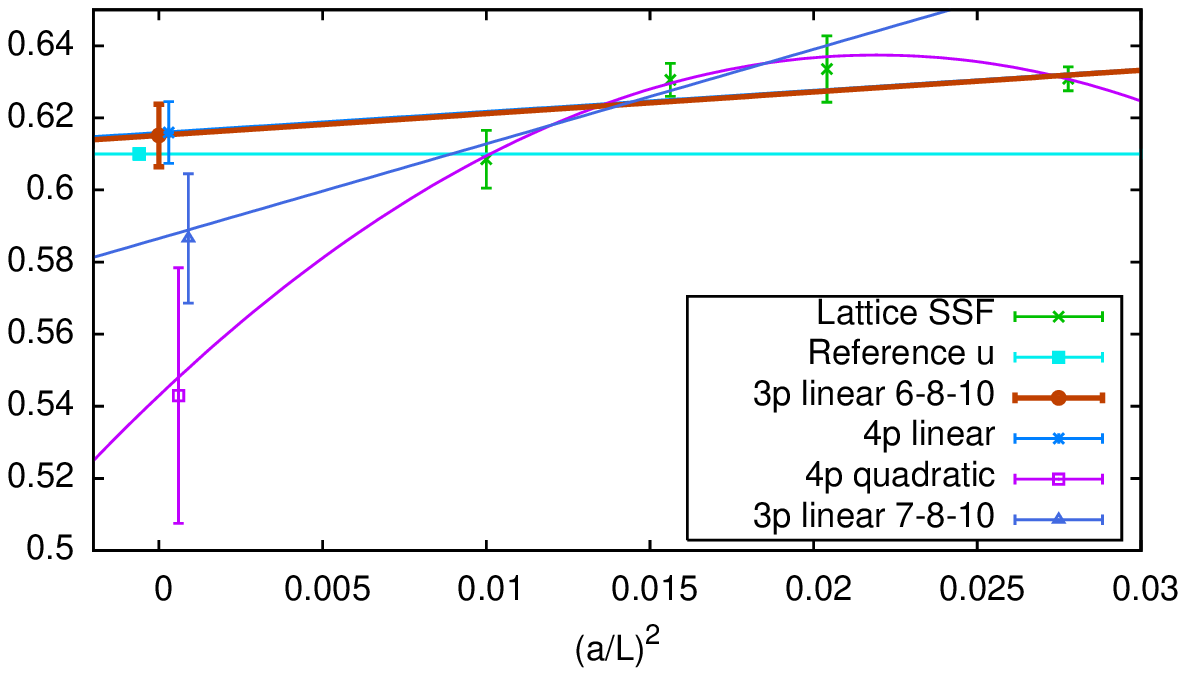}
\includegraphics[scale=0.65]{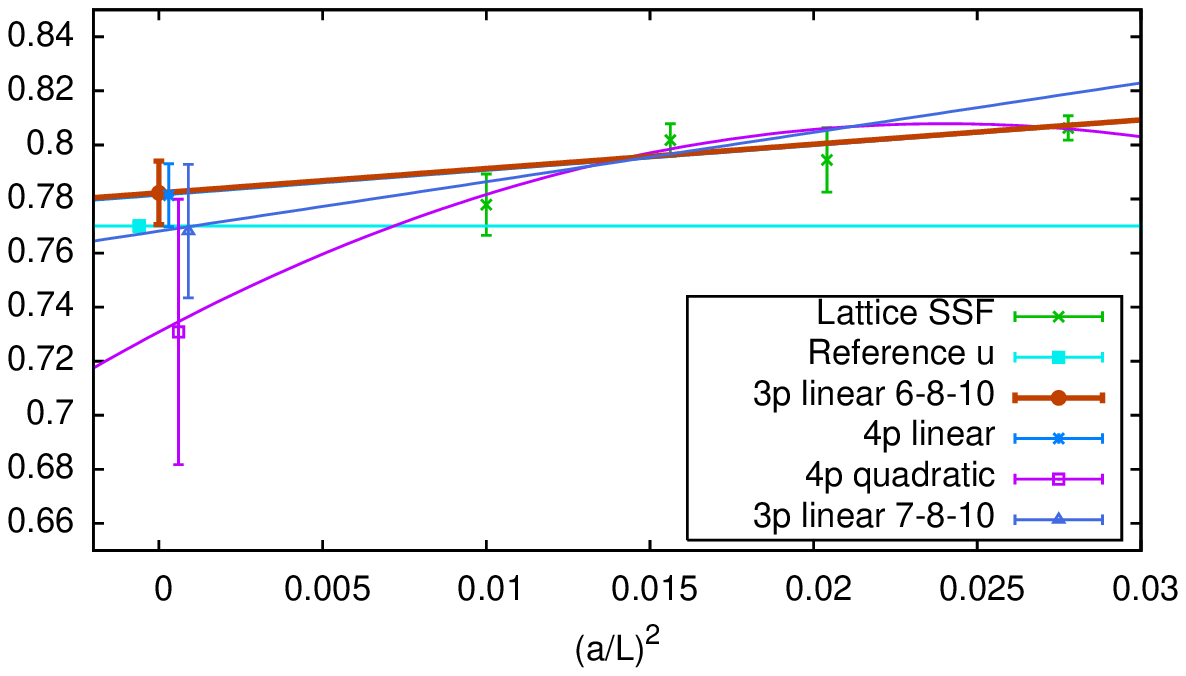}
\includegraphics[scale=0.65]{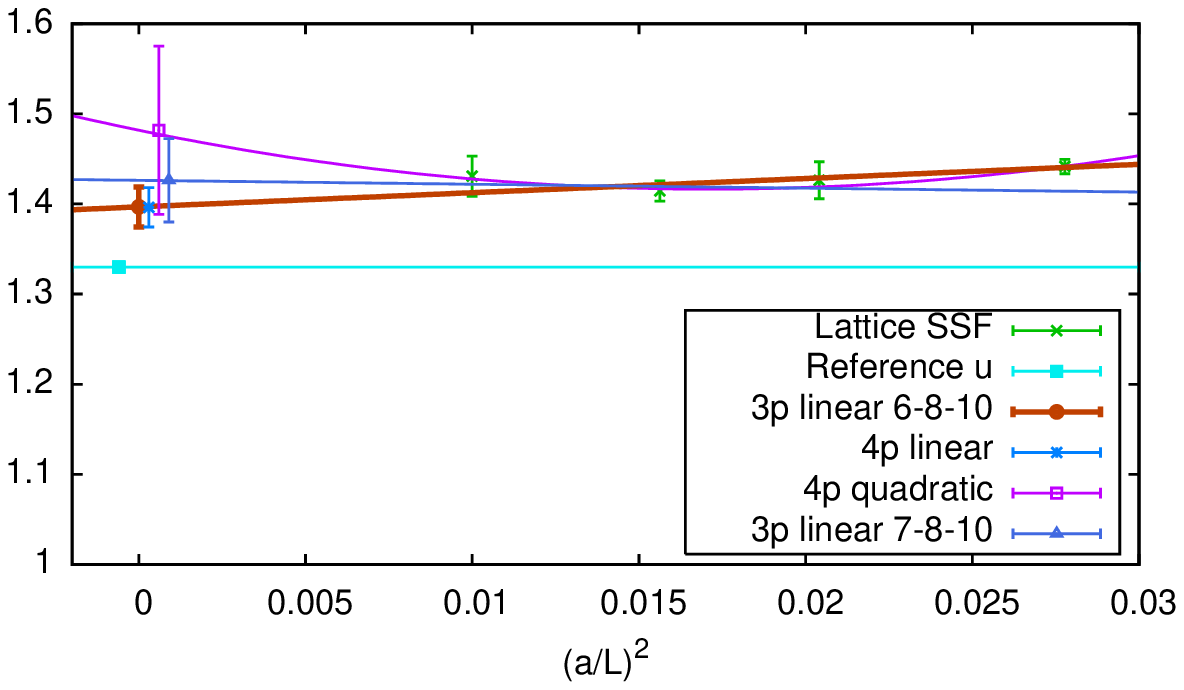}
\includegraphics[scale=0.65]{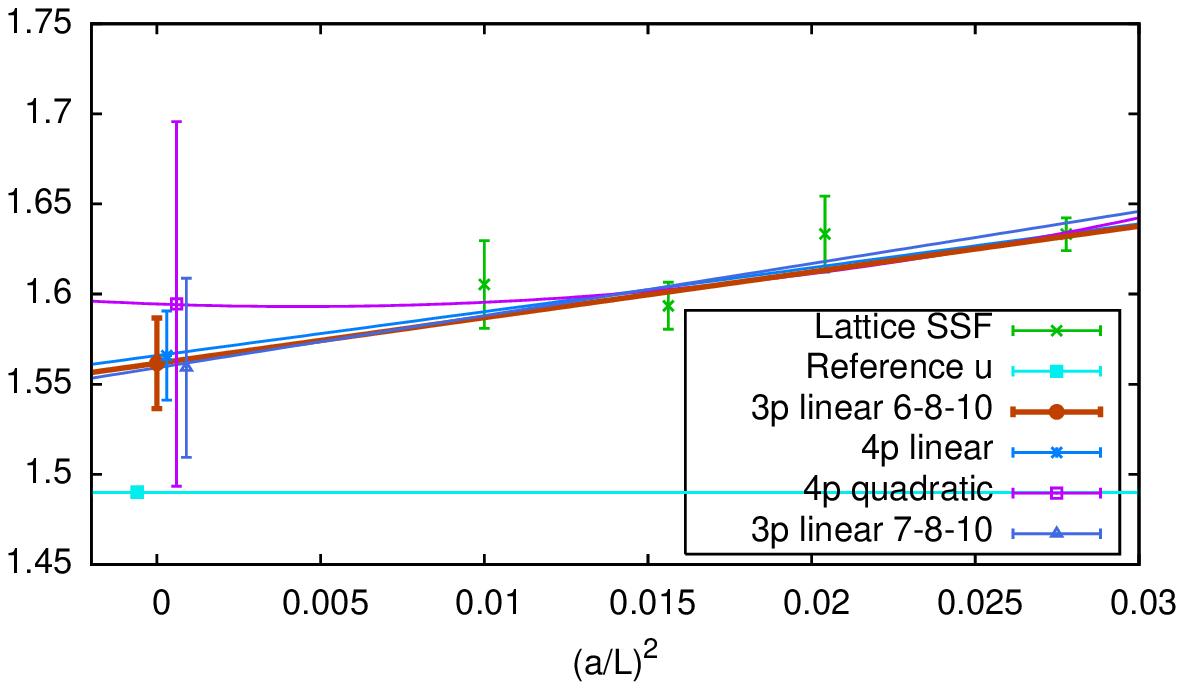}
\includegraphics[scale=0.65]{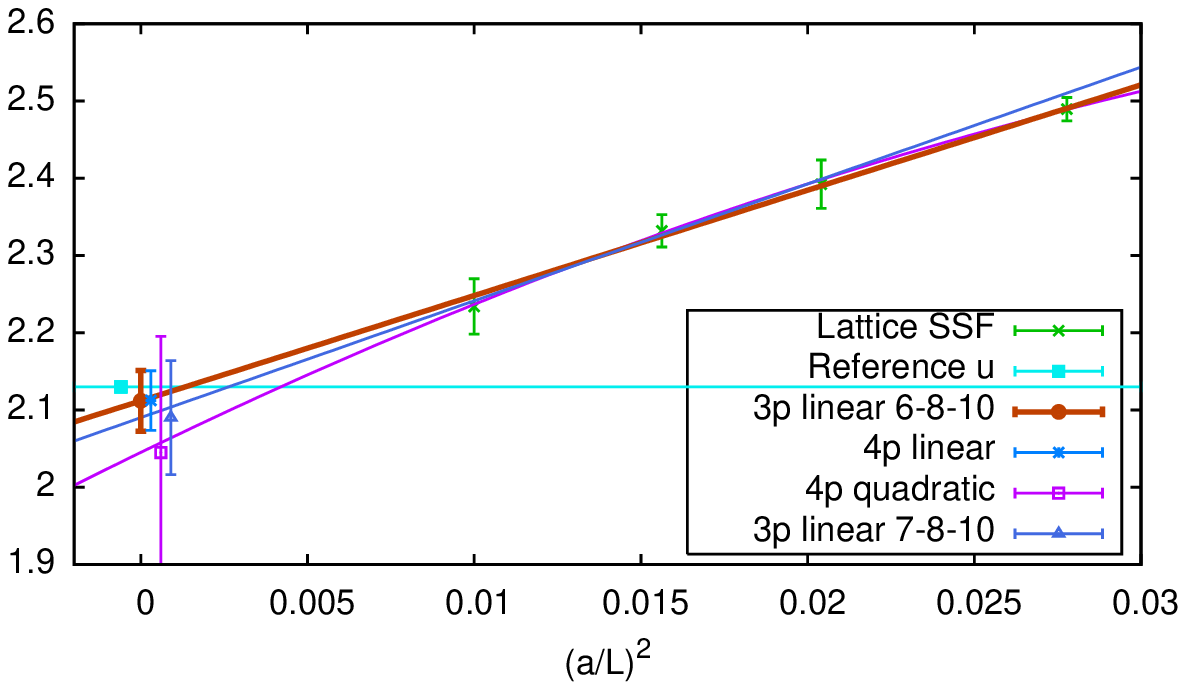}
\caption{Representative cases of the continuum extrapolation for the step-scaling functions
  using the procedures discussed in the main text.  The 3-point linear
extrapolation using data on $L/a=6, 8, 10$ is the central procedure.   The horizontal lines indicate the central values of the input
reference $u$.  As discussed in the main text, the quadratic
fit, and the 3-point linear fit using the $L/a=7,8,10$ data can lead
to unreliable results in the continuum limit in the weak-coupling
regime (top row).}
\label{fig:ssf_continuum_extrap}
\end{figure}
Figure~\ref{fig:ssf_continuum_extrap} shows representative plots of
the continuum extrapolation using the above procedures (quadratic fit
and the three linear fits).
From these plots, we observe that $\sigma_{l}$ and $\sigma_{q}$
are well consistent with each other at intermediate and strong
couplings.  In the weak-coupling regime (top row
of Figure~\ref{fig:ssf_continuum_extrap}), we notice that the
quadratic fit, and the 3-point linear fit using the $L/a=7,8,10$ data are
not consistent with the other two procedures.  They result in
$\sigma(u)$ smaller than $u$ after the continuum extrapolation.
However, we stress that in this regime, the lattice
step-scaling function, $\Sigma$, demonstrates very mild lattice-spacing
dependence, and is almost consistent with the input reference $u$.
This is the consequence of asymptotic freedom.  Furthermore, our data
do not show significant $O(a^{4})$ contributions in the continuum
extrapolation at strong and intermediate couplings (center and bottom rows
of Figure~\ref{fig:ssf_continuum_extrap}), where
the lattice artefacts are expected to be larger compared to the
small${-}u$ region.  In view of this, we conclude that the quadratic
fit, and the 3-point linear fit using the $L/a=7,8,10$ data can be
artificially amplifying statistical fluctuations and leading to
unreliable results in the weak-coupling regime.  In order to properly
address this issue, one has to generate data with very high
statistical accuracy ({\it e.g.}, $< 0.5\%$) at large $\beta$ values.  This is beyond the
scope of this work, since our main focus is on the existence of the
IRFP in the strong-coupling regime.

\section{Final results and discussion}
\label{sec:final_results}
\begin{figure}[t]
\includegraphics[scale=0.9]{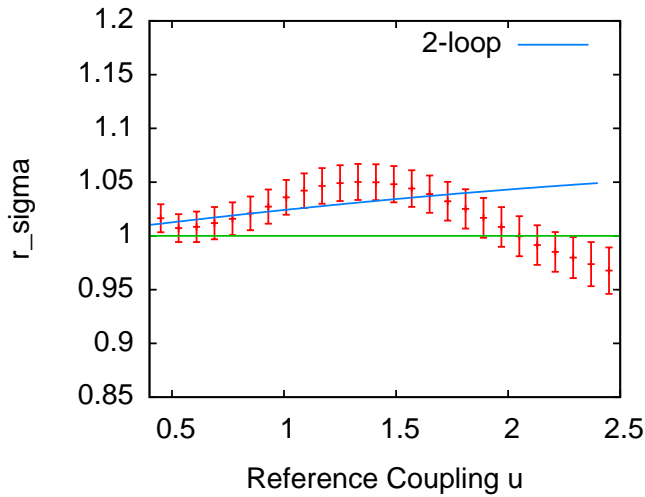}
\caption{$r_{\sigma}(u)$ from the central procedure.}
\label{fig:r_sigma_over_u_central}
\end{figure}

In this section, we present the final results of our analysis, and
discuss the estimation of systematic errors.   
We begin by showing the result from our central analysis
procedure for the ratio
$r_{\sigma} = \sigma(u) / u$, defined in Eq.~(\ref{eq:r_sigma_def}).
In performing this central-procedure analysis, we first interpolate in
the bare coupling, $\beta$, for simulation data obtained at $L/a = 6,
8, 10, 12, 16, 20$, using the non-decreasing polynomial function
in Eq.~(\ref{eq:non_decr_poly}) with the constraint from
Eq.~(\ref{eq:non_decr_constraint}), and the polynomial degrees and the
numbers of fit parameters presented in
Table~\ref{tab:chisq_beta_interpolation}.  We then carry out the step-scaling of
\beq
 L/a = (6, 8, 10) \longrightarrow 2L/a = (12, 16, 20),
\eeq
by extrapolating the step-scaling function to the continuum limit with
the linear form in $(a/L)^{2}$, Eq.~(\ref{eq:linear_continuum_extrap}).  Result
of this central analysis is shown in Fig.~\ref{fig:r_sigma_over_u_central},
which demonstrates evidence for the existence of an IRFP.  
\begin{figure}[t]
\includegraphics[scale=0.6]{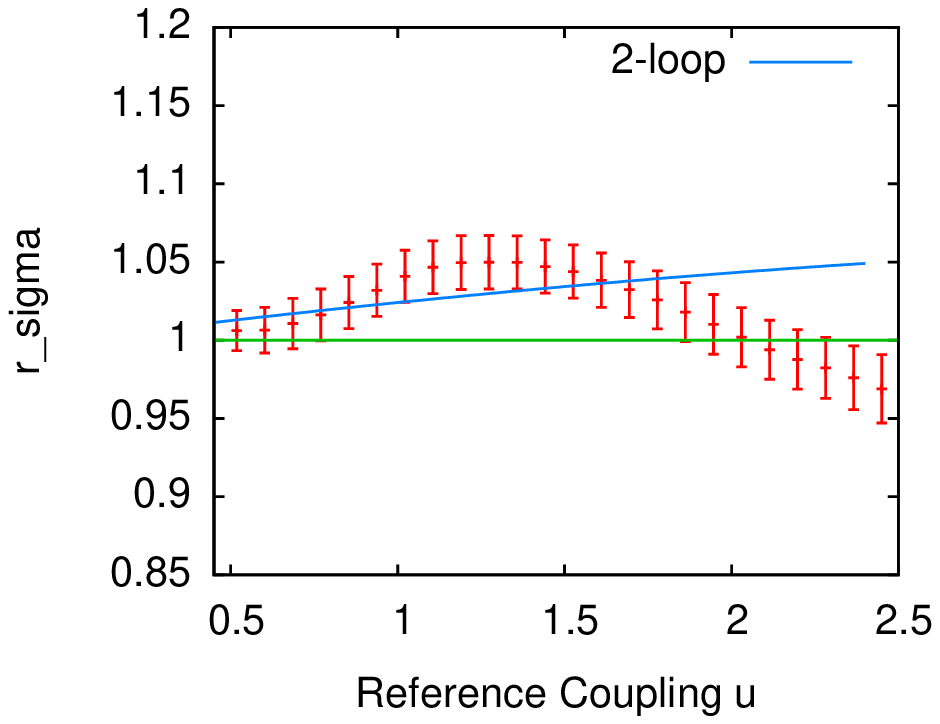}
\includegraphics[scale=0.6]{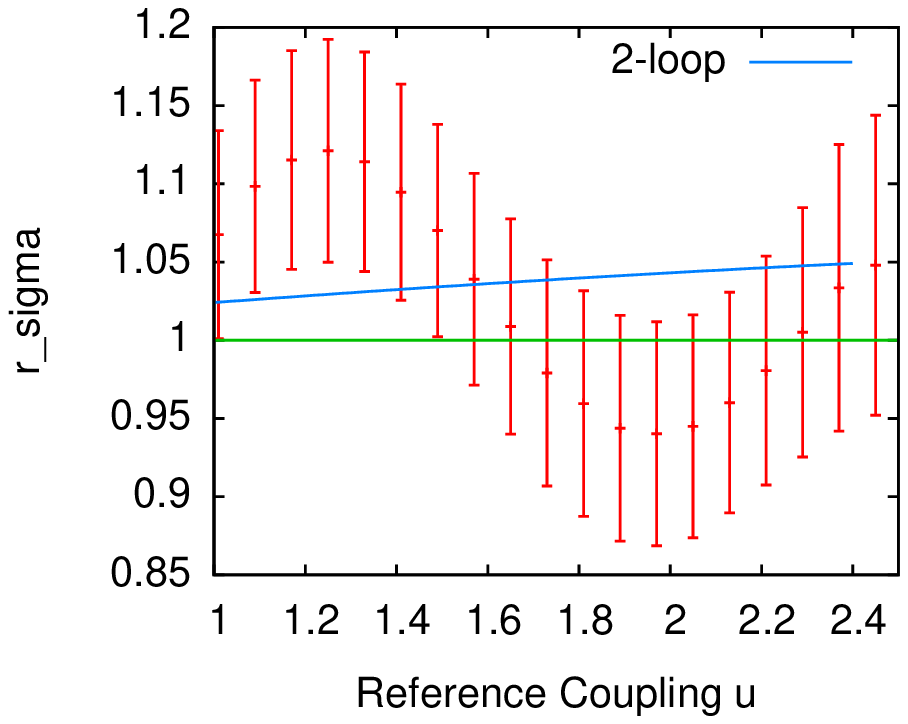}
\includegraphics[scale=0.6]{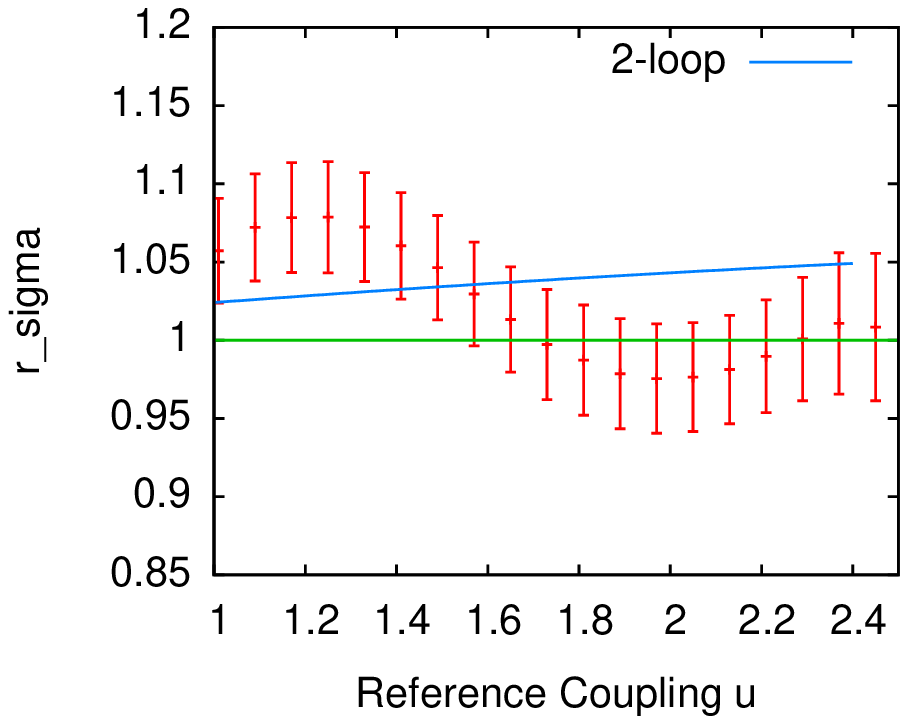}
\includegraphics[scale=0.6]{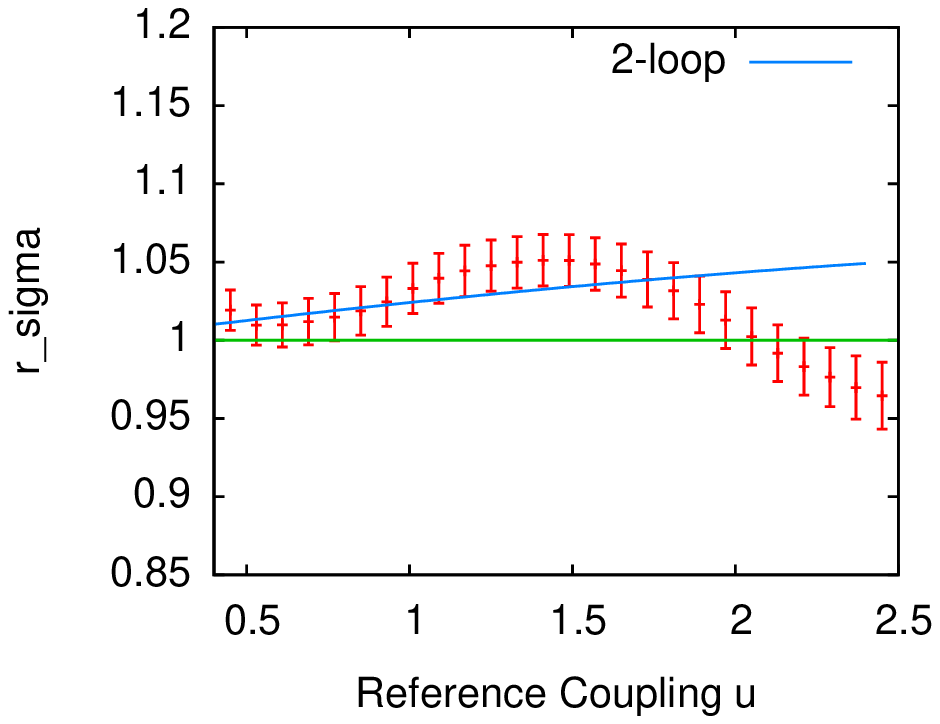}
\caption{Plots for $r_{\sigma}(u)$ obtained from our procedures for
  estimating systematic errors.
Top left:
$r_{\sigma}(u)$ from simple polynomial interpolation in
  $\beta$, Eq.~(\ref{eq:simple_polyn_beta}).  
Top right:
$r_{\sigma}(u)$ by performing the continuum extrapolation
  using quadratic function in $(a/L)^{2}$.  The
  rest is the same as the central procedure.
Bottom left:
$r_{\sigma}(u)$ by performing the continuum extrapolation
  using linear function in $(a/L)^{2}$, with $L/a = 7, 8, 10$.
Bottom right:
$r_{\sigma}(u)$ by performing the continuum extrapolation
  using linear function in $(a/L)^{2}$, with $L/a = 6, 7, 8, 10$.}
\label{fig:r_sigma_systematic}
\end{figure}
Next, we discuss the estimation of systematic effects arising from the 
$\beta{-}$value (bare-coupling) interpolation and the continuum extrapolation.
For this purpose, we perform the changes in the central
procedure.  These changes are carried out independently, {\it i.e.},
we vary one component in the central procedure, while keeping the
other fixed. 

We begin by varying the $\beta{-}$interpolation in the central procedure.  This is carried out by
changing the non-decreasing fit function in
Eq.~(\ref{eq:non_decr_poly}), to the simple polynomial form in
Eq.~(\ref{eq:simple_polyn_beta}) with the constraint of
Eq.~(\ref{eq:pt_constraint_simple_polyn}) and the numbers of parameters
reported in Table~\ref{tab:chisq_beta_interpolation}.  The result of this procedure
is shown in the top-left panel of Fig.~\ref{fig:r_sigma_systematic}.


In order to estimate systematic errors associate with the
continuum extrapolation, we perform various fits with the inclusion of
the $L/a=7$ and $L/a=14$ data.
First, we perform the quadratic fit using
Eq.~(\ref{eq:quadratic_continuum_extrap}).  This leads to the result
for $r_{\sigma}(u)$ as depicted in the top-right panel of
Fig.~\ref{fig:r_sigma_systematic}.  As expected, this
extrapolation strategy results in large statistical errors.
In addition to the quadratic fit, we also carry out the two
linear continuum extrapolations discussed in
Sec.~\ref{sec:continuum_extrapolation},
\bea
 L/a = (7, 8, 10) &\longrightarrow& 2L/a = (14, 16, 20), \\ \nonumber
 L/a = (6, 7, 8, 10) &\longrightarrow& 2L/a = (12, 14, 16, 20) .
\eea
The result from the first these procedures is presented in the 
bottom-left panel of
Fig.~\ref{fig:r_sigma_systematic}, while that from the
second one is shown in the bottom-right panel of
Fig.~\ref{fig:r_sigma_systematic}.
As discussed at the end of Sec.~\ref{sec:continuum_extrapolation}, the quadratic
fit, and the 3-point linear fit using the $L/a=7,8,10$ data can lead
to unreliable continuum extrapolations in the weak-coupling regime.
Therefore, in  Fig.~\ref{fig:r_sigma_systematic} we only show results at intermediate and large $u$ for
these two procedures.

\begin{figure}[t]
\includegraphics[scale=0.65]{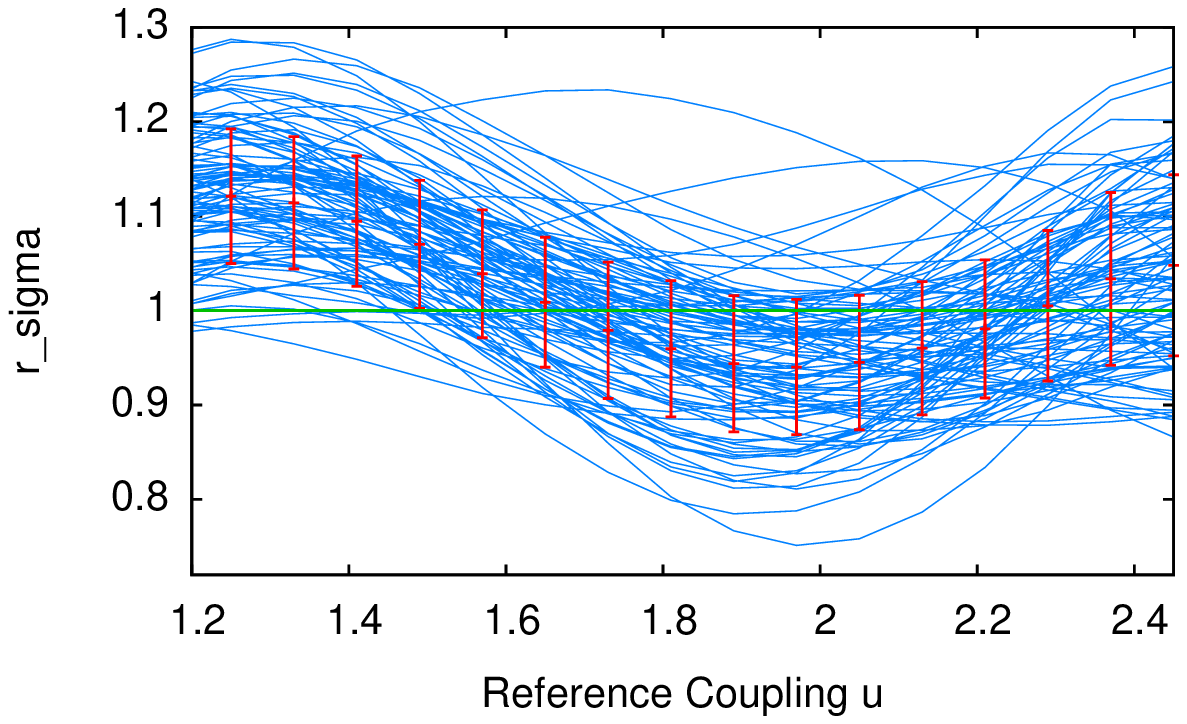}
\includegraphics[scale=0.65]{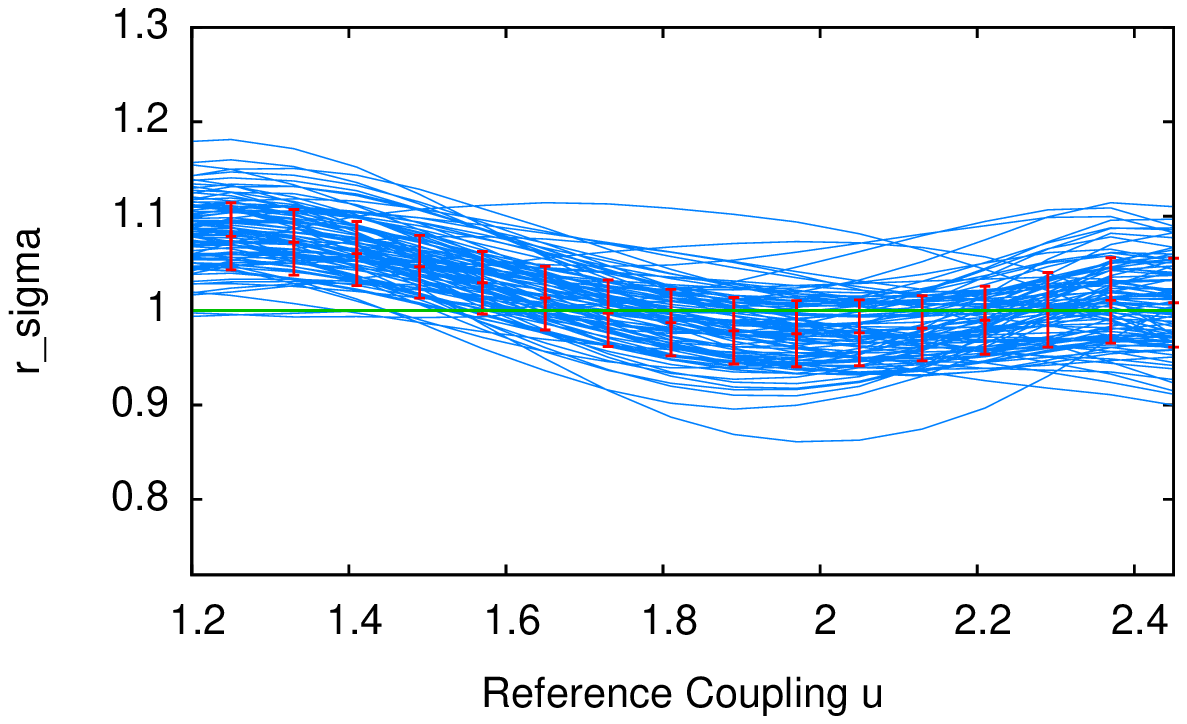}
\caption{$r_{\sigma}(u)$ in 100 bootstrap samples from quadratic
  continuum extrapolation (left), and 3-point linear continuum
  extrapolation using the $L/a = 7, 8, 10$ data (right).}
\label{fig:r_sigma_boot_samples}
\end{figure}

In Fig.~\ref{fig:r_sigma_over_u_central}, and in the top-left and
the bottom-right plots in Fig.~\ref{fig:r_sigma_systematic},
it is observed that these procedures lead to $r_{\sigma}(u)$
consistent with one in the UV and the IR, while statistically
different from this value between these two regimes.  This suggests that 
there exists an IRFP in SU(3) gauge theory with twelve flavours.
However, the continuum extrapolations using 4-point
quadratic fit (the top-right plot in Fig.~\ref{fig:r_sigma_systematic}) and 3-point linear fit without the $L/a=6$ data
(the bottom-left plot of Fig.~\ref{fig:r_sigma_systematic}) lead
to weaker evidence for the IR conformal behaviour.   For these two
procedures, in addition to
the difficulty in the continuum extrapolations in the weak-coupling
regime (discussed at the end of
Sec.~\ref{sec:continuum_extrapolation}), 
we also observe large errors in the IR
regime, leading to no apparent
feature that $r_{\sigma}(u)$ crosses one.  This phenomenon is actually
the consequence of the ``double crossing'' behaviour in some bootstrap
samples.  Namely, in these samples, $r_{\sigma}(u)$ crosses one from
above, and then turns around to cross the same value from below in a
slightly larger $u$.  We stress that out of 1000 bootstrap samples we
have created in this work, $r_{\sigma}(u)$ in more than 680 ($1\sigma$) of them cross the
unity from above in the IR regime, when the continuum extrapolations
are performed with the quadratic fit or the 3-point linear fit without
the $L/a=6$ data.  This leads to 
hints of the existence of an IRFP using these analysis procedures.  In
order to illustrate this point, in Fig.~\ref{fig:r_sigma_boot_samples} 
we plot 100 bootstrap samples in the intermediate${-}$ and
strong${-}u$ regions in these two procedures.

In Table \ref{tab:gstar}, we summarise the values of $g_{\ast}^{2}$ obtained from the above
procedures. 
%
%
\begin{table}
\begin{tabular}{|c|c|c|}\hline
$\beta$ interpolation & continuum extrapolation & $g^{2}_{\ast}$ \\\hline
non-decreasing polynomial & 3 point linear, $L/a = 6, 8, 10$         & 2.02(18) \\\hline
simple Polynomial  & 3 point linear, $L/a = 6, 8, 10$  & 2.02(21) \\\hline
non-decreasing polynomial &  4 point linear                       & 2.06(15) \\\hline
non-decreasing polynomial &  3 point linear, $L/a = 7, 8, 10$   & $> 1.66$ \\\hline
non-decreasing polynomial &  4 point quadratic
&  $> 1.62$ \\\hline
\end{tabular}
\caption{$g^{2}_{\ast}$ from various procedures.  The first row
  describes the central procedure.}
\label{tab:gstar}
\end{table}
%
%
%
%
Because it is challenging to precisely estimate systematic effects,
as discussed above, we take a conservative approach to conclude that in SU(3) gauge theory with twelve
flavours, our data suggest the existence of an IRFP around,
\beq
 g^{2}_{\ast} \sim 2.0 .
\label{eq:result_number}
\eeq
This result is similar to what we obtained with other collaborators
using a different analysis procedure~\cite{Aoyama:2011ry} by setting
the step size to be $1.5$.  Here we stress that it is more challenging
to control systematic effects and the correlation amongst data points in the procedure in
Ref.~\cite{Aoyama:2011ry}, because of the need for many interpolations
in lattice volumes when computing the lattice step-scaling function, $\Sigma$.

The result in Eq.~(\ref{eq:result_number}) is much smaller than that obtained in the SF
scheme~\cite{Appelquist:2007hu,Appelquist:2009ty},
\beq
 \left ( g^{({\rm SF})}_{\ast} \right )^{2} \sim 4.5 .
\eeq
The significant difference clearly indicates that the two schemes are
very different.  It should also be noted that in the TPL scheme, there
is an upper bound for the 
renormalised coupling constant, as discussed in Sec.~\ref{sec:twbc}. 
This may result in slower running behaviour compared to the SF scheme.

\section{Conclusion}
\label{sec:conclusion}
In this paper, we present our work on the lattice study of IR
behaviour in SU(3) gauge theory with twelve flavours.   We use the
step-scaling method to investigate the running coupling constant over
a large range of scale.  Our renormalisation scheme is defined via the
ratio of Polyakov loop correlators in the twisted and untwisted
directions.   In particular, we compute the ratio, $r_{\sigma}(u)$
defined in Eq.~(\ref{eq:r_sigma_def}), between the
step-scaling function and the input renormalised coupling.   In our
central analysis procedure, we perform the continuum extrapolation
using the 3-point linear fit with the reference coupling computed on
the $L/a=6, 8, 10$ lattices.  Data on these lattices are free of
volume interpolation.  Using this procedure, we find that this theory
contains an IRFP at around $g^{2}_{\ast} \sim 2$.

In this work, we have investigated systematic
errors in the bare coupling interpolation for each
lattice volume, and the continuum extrapolation.  We have performed
reasonable variations on these interpolation and extrapolation, and
carefully examined possible correlation amongst data points used
in the continuum extrapolation.  We find that the dominant systematic
effect arises from the continuum extrapolation.   To gain information
about possible errors in this extrapolation, we compute the
step-scaling function on the $L/a=14$ lattice, obtain the reference
input renormalised coupling for the $L/a=7$ lattice using an
interpolation procedure, and then study the continuum limit using the 4-point
linear and quadratic fits, as well as the 3-point linear fit without the
$L/a=6$ data.  We find that all our analysis procedures result in evidence
for the existence of an IRFP, although the latter two
continuum-extrapolation methods result in significant errors.
In view of this, the result of our work suggests
that SU(3) gauge theory with twelve fermions in
the fundamental representation contains an IRFP.

Our finding shows that the conformal window for SU(3) gauge theories
with fundamental fermions may lie below $N_{f}=12$.  Although this
conclusion agrees with most other
studies~\cite{Appelquist:2007hu,Appelquist:2009ty,Appelquist:2009ka,Appelquist:2011dp,Appelquist:2012sm,DeGrand:2011cu,Deuzeman:2009mh,Miura:2011mc},
the result in Ref.~\cite{Fodor:2009wk,Fodor:2011tu} leads to the
opposite conclusion.  
Combining this information with the recent result from the $N_{f}=10$  
calculation~\cite{Appelquist:2012nz},  this can indicate that the $N_{f}=12$ is
already very close to the lower bound of the conformal window.

\section*{Acknowledgments}
We are indebted to Tatsumi~Aoyama, Hiroaki~Ikeda, Etsuko~Itou,
Masafumi~Kurachi, Hideo~Matsufuru, Tetsuya~Onogi, and
Takeshi~Yamazaki, for their important contributions to many aspects of
this work, and their collaboration which led to the publication of Ref.~\cite{Aoyama:2011ry}.
We warmly thank Luigi Del Debbio, George Fleming, Ron Horgan, Kei-Ichi
Nagai, Maurizio Piai, Eibun Senaha, David~Schaich, Y.~Taniguchi, and N.~Yamada for discussions.  We are grateful to Stefan Meinel for
providing us with the computer code for generating the plots in
Fig.~\ref{fig:L7_L12_L10_L8_L6_corr}.   
Numerical simulation was carried out on
NEC SX-8 and Hitachi SR16000 at YITP, Kyoto University,
NEC SX-8R at RCNP, Osaka University,
and Hitachi SR11000 and IBM System Blue Gene Solution at KEK 
under its Large-Scale Simulation Program
(No.~09/10-22 and 10-16), as well as on the GPU cluster at
Taiwanese National Centre for High-performance Computing.
We acknowledge Japan Lattice Data Grid for data
transfer and storage.
C.-J.D.L. is supported by Taiwanese National Science Council
(NSC) via grant 99-2112-M-009-004-MY3.
H.O. acknowledges supports from the JSPS Grant-in-Aid for Scientific Research (S) number 22224003. 
K.O. acknowledges supports from NSC grant 099-2811-M-009-029, and the Special Project Grant from National
Chiao-Tung University during the progress of this work.
E.S. acknowledges the Grant-in-Aid number 21105508 and 23105714 from the Japanese Ministry 
of Education.

\appendix
\section{Low-lying eigenvalues of the Dirac operator}
\label{sec:eigenvalues}
To check the effects of taste-symmetry breaking in staggered fermions,
we study positive low-lying eigenvalues of the Dirac operator.  In
this section, we present a typical case of taste-symmetry
restoration when approaching the continuum limit at fixed physical
volume.  For this purpose, we compare the following two cases:
\begin{enumerate}
 \item $L/a=10$, $\beta=20.13$, in which $\bar{g}_{\rm latt}^{2} = 0.4031(76)$.
 \item $L/a=20$, $\beta=20.00$, in which $\bar{g}_{\rm latt}^{2} =
   0.4064(76)$.
\end{enumerate}
\begin{figure}
\includegraphics*[width=0.45\textwidth,angle=0]{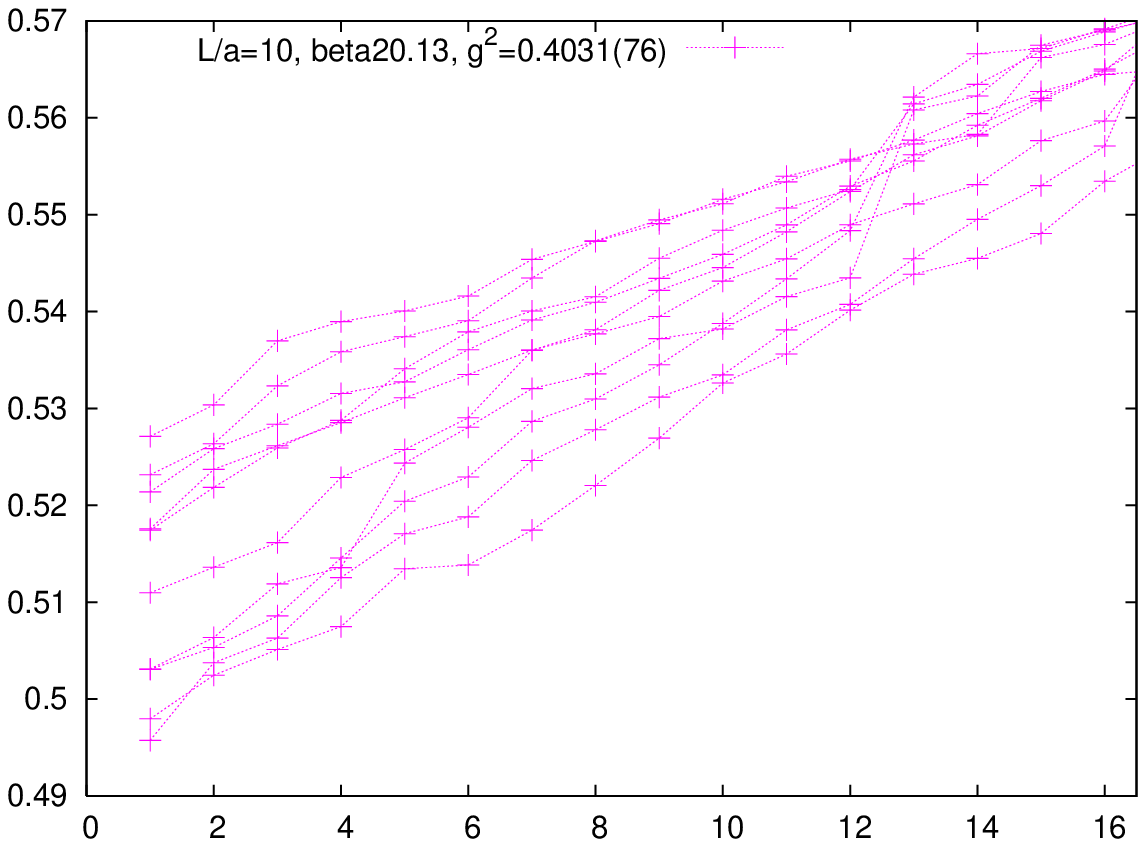}
\includegraphics*[width=0.45\textwidth,angle=0]{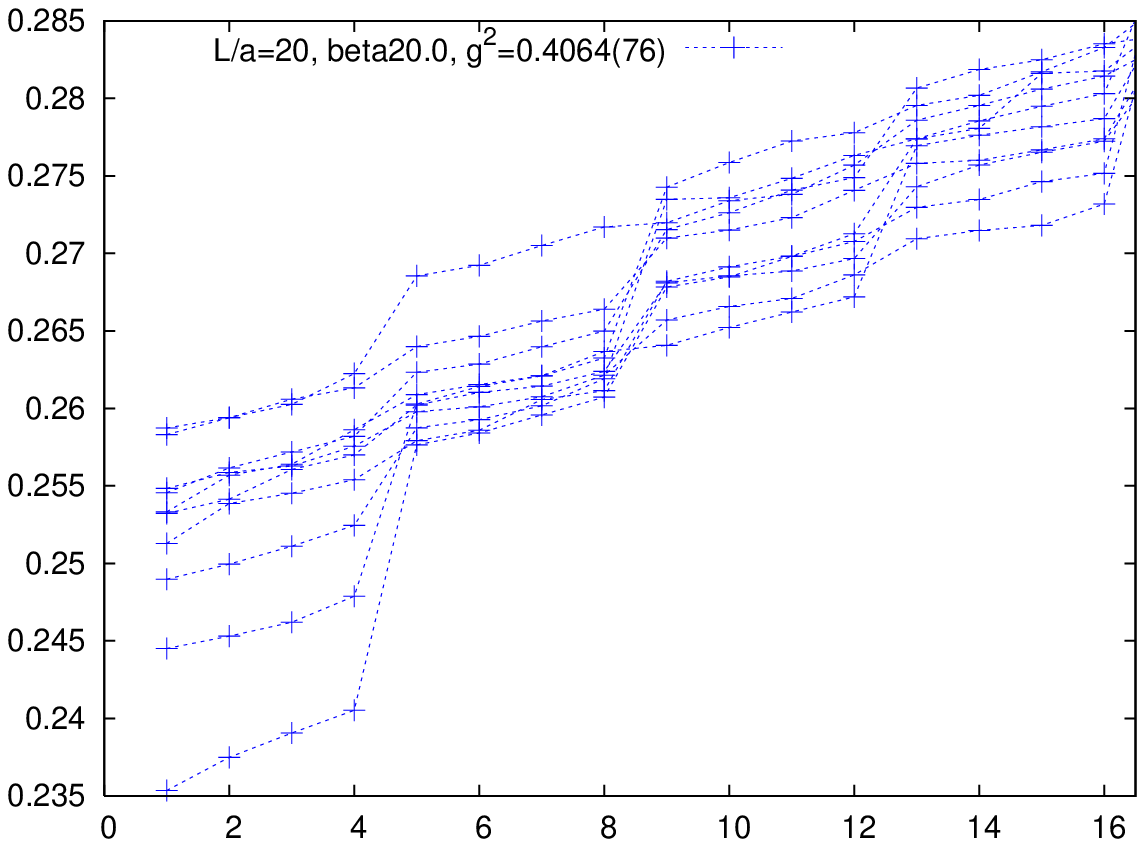}
\caption{Plots of the 16 lowest-lying positive eigenvalues (in lattice
  units) of the staggered
  fermion operator in this work.  The two plots show a coarse (left)
  and a fine (right) lattices of the same physical volume.  Each line
  in these plots connects the eigenvalues computed on the same gauge configuration.
}
\label{fig:eigenvalues}
\end{figure}
The renormalised coupling for these two cases, as shown in the table
for the raw data in App.~\ref{sec:plaquette_values_raw_data}, are well
consistent within statistical error.  This means that the physical
volumes are almost the same, while the lattice spacing of the second
case is half of that of the first case.

In Fig.~\ref{fig:eigenvalues}, we show the lowest-lying 16 eigenvalues
on 10 gauge configurations 
for each of the above two cases.  Every line in these plots connects
all the 16 eigenvalues in one configuration.  It is evident that on the finer lattice, the
4-fold degeneracy appears, while it is much less clear for the coarser lattice.
Although taste-symmetry restoration appears on fine lattices in
our work,  from 
Fig.~\ref{fig:eigenvalues}, it is indicated that such restoration does
not show up on the coarse lattices in our simulations.  Such effects are expected, since
unimproved staggered fermions are implemented.  This necessitates good
control of the continuum extrapolation, which is addressed in
detail in Sec.~\ref{sec:continuum_extrapolation}.

\section{Values of plaquette and TPL coupling constant raw data}
\label{sec:plaquette_values_raw_data}
In this appendix, we present details for the plaquette values and the
TPL scheme renormalised coupling constants
obtained at various lattice volumes, $L/a$, and bare couplings,
$\beta$.  We also give the numbers of HMC trajectories for the
computation of the TPL coupling.  
\begin{table}
\begin{tabular}{|c|c|c|}\hline
 $L/a$  &     $\beta$  &       Plaquette  \\ \cline{1-3}
6 &       5.36 &  $     0.6048877 (         72 )  $ \\ \cline{1-3}
6 &       5.53 &  $     0.6177253 (         79 )  $ \\ \cline{1-3}
6 &       5.81 &  $     0.6371712 (         82 )  $ \\ \cline{1-3}
6 &       6.12 &  $     0.6565360 (         71 )  $ \\ \cline{1-3}
6 &       6.76 &  $     0.6906691 (         92 )  $ \\ \cline{1-3}
6 &       7.82 &  $     0.7343932 (         75 )  $ \\ \cline{1-3}
6 &       8.45 &  $     0.7549697 (         77 )  $ \\ \cline{1-3}
6 &       9.42 &  $     0.7810709 (         57 )  $ \\ \cline{1-3}
6 &      11.15 &  $     0.8160114 (         69 )  $ \\ \cline{1-3}
6 &      13.85 &  $     0.8526629 (         54 )  $ \\ \cline{1-3}
6 &      15.23 &  $     0.8662760 (         52 )  $ \\ \cline{1-3}
6 &      17.55 &  $     0.8842610 (         44 )  $ \\ \cline{1-3}
6 &      20.13 &  $     0.8992980 (         39 )  $ \\ \cline{1-3} 
\end{tabular}
\hspace{0.5cm}
\begin{tabular}{|c|c|c|}\hline
 $L/a$  &     $\beta$  &       Plaquette  \\ \cline{1-3}
8 &       5.36 &  $      0.604824 (         17 )  $ \\ \cline{1-3}
8 &       5.53 &  $     0.6176221 (         88 )  $ \\ \cline{1-3}
8 &       5.81 &  $     0.6370758 (         70 )  $ \\ \cline{1-3}
8 &       6.12 &  $     0.6564552 (         52 )  $ \\ \cline{1-3}
8 &       6.47 &  $     0.6760071 (         31 )  $ \\ \cline{1-3}
8 &       6.76 &  $     0.6906010 (         54 )  $ \\ \cline{1-3}
8 &       7.11 &  $     0.7065526 (         34 )  $ \\ \cline{1-3}
8 &       7.82 &  $     0.7343407 (         47 )  $ \\ \cline{1-3}
8 &       8.45 &  $     0.7549265 (         23 )  $ \\ \cline{1-3}
8 &       9.42 &  $     0.7810425 (         38 )  $ \\ \cline{1-3}
8 &      11.15 &  $     0.8159758 (         68 )  $ \\ \cline{1-3}
8 &      13.85 &  $     0.8526510 (         50 )  $ \\ \cline{1-3}
8 &      15.23 &  $     0.8662602 (         54 )  $ \\ \cline{1-3}
8 &      17.55 &  $     0.8842492 (         48 )  $ \\ \cline{1-3}
8 &      20.13 &  $     0.8992908 (         17 )  $ \\ \cline{1-3}
\end{tabular}
\hspace{0.5cm}
\begin{tabular}{|c|c|c|}\hline
 $L/a$ &    $\beta$  &       Plaquette  \\ \cline{1-3}
10  &       5.36 &  $      0.604796 (         14 )  $ \\ \cline{1-3}
10  &       5.53 &  $     0.6176175 (         64 )  $ \\ \cline{1-3}
10  &       5.81 &  $     0.6370675 (         56 )  $ \\ \cline{1-3}
10  &       6.12 &  $     0.6564540 (         48 )  $ \\ \cline{1-3}
10  &       6.47 &  $     0.6759944 (         48 )  $ \\ \cline{1-3}
10  &       6.76 &  $     0.6905794 (         18 )  $ \\ \cline{1-3}
10  &       7.11 &  $     0.7065347 (         44 )  $ \\ \cline{1-3}
10  &       7.82 &  $     0.7343367 (         29 )  $ \\ \cline{1-3}
10  &       8.45 &  $     0.7549194 (         25 )  $ \\ \cline{1-3}
10  &       9.42 &  $     0.7810298 (         27 )  $ \\ \cline{1-3}
10  &      11.15 &  $     0.8159762 (         41 )  $ \\ \cline{1-3}
10  &      13.85 &  $     0.8526444 (         36 )  $ \\ \cline{1-3}
10  &      15.23 &  $     0.8662684 (         35 )  $ \\ \cline{1-3}
10  &      17.55 &  $     0.8842406 (         33 )  $ \\ \cline{1-3}
10  &      20.13 &  $     0.8992850 (         27 )  $ \\ \cline{1-3}
\end{tabular}
\hspace{0.5cm}
\begin{tabular}{|c|c|c|}\hline
 $L/a$  &     $\beta$  &       Plaquette  \\ \cline{1-3}
12  &       5.36 &  $     0.6047811 (         33 )  $ \\ \cline{1-3}
12  &       5.53 &  $     0.6176126 (         46 )  $ \\ \cline{1-3}
12  &       5.81 &  $     0.6370591 (         35 )  $ \\ \cline{1-3}
12  &       6.12 &  $     0.6564331 (         30 )  $ \\ \cline{1-3}
12  &       6.47 &  $     0.6759845 (         26 )  $ \\ \cline{1-3}
12  &       6.76 &  $     0.6905844 (         30 )  $ \\ \cline{1-3}
12  &       7.11 &  $     0.7065395 (         26 )  $ \\ \cline{1-3}
12  &       7.82 &  $     0.7343314 (         23 )  $ \\ \cline{1-3}
12  &       8.45 &  $     0.7549160 (         17 )  $ \\ \cline{1-3}
12  &       9.42 &  $     0.7810290 (         19 )  $ \\ \cline{1-3}
12  &      11.15 &  $     0.8159771 (         21 )  $ \\ \cline{1-3}
12  &      13.85 &  $     0.8526457 (         28 )  $ \\ \cline{1-3}
12  &      15.23 &  $     0.8662619 (         28 )  $ \\ \cline{1-3}
12  &      17.55 &  $     0.8842412 (         23 )  $ \\ \cline{1-3}
12  &      20.13 &  $     0.8992891 (         19 )  $ \\ \cline{1-3}
\end{tabular} 
\\ \ \\ \ \\ \ \\ \ \\ \ \\
\begin{tabular}{|c|c|c|}\hline
$L/a $&    $\beta$ &       Plaquette  \\ \cline{1-3}
14 &       5.36 &  $     0.6047816 (         30 )  $ \\ \cline{1-3}
14 &       5.53 &  $     0.6176160 (         32 )  $ \\ \cline{1-3}
14 &       5.81 &  $     0.6370623 (         33 )  $ \\ \cline{1-3}
14 &       6.12 &  $     0.6564415 (         30 )  $ \\ \cline{1-3}
14 &       6.47 &  $     0.6759856 (         28 )  $ \\ \cline{1-3}
14 &       6.76 &  $     0.6905864 (         26 )  $ \\ \cline{1-3}
14 &       7.11 &  $     0.7065419 (         26 )  $ \\ \cline{1-3}
14 &       7.82 &  $     0.7343298 (         22 )  $ \\ \cline{1-3}
14 &       8.45 &  $     0.7549137 (         18 )  $ \\ \cline{1-3}
14 &       9.42 &  $     0.7810323 (         17 )  $ \\ \cline{1-3}
14 &      11.15 &  $     0.8159718 (         26 )  $ \\ \cline{1-3}
14 &      13.85 &  $     0.8526501 (         22 )  $ \\ \cline{1-3}
14 &      15.23 &  $     0.8662675 (         24 )  $ \\ \cline{1-3}
14 &      17.55 &  $     0.8842430 (         19 )  $ \\ \cline{1-3}
14 &      20.13 &  $     0.8992948 (         17 )  $ \\ \cline{1-3} 
\end{tabular} 
\hspace{0.9cm}
\begin{tabular}{|c|c|c|}\hline
 $L/a$  &     $\beta$  &       Plaquette  \\ \cline{1-3}
16 &       5.36 &  $     0.6047781 (         27 )  $ \\ \cline{1-3}
16 &       5.53 &  $     0.6176112 (         25 )  $ \\ \cline{1-3}
16 &       5.81 &  $     0.6370620 (         24 )  $ \\ \cline{1-3}
16 &       6.12 &  $     0.6564355 (         22 )  $ \\ \cline{1-3}
16 &       6.47 &  $     0.6759890 (         19 )  $ \\ \cline{1-3}
16 &       6.76 &  $     0.6905800 (         26 )  $ \\ \cline{1-3}
16 &       7.11 &  $     0.7065370 (         18 )  $ \\ \cline{1-3}
16 &       7.82 &  $     0.7343300 (         18 )  $ \\ \cline{1-3}
16 &       8.45 &  $     0.7549165 (         11 )  $ \\ \cline{1-3}
16 &       9.42 &  $     0.7810290 (         14 )  $ \\ \cline{1-3}
16 &      11.15 &  $     0.8159751 (         21 )  $ \\ \cline{1-3}
16 &      13.85 &  $     0.8526473 (         16 )  $ \\ \cline{1-3}
16 &      15.23 &  $     0.8662636 (         14 )  $ \\ \cline{1-3}
16 &      17.55 &  $     0.8842414 (         12 )  $ \\ \cline{1-3}
16 &      20.13 &  $     0.8992913 (         11 )  $ \\ \cline{1-3} 
\end{tabular} 
\hspace{0.9cm}
\begin{tabular}{|c|c|c|}\hline
 $L/a$  &     $\beta$  &       Plaquette  \\ \cline{1-3}
20  &       5.70 &  $    0.62965511 (         64 )  $ \\ \cline{1-3}
20  &       6.00 &  $     0.6491829 (         14 )  $ \\ \cline{1-3}
20  &       6.50 &  $     0.6775588 (         10 )  $ \\ \cline{1-3}
20  &       7.00 &  $     0.7017067 (         25 )  $ \\ \cline{1-3}
20  &       8.00 &  $    0.74055770 (         90 )  $ \\ \cline{1-3}
20  &       9.00 &  $    0.77044016 (         58 )  $ \\ \cline{1-3}
20  &      10.00 &  $    0.79413900 (         68 )  $ \\ \cline{1-3}
20  &      12.00 &  $    0.82934963 (         65 )  $ \\ \cline{1-3}
20  &      16.00 &  $    0.87281842 (         60 )  $ \\ \cline{1-3}
20  &      18.00 &  $    0.88718074 (         65 )  $ \\ \cline{1-3}
20  &      20.00 &  $     0.8986222 (         23 )  $ \\ \cline{1-3}
20  &      50.00 &  $     0.9597892 (         11 )  $ \\ \cline{1-3}
\end{tabular} 
\caption{The expectation values of the plaquette in a significant
  fraction of our simulations.}
\label{tab:plaquette}
\end{table}

\clearpage

%
\begin{table}
\begin{tabular}{||c|c|c|c||}
\hline
$L/a$ & $\beta$ & $\bar{g}^{2}_{\rm latt}$  & \# of traj. \\ \hline
6&     4.00 & 2.885 ( 49 ) &      69000  \\ \hline
6&     4.30 & 2.942 ( 38 ) &      94000  \\ \hline
6&     4.50 & 2.808 ( 37 ) &      108000 \\ \hline
6&     4.70 & 2.789 ( 38 ) &      78000  \\ \hline
6&     5.00 & 2.716 ( 33 ) &      96000  \\ \hline
6&     5.36 & 2.488 ( 10 ) &      696720 \\ \hline
6&     5.50 & 2.434 ( 33 ) &      72000  \\ \hline
6&     5.53 & 2.408 ( 11 ) &      718616 \\ \hline
6&     5.81 & 2.248 ( 12 ) &      530243 \\ \hline
6&     6.00 & 2.205 ( 26 ) &      90000  \\ \hline
6&     6.12 & 2.143 ( 10 ) &      603007 \\ \hline
6&     6.50 & 1.969 ( 30 ) &      54000  \\ \hline
6&     6.76 & 1.869 ( 11 ) &      306497 \\ \hline
6&     7.00 & 1.810 ( 27 ) &      54000  \\ \hline
6&     7.82 & 1.530 (  9 ) &      383859 \\ \hline
6&     8.00 & 1.531 ( 19 ) &      78000  \\ \hline
6&     8.45 & 1.348 (  9 ) &      289118 \\ \hline
6&     9.00 & 1.224 ( 16 ) &      78000  \\ \hline
6&     9.42 & 1.144 (  6 ) &      389334 \\ \hline
6&     10.00 & 1.050 ( 14 ) &     54000  \\ \hline
6&     11.15 & 0.8819 ( 60 ) &    330175 \\ \hline
6&     12.00 & 0.7844 ( 69 ) &    90000  \\ \hline
6&     13.85 & 0.6425 ( 33 ) &    352374 \\ \hline
6&     14.00 & 0.6273 ( 47 ) &    90000  \\ \hline
6&     15.23 & 0.5646 ( 28 ) &    339500 \\ \hline
6&     16.00 & 0.5158 ( 32 ) &    108000 \\ \hline
6&     17.55 & 0.4645 ( 18 ) &    353866 \\ \hline
6&     18.00 & 0.4511 ( 30 ) &    60000  \\ \hline
6&     20.00 & 0.3895 ( 21 ) &    72000  \\ \hline
6&     20.13 & 0.3891 ( 16 ) &    330238 \\ \hline
6&     50.00 & 0.1322 (  5 ) &    44250  \\ \hline
\end{tabular}
\hspace{0.5cm}
\begin{tabular}{||c|c|c|c||}
\hline
$L/a$ & $\beta$ & $\bar{g}^{2}_{\rm latt}$  & \# of traj. \\ \hline
8&  4.50 & 3.218 ( 51 ) &      113000 \\ \hline
8&  4.70 & 3.098 ( 52 ) &      85000  \\ \hline
8&  5.00 & 2.918 ( 57 ) &      94250  \\ \hline
8&  5.36 & 2.692 ( 70 ) &      42935  \\ \hline
8&  5.50 & 2.655 ( 50 ) &      75500  \\ \hline
8&  5.53 & 2.676 ( 29 ) &      471893 \\ \hline
8&  5.81 & 2.471 ( 21 ) &      415827 \\ \hline
8&  6.00 & 2.382 ( 41 ) &      95000  \\ \hline
8&  6.12 & 2.307 ( 17 ) &      584764 \\ \hline
8&  6.47 & 2.136 ( 10 ) &      129309 \\ \hline
8&  6.50 & 2.110 ( 31 ) &      99000  \\ \hline
8&  6.76 & 2.004 ( 19 ) &      356603 \\ \hline
8&  7.00 & 1.923 ( 22 ) &      153000 \\ \hline
8&  7.11 & 1.842 ( 11 ) &      720570 \\ \hline
8&  7.82 & 1.602 ( 13 ) &      344514 \\ \hline
8&  8.00 & 1.571 ( 34 ) &      63500  \\ \hline
8&  8.45 & 1.420 (  7 ) &      987652 \\ \hline
8&  9.00 & 1.280 ( 16 ) &      130750 \\ \hline
8&  9.42 & 1.192 ( 10 ) &      317269 \\ \hline
8&  10.00 & 1.073 ( 18 ) &     72250  \\ \hline
8&  11.15 & 0.8978 ( 99 ) &    164137 \\ \hline
8&  12.00 & 0.7919 ( 74 ) &    126750 \\ \hline
8&  13.85 & 0.6522 ( 52 ) &    190057 \\ \hline
8&  14.00 & 0.6492 ( 57 ) &    95500  \\ \hline
8&  15.23 & 0.5733 ( 48 ) &    170455 \\ \hline
8&  16.00 & 0.5284 ( 50 ) &    78500  \\ \hline
8&  17.55 & 0.4660 ( 35 ) &    166701 \\ \hline
8&  18.00 & 0.4565 ( 39 ) &    83500  \\ \hline
8&  20.00 & 0.3910 ( 33 ) &    111073 \\ \hline
8&  20.13 & 0.3908 ( 25 ) &    188895 \\ \hline
8&  50.00 & 0.1308 (  8 ) &    49750  \\ \hline
8&  99.00 & 0.06368 ( 26 ) &   59750  \\ \hline
\end{tabular}
\hspace{0.5cm}
\begin{tabular}{||c|c|c|c||}
\hline
$L/a$ & $\beta$ & $\bar{g}^{2}_{\rm latt}$  & \# of traj. \\ \hline
10  &  4.50 & 3.600 ( 71 ) &      220400 \\ \hline
10  &  5.00 & 3.149 ( 62 ) &      95000  \\ \hline
10  &  5.36 & 2.823 ( 62 ) &      89439  \\ \hline
10  &  5.50 & 2.808 ( 53 ) &      114800 \\ \hline
10  &  5.53 & 2.785 ( 45 ) &      177056 \\ \hline
10  &  5.81 & 2.605 ( 42 ) &      216874 \\ \hline
10  &  6.00 & 2.477 ( 42 ) &      130000 \\ \hline
10  &  6.12 & 2.432 ( 33 ) &      249705 \\ \hline
10  &  6.47 & 2.219 ( 26 ) &      307266 \\ \hline
10  &  6.50 & 2.230 ( 42 ) &      142400 \\ \hline
10  &  6.76 & 2.090 ( 24 ) &      305980 \\ \hline
10  &  7.00 & 1.988 ( 26 ) &      208000 \\ \hline
10  &  7.11 & 1.960 ( 28 ) &      256781 \\ \hline
10  &  7.82 & 1.651 ( 16 ) &      454309 \\ \hline
10  &  8.00 & 1.613 ( 39 ) &      68000  \\ \hline
10  &  8.45 & 1.445 ( 13 ) &      503970 \\ \hline
10  &  9.00 & 1.351 ( 26 ) &      80000  \\ \hline
10  &  9.42 & 1.238 ( 15 ) &      274746 \\ \hline
10  &  10.00 & 1.128 ( 25 ) &     83750  \\ \hline
10  &  11.15 & 0.939 ( 17 ) &     112614 \\ \hline
10  &  12.00 & 0.821 ( 14 ) &     80000  \\ \hline
10  &  13.85 & 0.6563 ( 94 ) &    83893  \\ \hline
10  &  14.00 & 0.6363 ( 74 ) &    120000 \\ \hline
10  &  15.23 & 0.5672 ( 82 ) &    91641  \\ \hline
10  &  16.00 & 0.5359 ( 64 ) &    88500  \\ \hline
10  &  17.55 & 0.4741 ( 58 ) &    88444  \\ \hline
10  &  18.00 & 0.4517 ( 55 ) &    74000  \\ \hline
10  &  20.00 & 0.3825 ( 52 ) &    49000  \\ \hline
10  &  20.13 & 0.3977 ( 45 ) &    85527  \\ \hline
10  &  50.00 & 0.1334 ( 10 ) &    65500  \\ \hline
10  &  99.00 & 0.06387 ( 40 ) &   39500  \\ \hline
 \end{tabular}
\end{table}
\begin{table}
\begin{tabular}{||c|c|c|c||}
\hline
$L/a$ & $\beta$ & $\bar{g}^{2}_{\rm latt}$   & \# of traj.  \\ \hline
12&  4.50 & 3.64 ( 16 ) &       154400 \\ \hline
12&  4.70 & 3.718 ( 99 ) &      148300 \\ \hline
12&  5.00 & 3.249 ( 73 ) &      160400 \\ \hline
12&  5.30 & 2.953 ( 60 ) &      129700 \\ \hline
12&  5.36 & 3.029 ( 46 ) &      272639 \\ \hline
12&  5.50 & 3.123 ( 65 ) &      154700 \\ \hline
12&  5.53 & 2.951 ( 55 ) &      233669 \\ \hline
12&  5.81 & 2.723 ( 44 ) &      269109 \\ \hline
12&  6.00 & 2.510 ( 47 ) &      167200 \\ \hline
12&  6.12 & 2.563 ( 38 ) &      283205 \\ \hline
12&  6.47 & 2.278 ( 28 ) &      330998 \\ \hline
12&  6.76 & 2.096 ( 32 ) &      265744 \\ \hline
12&  7.00 & 2.058 ( 40 ) &      146400 \\ \hline
12&  7.11 & 1.966 ( 27 ) &      256821 \\ \hline
12&  7.82 & 1.671 ( 25 ) &      262368 \\ \hline
12&  8.00 & 1.569 ( 30 ) &      139200 \\ \hline
12&  8.45 & 1.471 ( 20 ) &      397273 \\ \hline
12&  9.00 & 1.316 ( 22 ) &      160500 \\ \hline
12&  9.42 & 1.264 ( 19 ) &      256230 \\ \hline
12&  10.00 & 1.134 ( 22 ) &     159000 \\ \hline
12&  11.15 & 0.914 ( 12 ) &     173714 \\ \hline
12&  12.00 & 0.844 ( 15 ) &     102000 \\ \hline
12&  13.85 & 0.673 ( 14 ) &     79126  \\ \hline
12&  14.00 & 0.647 ( 13 ) &     79200  \\ \hline
12&  15.23 & 0.589 ( 11 ) &     75219  \\ \hline
12&  16.00 & 0.5467 ( 82 ) &    84600  \\ \hline
12&  17.55 & 0.4658 ( 71 ) &    85184  \\ \hline
12&  18.00 & 0.4463 ( 64 ) &    90000  \\ \hline
12&  20.00 & 0.3928 ( 50 ) &    86400  \\ \hline
12&  20.13 & 0.3982 ( 64 ) &    83045  \\ \hline
12&  50.00 & 0.1315 ( 11 ) &    65182  \\ \hline
12&  99.00 & 0.06386 ( 50 ) &   36400  \\ \hline
\end{tabular}
\hspace{2.5cm}
\begin{tabular}{||c|c|c|c||}
\hline
$L/a$ & $\beta$ & $\bar{g}^{2}_{\rm latt}$  & \# of traj. \\ \hline
14 &  5.36 & 3.295 ( 69 ) &     199385 \\ \hline
14 &  5.53 & 2.837 ( 75 ) &     124117 \\ \hline
14 &  5.81 & 2.675 ( 67 ) &     125696 \\ \hline
14 &  6.12 & 2.610 ( 72 ) &     129106 \\ \hline
14 &  6.47 & 2.287 ( 57 ) &     128286 \\ \hline
14 &  6.76 & 2.201 ( 56 ) &     143854 \\ \hline
14 &  7.11 & 2.125 ( 49 ) &     140251 \\ \hline
14 &  7.82 & 1.639 ( 37 ) &     144056 \\ \hline
14 &  8.45 & 1.536 ( 34 ) &     169990 \\ \hline
14 &  9.42 & 1.257 ( 30 ) &     146017 \\ \hline
14 &  11.15 & 0.909 ( 27 ) &    50262  \\ \hline
14 &  13.85 & 0.666 ( 15 ) &    52658  \\ \hline
14 &  15.23 & 0.612 ( 14 ) &    52301  \\ \hline
14 &  17.55 & 0.4683 ( 96 ) &   53082  \\ \hline
14 &  20.13 & 0.4036 ( 79 ) &   49930  \\ \hline
\end{tabular}
\end{table}
\begin{table}
\begin{tabular}{||c|c|c|c||}
\hline
$L/a$ & $\beta$ & $\bar{g}^{2}_{\rm latt}$  & \# of traj. \\ \hline
16 &  5.30 & 3.065 ( 71 ) &       321200\\ \hline
16 &  5.36 & 3.06 ( 11 ) &        187232\\ \hline
16 &  5.50 & 2.950 ( 67 ) &       256050\\ \hline
16 &  5.53 & 2.953 ( 83 ) &       191286\\ \hline
16 &  5.70 & 2.851 ( 63 ) &       235080\\ \hline
16 &  5.81 & 2.728 ( 70 ) &       186009\\ \hline
16 &  6.12 & 2.490 ( 65 ) &       183776\\ \hline
16 &  6.47 & 2.387 ( 44 ) &       273140\\ \hline
16 &  6.50 & 2.259 ( 57 ) &       286230\\ \hline
16 &  6.76 & 2.165 ( 66 ) &       136446\\ \hline
16 &  7.11 & 1.997 ( 40 ) &       244791\\ \hline
16   &  7.82 & 1.697 ( 47 ) &     136365\\ \hline
16 &  8.00 & 1.725 ( 50 ) &       141570\\ \hline
16 &  8.45 & 1.520 ( 24 ) &       368201\\ \hline
16 &  9.00 & 1.379 ( 41 ) &       114100\\ \hline
16 &  9.42 & 1.229 ( 28 ) &       147603\\ \hline
16  &  11.15 & 0.964 ( 26 ) &     72562 \\ \hline
16 &  12.00 & 0.836 ( 17 ) &      118000\\ \hline
16 &  13.85 & 0.700 ( 19 ) &      70801 \\ \hline
16 &  15.23 & 0.566 ( 13 ) &      80752 \\ \hline
16 &  16.00 & 0.5431 ( 89 ) &     116000\\ \hline
16 &  17.55 & 0.4785 ( 100 ) &    83657 \\ \hline
16 &  18.00 & 0.469 ( 13 ) &      40000 \\ \hline
16 &  20.00 & 0.3902 ( 86 ) &     44700 \\ \hline
16 &  20.13 & 0.4135 ( 77 ) &     79816 \\ \hline
16 &  50.00 & 0.1327 ( 16 ) &     60900 \\ \hline
16 &  99.00 & 0.06326 ( 68 ) &    28050 \\ \hline
\end{tabular}
\hspace{2.5cm}
\begin{tabular}{||c|c|c|c||}
\hline
$L/a$ & $\beta$ & $\bar{g}^{2}_{\rm latt}$   & \# of traj. \\ \hline
20 &       5.70 &  $        2.940 (         58 )   $ &1892896\\ \hline
20 &       6.00 &  $        2.663 (         67 )   $ &443775\\ \hline
20 &       6.50 &  $        2.401 (         54 )   $ &301480\\ \hline
20 &       7.00 &  $        2.108 (         45 )   $ &430782\\ \hline
20 &       8.00 &  $        1.725 (         38 )   $ &295316\\ \hline
20 &       9.00 &  $        1.450 (         33 )   $ &322420\\ \hline
20 &      10.00 &  $        1.187 (         24 )   $ &263795\\ \hline
20 &      12.00 &  $        0.8437 (         17 ) $ &258279\\ \hline
20 &      14.00 &  $        0.6450 (         14 ) $ &125942\\ \hline
20 &      16.00 &  $        0.5545 (         11 ) $ &155575\\ \hline
20 &      18.00 &  $        0.4565 (         80 ) $ &148488\\ \hline
20 &      20.00 &  $        0.4064 (         76 ) $ &123948\\ \hline
20 &      50.00 &  $        0.1352 (         12 ) $ &147168\\ \hline
\end{tabular}
\label{tab:raw_data_TPL_coupling}
\caption{ Raw data for the renormalised coupling in the TPL scheme.}
\end{table}
%


\clearpage

\bibliographystyle{apsrev.bst}
\bibliography{refs} 

\end{document}